\documentclass[11pt]{article}

\pdfoutput=1 

\usepackage[dvipsnames]{xcolor}
\usepackage{amssymb,amsmath,amscd}
\usepackage{epsfig}
\usepackage{epstopdf}
\usepackage{latexsym}
\usepackage{graphicx}
\usepackage{booktabs}
\usepackage{bbm}
\usepackage[margin=15pt,small]{caption}
\usepackage{subcaption}
\usepackage{enumitem}
\usepackage{float}
\usepackage[toc]{appendix}
\usepackage[numbers,sort&compress]{natbib}
\usepackage{tikz}
\usepackage[labelfont=bf]{caption}
\usepackage[export]{adjustbox}
\usepackage{nicematrix}
\usepackage{braket,cellspace}
\usepackage{mathrsfs}
\usepackage{braket}
\usepackage{chngpage}
\usepackage{soul}
\usetikzlibrary{positioning} 
\usepackage{xfrac} 
\usepackage{mathrsfs}
\definecolor{darkspringgreen}{rgb}{0.09, 0.45, 0.27} 
\usepackage[
      colorlinks=true,
      linkcolor=darkspringgreen,  
      urlcolor=darkspringgreen,    
      filecolor=darkspringgreen,     
      citecolor=darkspringgreen,
      linktocpage=true,
      pdfstartview=FitV,
      bookmarksopen=true     
      ]{hyperref} 

\numberwithin{equation}{section} 
\allowdisplaybreaks 
\setlist{nolistsep} 
\let\oldbibliography\thebibliography
\renewcommand{\thebibliography}[1]{\oldbibliography{#1}
\setlength{\itemsep}{0pt}} 

\newcommand{\scri}{\mathscr{I}}

\begin{document}  

\begin{titlepage}
\begin{center} 

\vspace*{20mm}

{\LARGE \bf
A Matrix Model for Flat Space Quantum Gravity
}

\vspace*{15mm}
 
{\bf Arjun Kar, Lampros Lamprou, Charles Marteau, and Felipe Rosso  }
\vskip 2mm
Department of Physics and Astronomy 
\\
University of British Columbia \\
Vancouver, BC V6T 1Z1, Canada  

\bigskip
\tt{arjunkar@phas.ubc.ca, llamprou90@gmail.com \\
marteau.charles.75@gmail.com, feliperosso6@gmail.com}  

\end{center}

\bigskip

\begin{abstract}
\noindent \noindent We take a step towards the non-perturbative description of a two-dimensional  dilaton-gravity theory which has a vanishing cosmological constant and contains black holes. This is done in terms of a double-scaled Hermitian random matrix model which non-perturbatively computes the partition function for the asymptotic Bondi Hamiltonian. To arrive at this connection we first construct the gauge-invariant asymptotic phase space of the theory and determine the relevant asymptotic boundary conditions, compute the classical S-matrix and, finally, shed light on the interpretation of the Euclidean path integral defined in previous works. We then construct a matrix model that matches the topological expansion of the latter, to all orders. This allows us to compute the fine-grained Bondi spectrum and other late time observables and to construct asymptotic Hilbert spaces. We further study aspects of the semi-classical dynamics of the finite cut-off theory coupled to probe matter and find evidence of maximally chaotic behavior in out-of-time-order correlators. We conclude with a strategy for constructing the non-perturbative S-matrix for our model coupled to probe matter and comment on the treatment of black holes in celestial holography.

\end{abstract}

\vfill
\end{titlepage}

\setcounter{tocdepth}{2}

\tableofcontents


\setlength{\parskip}{0.75 em} 

\newpage 

\section{Introduction}
\label{sec:intro}
The quantum dynamics of gravity in a Universe with vanishing cosmological constant is of indisputable theoretical interest; gravitational phenomena confined to distance scales shorter than the Hubble radius can be formulated as scattering processes in an asymptotically Minkowski spacetime. The pursuit of a theory of ``flat'' quantum gravity has led to the construction of pillars of modern theoretical physics, from perturbative superstring theory to the BFSS matrix model \cite{Banks:1996vh}. The depth of these insights, however, came along with a practical shortcoming: The lack of a description that was both non-perturbatively complete and computationally tractable.

Recent breakthroughs in AdS/CFT accentuated the capacity of two-dimensional dilaton-gravity theories to bring non-perturbative explorations of quantum gravity within reach \cite{Maldacena:2016upp,Jensen:2016pah,Engelsoy:2016xyb}. The insight that fueled this success was that the Euclidean path integral of such AdS$_2$ theories is secretly computing expectation values of certain observables in a dual ensemble of random matrices \cite{Saad:2019lba}. Encouraged by this progress, we embark on an attempt to import this technology to the study of quantum gravity with vanishing cosmological constant. The goal of this paper is to present a Hermitian matrix ensemble that provides a (non-unique) non-perturbative description of flat quantum gravity observables in two dimensions, via a dictionary that we begin to systematically study. Our investigation is a continuation of the works \cite{Cangemi:1992bj,Afshar:2019axx,Afshar:2019tvp,Godet:2020xpk,Godet:2021cdl} and the main results were concisely summarized in our companion paper \cite{Kar:2022vqy}.

The protagonist of this paper is a dilaton-gravity theory we dub Cangemi--Jackiw (CJ) gravity, after the authors of \cite{Cangemi:1992bj}. It is a close cousin of the pure gravity sector of the famous CGHS model \cite{Callan:1992rs}, with the distinction that the ``vacuum energy'' is promoted to a dynamical variable via the inclusion of a topological gauge field\footnote{For this reason, this model has also sometimes been called $\widehat{\mathrm{CGHS}}$ in the literature \cite{Callan:1992rs,Afshar:2019axx}.} ---a deformation with deep quantum mechanical consequences. Since quantum gravity in flat space is, ultimately, the theory of an S-matrix, our first task is to characterize the phase space by prescribing asymptotic boundary conditions (Section \ref{SectionLorentz}). A long list of recent works \cite{Strominger:2013jfa,Raclariu:2021zjz,Pasterski:2021rjz} established a detailed manual for this procedure in higher dimensions. Our adaptation to CJ gravity starts with freezing both the value and the derivative of the dilaton near null infinity ${\scri}^\pm$. This enables us to define a preferred Bondi coordinate frame in the asymptotic region and then restrict metric fluctuations to those preserving the asymptotic Bondi form. The resulting degrees of freedom consist of a pair of boundary modes controlling the induced metric on the equipotential dilaton curves\footnote{Unlike JT gravity in AdS where the induced metric is fixed on an asymptotic cut-off surface $\Phi=1/\epsilon$, the induced metric on such cut-offs in CJ gravity is allowed to fluctuate. The rate of the physical boundary clock is instead set by the normal derivative of the dilaton along the boundary which we indeed keep fixed.  } and the gauge field Wilson line, and they form the coadjoint orbit of an infinite dimensional asymptotic symmetry group. Their classical dynamics is organized in a boundary action, first derived in \cite{Afshar:2019axx},\footnote{See also \cite{Godet:2021cdl} for an alternative derivation more aligned with our approach in this paper.} whose equations of motion enforce the conservation of the two asymptotic charges of CJ gravity in Bondi time. Our Lorentzian analysis is systematic and physically justifies a number of choices made in these previous works.

We quantize the theory in Section \ref{sec:Euclidean} by constructing its Euclidean path integral. This is defined by Wick rotating the asymptotic Bondi time $u \to i \tau$ and integrating over all configurations with $\tau \sim \tau + \beta$\footnote{Once again, in contrast to JT in AdS, this does not correspond to fixing the proper length of a regulated Euclidean boundary, but to fixing the periodicity of the orbit of the asymptotic Killing vector $\varepsilon^{\mu\nu}\partial_\mu \Phi \partial_\nu $ which is held fixed by our boundary conditions.} that respect the asymptotic conditions of Section \ref{SectionLorentz}, order by order in the topological expansion. The path integral serves a double purpose: (a) It defines the partition function of the Bondi Hamiltonian allowing us to study its quantum mechanical spectrum, and (b) it selects a natural inner product between asymptotic states and hence defines a Hilbert space at $\scri^\pm$. Its computation was previously performed in \cite{Afshar:2019tvp,Godet:2020xpk} and in Appendix \ref{zapp:3} we put it on more solid ground by leveraging the BF formulation of CJ gravity \cite{Cangemi:1992bj} to rigorously determine the path integral measure. The Euclidean path integral can be generalized to include an arbitrary number $n$ of asymptotic boundaries with Bondi periods $\beta_i$, $i=1,\dots, n$ and the output for the connected contributions is
\begin{equation} 
Z(\beta)\simeq 
\frac{2(2\pi)^4}{\pi(\gamma\beta)^2}
e^{S_0} \ ,
\qquad
Z(\beta_1,\beta_2)\simeq 
\frac{1}{\gamma(\beta_1+\beta_2)}\ ,
\qquad
Z(\beta_1,\dots,\beta_n)\simeq 0\ , \label{Zsummary}
\end{equation}
where $\gamma$ is an arbitrary unit of inverse length introduced for dimensional consistency and $S_0 \in \mathbb{R}_+$ is the parameter that multiplies the Euler characteristic in the action (\ref{eq:22}). The symbol $\simeq$ means the equality holds to all orders in $e^{-S_0}$ perturbation theory.

These path integrals strongly hint at an interpretation as connected correlation functions of Bondi partition functions $\langle \prod_{i=1}^n {\rm Tr}\,e^{-\beta_i H} \rangle$, where the Bondi Hamiltonian is a random variable selected from a matrix ensemble, that we call a \textit{celestial matrix model} \cite{Kar:2022vqy}, in accordance with the JT discussion \cite{Saad:2019lba}. In Section \ref{sec:Euclidean}, we confirm this intuition by establishing that in the ensemble of $N\times N$ Hermitian random matrices with a probability measure determined by the simple potential
\begin{equation}\label{eq:pot}
    V(M) = -M^2 +\frac{1}{4}M^4\ ,
\end{equation}
there exists a single trace operator $\mathbb{O}(\beta)$ whose correlation functions match (\ref{Zsummary}) in $1/N$ perturbation theory, in a specific double scaling limit we make precise in the main text and Appendix \ref{zapp:4}
\begin{equation}
Z(\beta_1,\dots,\beta_n)\simeq \langle \mathbb{O}(\beta_1)\dots \mathbb{O}(\beta_n) \rangle_c\ . \label{dictionarysummary}
\end{equation}
The relevant observable is uniquely determined and reads
\begin{equation}
    \mathbb{O}(\beta) =
    \int_{-\infty}^{+\infty}
    \frac{dp}{\sqrt{\gamma}}
    {\rm Tr}\,e^{-\beta (\bar{M}^2+p^2)}\ , \label{matrixobssummary}
\end{equation}
suggesting that our Bondi Hamiltonian contains contributions from two decoupled sectors: A matrix model sector and a free non-relativistic particle in one dimension. While the factorized form of (\ref{matrixobssummary}) is indicative of an analogous factorization at the level of the semi-classical CJ action, no such decomposition was identified, rendering the physical meaning of (\ref{matrixobssummary}) somewhat mysterious at this point.\footnote{At the technical level, we do find several clues that connect the free particle contribution in (\ref{matrixobssummary}) to a central extension of the Poincar\'e algebra, the Maxwell algebra (\ref{eq:Maxwell}), which plays a crucial role in the construction of CJ gravity \cite{Cangemi:1992bj,Grumiller:2020elf}.}

Despite its obvious appeal, the true power of the match between the CJ gravity partition function and the celestial matrix model in $e^{-S_0}$ perturbation theory (\ref{dictionarysummary}) comes from unblocking the road to the study of non-perturbative quantum gravity effects $\mathcal{O}(e^{-e^{S_0}})$. While no technique for calculating such non-perturbative contributions from the action formulation of CJ gravity currently exists, they can be explicitly and rigorously computed using our celestial matrix model, which is non-perturbatively stable.\footnote{The stability is immediately implied by the asymptotic behavior $V(\pm \infty)=+\infty$ of the matrix potential (\ref{eq:pot}). This is in contrast to the Hermitian random matrix model constructed in \cite{Saad:2019lba} for JT gravity, which is non-perturbatively unstable. However, a stable completion of JT gravity was given in \cite{Johnson:2019eik,Johnson:2021tnl}, using complex instead of Hermitian matrices.} In this work we shall assume the (non-unique) non-perturbative completion of CJ gravity provided by the matrix model, which practically speaking means replacing the symbol~$\simeq$ in (\ref{dictionarysummary}) by a strict equality.\footnote{Non-perturbative effects shift the higher point correlation functions $Z(\beta_1,\dots \beta_n)$ for $n\geq 3$ to a positive non-zero value. As discussed in \cite{Godet:2021cdl}, this is a necessary requirement for the density of states to be positive definite for all energies.} This allows us to study a plethora of observables in CJ quantum gravity that are well beyond the reach of perturbation theory. In Section \ref{sec:Nonperturbative} we characterize the fine grained spectrum of the Bondi Hamiltonian (Figure \ref{fig:5}), the late time behavior of the spectral form factor (Figure \ref{fig:6}) and low temperature thermodynamics through the quenched free energy (Figure \ref{fig:9}). All of these observables exhibit unusual features when compared to analogous results for JT gravity \cite{Johnson:2020exp,Johnson:2021zuo,Johnson:2021rsh}, which we attribute to the fact that CJ gravity is a theory of flat instead of AdS space-times.

In Section \ref{sec:Scrambling}, we initiate the study of CJ theory coupled to probe matter, by including insertions of shockwaves at the asymptotic boundary. The dynamics can be mapped to that of particle emission from a pair of massive oppositely-charged particles in a constant electric field. We show that two-point correlators, computed in the geodesic approximation in a state containing a shock exhibit Lyapunov behaviour, implying that the microscopic CJ dual is a fast scrambler. This analysis relies crucially on the aforementioned proper identification of the holographic time.

We end this paper with a discussion of the CJ gravity S-matrix, in Section \ref{sec:Discussion}. One of the goals of the celestial holography program \cite{Raclariu:2021zjz,Pasterski:2021rjz}  is to generate a non-perturbative S-matrix, using ``holographic'' degrees of freedom. A question of particular interest is how to incorporate black holes in this framework and describe scattering in their presence. Our low-dimensional toy model provides beginnings of answers to these questions. Inspired by the recent non-perturbative treatment of holographic correlators in JT gravity, we make a proposal for incorporating non-perturbative effects in the S-matrix about the two-sided black hole, using our celestial matrix model. Finally we discuss some implications of our findings in CJ gravity for higher-dimensional theories.

\section{Two-dimensional Lorentzian asymptotically flat gravity}
\label{SectionLorentz}

Our goal in this Section is to discuss the classical CJ theory and the flat space gravitational dynamics it describes. We carefully explain the physical rationale behind the quite subtle asymptotic ``fall-off'' conditions we impose on the CJ configuration space and characterize the phase space of asymptotic states of the Lorentzian theory, the asymptotic charges and the matching conditions needed for the classical S-matrix. We finally show that the dynamics of CJ gravity can be recast in terms of an effective action for a pair of functions on $\scri^+\cup \scri^-$ describing a dynamical asymptotic frame, whose equations of motion enforce the conservation of Bondi energy. This action is analogous to the ``boundary particle'' action of JT gravity and will be of central importance when studying the partition function of the Bondi Hamiltonian in Sections \ref{sec:Euclidean} and \ref{sec:Nonperturbative}.

\subsection{Classical CJ gravity}

The action defining the classical theory reads:
\begin{equation}\label{eq:CJaction}
I_{\rm CJ}=I_{\rm topological}+I_{\rm bulk}+I_\partial\ ,
\end{equation}
with the various terms given explicitly by
\begin{equation}\label{eq:CJterms}
\begin{aligned}
I_{\rm topological} & =
\frac{S_0}{4\pi}\bigg[ 
\int_{\mathcal{M}}d^2x\sqrt{-g}R
+2\int_{\partial \mathcal{M}}
dy\sqrt{-h}K
\bigg]=S_0\chi(\mathcal{M})\ , \\[4pt]
I_{\rm bulk} & =
\frac{1}{2}
\int_{\mathcal{M}} d^2x\sqrt{-g}\left(
\Phi R+2\Psi-2\Psi \varepsilon^{\mu \nu}\partial_\mu A_\nu
\right)
\ , \\[4pt]
I_\partial & =
\frac{1}{2}
\int_{\partial \mathcal{M}}
dy\sqrt{-h}\big[ 
2\Phi K-n^\mu \nabla_\mu \Phi
\big]\ ,
\end{aligned}
\end{equation}
where $\mathcal{M}$ is the two-dimensional manifold where the theory is defined, $\Phi$ a dilaton field, $\Psi$ a ``vacuum energy'' scalar field and $A_\mu$ a gauge field. The boundary terms in $I_{\partial {\cal M}}$ are included to ensure a well-posed variational problem and a finite on-shell action, as we will discuss shortly. This action is closely related to the more widely known CGHS model \cite{Callan:1992rs}. The important distinction is that the ``vacuum energy'' $\Psi$ is a coupling constant in CGHS but is promoted to a dynamical field in CJ gravity via the inclusion of a dynamical gauge field $A_\mu$. It may also be thought of as the flat space limit of Jackiw-Teitelboim dilaton gravity coupled to the simplest possible matter sector: A topological gauge field $A_\mu$ whose canonical momentum is the vacuum energy field $\Psi$.

While the field content of CJ gravity might seem intricate, the model has a very natural origin as the ``simplest" theory of two-dimensional flat space gravity analogous to JT gravity. Two of the most important features of JT gravity which make it solvable are the linear dilaton in the action, which enforces constant curvature solutions $R=-2/\ell^2$, and its equivalence to a BF gauge theory with gauge group ${\rm SL}(2,\mathbb{R})$. Having the BF description not only means the theory is topological, i.e. contains only boundary degrees of freedom, but also allows for a rigorous computation of its path integral \cite{Witten:1991we}.

A naive way of obtaining a flat space theory is to take the limit $\ell\rightarrow \infty$ of JT gravity. In this limit the algebra $\mathfrak{sl}(2,\mathbb{R})$ becomes the Poincar\'e algebra $\mathfrak{iso}(2)$. This is problematic for the BF formulation given that the Poincar\'e algebra does not admit a ad-invariant non-degenerate bilinear form, which is necessary for the usual formulation of the gauge theory \cite{Grumiller:2020elf}. Cangemi and Jackiw solved this problem by insteading considering the minimal extension of the Poincar\'e algebra that admits such a bilinear form, the \textit{Maxwell algebra}:\footnote{See \cite{Grumiller:2020elf} for a more detailed discussion as to why this is the minimal extension of the Poincar\'e algebra useful for the BF theory definition. The generators $K$ and $Q$ are rescaled compared to the ones defined in Appendix \ref{zapp:3}, see (\ref{eq:142}).}
\begin{equation}\label{eq:Maxwell}
[K,P_\pm]=\mp P_\pm\ ,
\qquad \qquad
[P_+,P_-]=Q\ .
\end{equation}
Apart from the two null translations $P_\pm$ and the boost $K$ there is a central extension $Q$ which ultimately ensures the existence of the non-degenerate bilinear form. As shown by Cangemi and Jackiw \cite{Cangemi:1992bj} the bulk action in (\ref{eq:CJaction}) is equivalent to a BF theory whose gauge group is generated by the Maxwell algebra (see Appendix \ref{zapp:3}). We shall see the algebra (\ref{eq:Maxwell}) make its appearance in several instances when studying various features of CJ gravity. In particular, there are some hints that point towards a connection between the central extension $Q$ and the free particle (\ref{matrixobssummary}) that appears in the non-perturbative completion provided by the celestial matrix model.

\subsection*{CJ gravity on-shell} 

Our first step is to understand the general classical solution of CJ gravity. The equations of motion are:
\begin{equation}\label{eq:eom}
\delta I_{\rm CJ}=0
\qquad \Longrightarrow \qquad
\begin{cases}
\begin{aligned}
R & =0\ , \hspace{37 pt} \nabla_\mu \nabla_\nu \Phi  = \Psi g_{\mu \nu}\ , \\
\,\,
\varepsilon^{\mu \nu}\partial_\mu A_\nu & =1 \ ,
\hspace{55 pt}
\partial_\mu \Psi=0 \ ,
\end{aligned}
\end{cases}
\end{equation}
The $R=0$ equation implies that the spacetime metric does not fluctuate on-shell so we can choose a coordinate frame in which it simply reads $ds^2= dx^+ dx^-$. This fixes completely the diffeormorphism gauge, up to overall translations and boosts which are, of course, global symmetries of the solution. The remaining equations can be solved exactly about this background and give
\begin{equation}\label{vacsolution}
\begin{aligned}
ds^2 & = dx^+dx^-\ , \hspace{100pt} 
\Phi= \frac{\Lambda}{2}(x^+ -c_1)(x^- -c_2) +\phi_h\ , \\
A & =\frac{1}{2}x^+ dx^- +dg(x^+,x^-)\ ,
\hspace{33pt} 
\Psi = \Lambda\ ,
\end{aligned}
\end{equation}
where $x^\pm$ are ordinary null coordinates in Minkowski and $(\Lambda, c_1,c_2,\phi_h)$ are integration constants.\footnote{Notice that unlike the metric, where we completely fixed the gauge, we chose to keep the gauge parameter $g$ explicit in $A$ since the large gauge transformations of $A$ will be important in the discussion of boundary conditions below.} Two of them can be set to zero by a redefinition of the origin of our Minkowski coordinate system, $c_1=c_2=0$. In contrast, $(\Lambda, \phi_h)$ label distinct physical configurations, hence they are dynamical variables. Furthermore, the physical information contained in the gauge parameter $g(x^+,x^-)$ is captured by the Wilson line degree of freedom $W= \int A_\mu dx^\mu$ which depends on the difference $g_{LR}$ of the gauge parameter in the two asymptotic regions. Lastly, the phase space is completed by the inclusion of the ``gravitational Wilson line'', measuring the relative time between the two asymptotic regions of spacetime, as we will see momentarily.

\subsection*{Gravitational physics in the dilaton frame}

The spacetime metric of CJ gravity is fixed and the gravitational dynamics is encoded  entirely in the profile of the dilaton $\Phi$; this situation is familiar from Jackiw-Teitelboim gravity in AdS. In order to understand the physics of the solution (\ref{vacsolution}) in the gravitational language, it is useful to switch to a coordinate frame that trivializes the dilaton profile and absorbs all dynamical variables in the metric. This dilaton frame will, of course, depend on the specific on-shell configuration and hence it will itself be dynamical. In particular, we will use the value of $\Phi$ as a radial coordinate $r\propto \Phi(x^+,x^-)$ and we will define a clock using the vector field\footnote{Where $\varepsilon^{\mu\nu} = g^{\mu\kappa}g^{\nu\lambda}\sqrt{-g}\,\epsilon_{\kappa\lambda} = \frac{\epsilon^{\mu\nu}}{\sqrt{-g}}$ with $\epsilon_{+-}=-\epsilon_{-+}=1=\epsilon^{+-}=-\epsilon^{-+}$ and $\epsilon_{++}=\epsilon_{--}=0=\epsilon^{++}=\epsilon^{--}$.} $\zeta^\mu \propto \varepsilon^{\mu\nu} \partial_\nu \Phi$ which is, in fact, a timelike Killing vector of the solution (\ref{vacsolution}) that preserves the equipotential $\Phi$ curves. 

In defining this dilaton-dependent coordinate frame, we are confronted by the absence of a natural scale in flat space quantum gravity, i.e. a dimensionful parameter that can serve as the unit of distance when converting the dimensionless dilaton value to a spacetime coordinate. We are, therefore, required to introduce a new dimensionful parameter $\gamma$ which will be taken to have dimensions of $\text{(length)}^{-1}$. We may now introduce coordinates  $\left(r(\Phi),t(\Phi)\right)$ with dimensions of length defined as\footnote{These relations do not quite determine a coordinate change uniquely on their own, but they do when combined with an asymptotic falloff condition on the metric.  We explain this point in more detail under \eqref{generaldilatonframe}.}
\begin{align}
    r&=\frac{1}{\gamma}\Phi\, , 
    \qquad \qquad \qquad 
    \frac{\partial}{\partial t} = 
    \frac{1}{\gamma}
    \varepsilon^{\mu\nu}\partial_\nu \Phi \, \partial_\mu\ . \label{dilframe}
\end{align}

\begin{figure}
    \centering
    \includegraphics[scale=0.55]{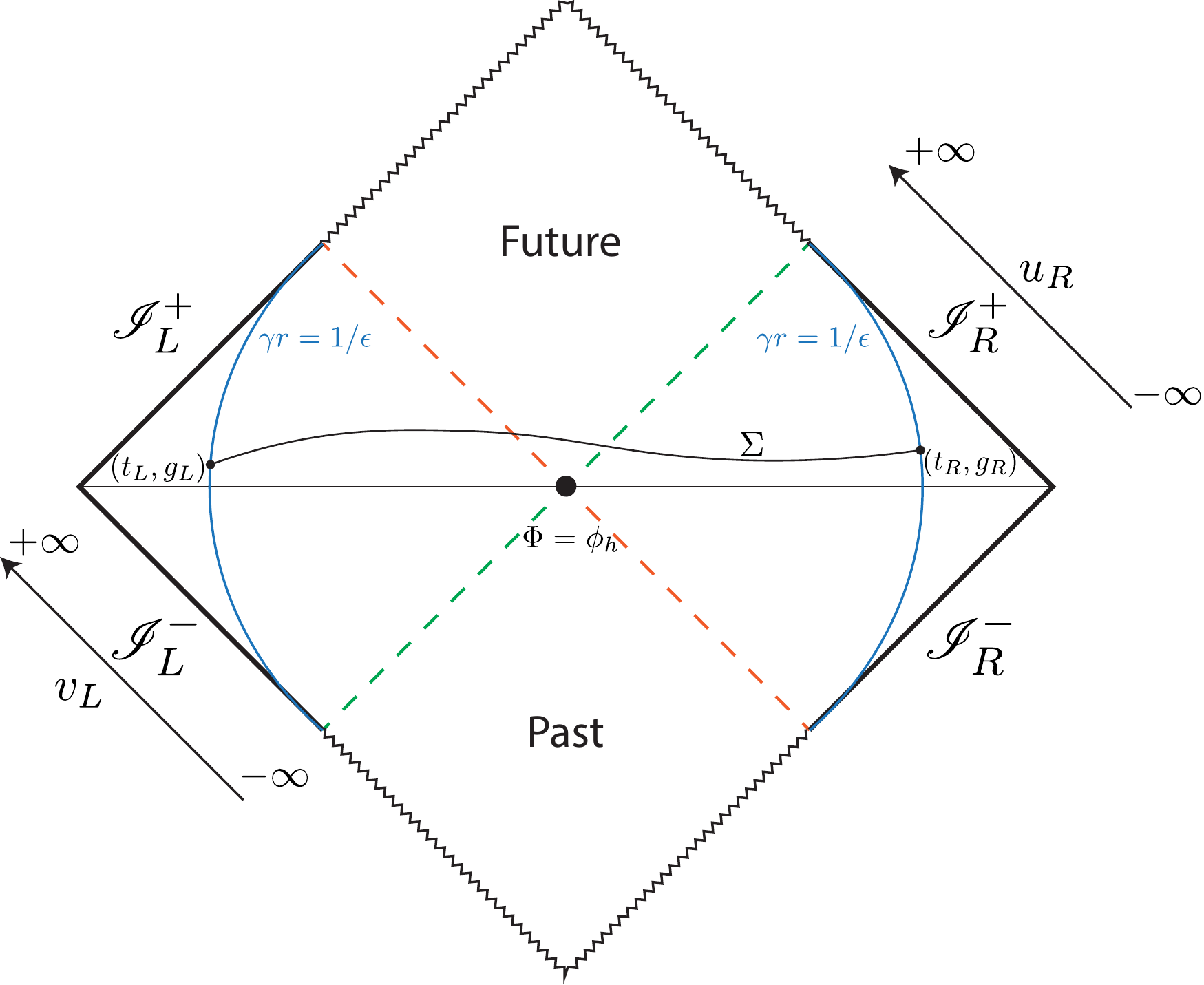}
    \caption{Penrose diagram of the Lorentzian solution: The causal structure is that of a two-sided black hole. The physical boundary is the locus of $\Phi=1/\varepsilon$, the horizon is where $\Phi=\phi_h$ while the singularity is where the dilaton diverges negatively.}
    \label{fig:Penrose}
\end{figure}

We define the physical asymptotic boundary to be at $\gamma r=\Phi=\frac{1}{\epsilon} \to +\infty$. This consists of two disconnected components, depending on whether $x^\pm >0$ or $x^\pm<0$. The explicit form of $(r,t)$ in terms of the global Minkowski frame $(x^+,x^-)$ is obtained by substituting (\ref{vacsolution}) in (\ref{dilframe}) and the solution consists of four patches (Figure~\ref{fig:Penrose})
\begin{align}\label{dilframeexplicit}
    x_R^\pm &=  \sqrt{\frac{2\gamma}{\Lambda}
    \Big(r-\frac{\phi_h}{\gamma}\Big)} \,\,e^{\pm \frac{\Lambda (t-t_R)}{\gamma}} \quad r>\frac{\phi_h}{\gamma}\ ,
    \qquad \quad 
    x_L^\pm = - \sqrt{\frac{2\gamma}{\Lambda}\Big(r-\frac{\phi_h}{\gamma}\Big)} \,\,e^{\pm \frac{\Lambda (t-t_L)}{\gamma}}\quad r>\frac{\phi_h}{\gamma}\ ,\nonumber
    \\[5pt]
    x_F^\pm &= \pm\sqrt{\frac{2\gamma}{\Lambda}\Big(\frac{\phi_h}{\gamma}-r\Big)} \,\,e^{\pm \frac{\Lambda (t-t_F)}{\gamma}}\quad r<\frac{\phi_h}{\gamma}\ ,
    \qquad\, x_P^\pm = \mp \sqrt{\frac{2\gamma}{\Lambda}\Big(\frac{\phi_h}{\gamma}-r\Big)} \,\,e^{\pm \frac{\Lambda (t-t_P)}{\gamma}}\quad r<\frac{\phi_h}{\gamma}\ .
\end{align}
Here, $(t_{R},t_L,t_F,t_P)$ are integration constants parametrizing our freedom to choose the initial moment of the dilaton clock $t(\Phi)$ in each patch. Not all of them are physical. In fact, the past and future patches do not contain parts of the asymptotic boundary $r=+\infty$ so the choice of clock there is pure gauge and $t_P,t_F$ can be fixed to be $t_P=t_F=t_R$. Moreover, the Lorentz boost symmetry can be used to set $t_R$ to $0$ at the expense of shifting $t_L$ to $t_{LR}=t_L-t_R$. The difference $t_{LR}$ between the Left and Right patch clocks, however, is physical and cannot be removed, since it implies a difference between the clocks on the two components of the asymptotic boundary $\scri^+$ (Figure~\ref{fig:Penrose}). Different values of $t_{LR}$ label different solutions, hence $t_{LR}$ is a phase space variable.

The ``new'' integration constant, $t_{LR}$,\footnote{This is not actually a new coordinate of the phase space but, instead, one that we neglected in our description of the classical solution (\ref{vacsolution}) above. The covariant phase space of the theory is formulated on a Cauchy slice. Selecting such a slice amounts to picking the time at which the slice meets the asymptotic boundary, up to trivial diffeomorphisms. The choice of boundary time is a phase space coordinate. It is this degree of freedom, discussed in detail in \cite{Harlow:2018tqv,Harlow:2019yfa}, that is being re-introduced in this step explicitly.} encodes the classical dynamics of our asymptotic coordinate frame; it is the 2D analog of the 4D supertranslations, i.e. the local choice of a $\scri^+$ coordinate along the celestial sphere, as becomes evident when our metric is expressed in Bondi gauge (\ref{LorentzianSol}). It is a non-trivial phase space coordinate because shifts of its value are generated by the large diffeomorphism $\delta_\xi g_{\mu\nu}= \nabla_{(\mu}\xi_{\nu)}$ with $\xi \overset{r\to \infty}{\longrightarrow} \partial_t $. 

In our dilaton frame, the general metric and gauge field solutions read:
\begin{equation}
ds^2 = -H(r)dt^2+\frac{dr^2}{H(r)}\ ,
\hspace{30pt}
A = -
\frac{\gamma}{2\Lambda}H(r)
\Big(dt -
\frac{dr}{H(r)}\Big) + dg(r,t)\ ,\label{vacsolutionrt}
\end{equation}
where $H(r)=(r-\phi_h/\gamma)\frac{2\Lambda}{\gamma}$. We have, therefore, traded the dynamics of the dilaton for dynamics of the metric, since  all configurations have $\Phi= \gamma\,r$ while three of the on-shell dynamical variables $\Lambda, \phi_h, t_{LR}$ appear in the metric;\footnote{The parameter $t_{LR}$ is of course implicit.} the fourth controls the gauge field configuration through the asymptotic behavior of the gauge parameter $g$, as we will see shortly. 

The asymptotic boundary is at $r\to \infty$ and, by analogy to JT gravity, we will associate the surface $r=\frac{\Phi}{\gamma}\to -\infty$ with a ``singularity''. The metric has a horizon at $r=\frac{\phi_h}{\gamma}$. This is a Rindler horizon originating from the fact that the dilaton solution defines an accelerated frame and its location is determined by our dynamical variable $\phi_h$. The acceleration and, by extension, the temperature of this Rindler metric can be obtained from the near horizon metric at $r-\frac{\phi_h}{\gamma} = \epsilon \ll 1$ by Wick rotating time $t=i\tau$ and finding the period $\tau \sim \tau +\beta$ for which the time-compactified Euclidean metric is smooth. This yields:
\begin{equation}\label{temperature}
    \beta = \frac{2\pi \gamma}{\Lambda}\ .
\end{equation}
The Rindler acceleration is determined by our other dynamical variable, the ``vacuum energy'' field $\Psi= \Lambda$. The ability to accommodate different temperatures is one of the reasons for considering the CJ gravity model instead of CGHS, in which the vacuum energy $\Lambda$ is a fixed parameter of the theory and, by extension, so is the Rindler temperature.

\paragraph{The physical meaning of the temperature:} In the discussion of the temperature of the solution above, we made a subtle but important departure from the familiar AdS$_2$ case. The temperature (\ref{temperature}) does not correspond to the proper length of the asymptotic boundary of the Euclidean solution which is, instead, a fluctuating variable in CJ gravity; $\beta$ is referring to the periodicity of the Euclidean orbit of the Killing vector $\zeta^\mu = \epsilon^{\mu\nu} \partial_\nu \Phi$. It is this vector field that will be held fixed asymptotically via our choice of boundary conditions in the CJ path integral. The periodicity of its Euclidean flow will be identified with the temperature of the dual matrix model. The choice of the asymptotic clock in (\ref{dilframe}) ought to be understood, therefore, as an element of the holographic dictionary of CJ quantum gravity which is distinct from its AdS$_2$ cousin.

\subsection{Asymptotic boundary conditions}
\label{subsec:abc}

From the analysis of the previous section, pure classical CJ gravity might seem of limited interest. 
What we are interested in is quantizing it and, ultimately, coupling it to matter in order to study the quantum gravitational contribution to flat space S-matrices. Both of these tasks require us to understand the allowed off-shell configuration space of pure CJ gravity. In the quantum theory, this selects the appropriate functional domain for the path integration, while in the presence of matter, it constrains the allowed matter configurations involved in scattering processes. Characterizing the off-shell configuration space amounts to taming the behavior of fields as they approach null infinity, i.e. imposing asymptotic boundary conditions.

We will do so by following the logic of the higher dimensional discussions and introducing Bondi coordinates $u_{L,R}=t_{L,R}-\frac{\gamma}{2\Lambda} \log \left(r_{L,R}-\frac{\phi_h}{\gamma}\right) $ that allow us to reach the null asymptotic boundary $\scri^+$ by taking $r\to \infty$ and $t\to \infty$ while keeping $u$ finite. In taking this limit for the pair Rindler wedges of Figure~\ref{fig:Penrose} there is a subtlety: We would like to approach the Left and Right null asymptotic boundaries while keeping fixed the relative synchronization of the Left and Right clocks, i.e. the gravitational Wilson line $t_{LR}$. This requires us that to simultaneously approach either $\scri^-_L\cup \scri^+_R$ or $\scri^+_L\cup \scri^-_R$. Since the analysis is identical for these two cases, we will refer to both of them abstractly as $\scri$ below.

Given a preferred asymptotic Bondi frame, the gravitational configuration space will be constructed by studying the expansion of the allowed configurations in $1/r$. The general classical solution in both the Left and Right patch of Figure~\ref{fig:Penrose} in Bondi gauge has the form:
\begin{equation}
\begin{aligned}
ds^2&= 
    -
    \frac{2\Lambda}{\gamma}
    \Big(r-\frac{\phi_h}{\gamma}\Big)du^2 - 2dudr \ ,
    \hspace{60 pt} \Phi= \gamma \,r\ ,\\
    A&=-
    \Big(r-\frac{\phi_h}{\gamma}\Big)du + dg(r,u)\ , \hspace{73 pt} \Psi=\Lambda\ .
\end{aligned}
\label{LorentzianSol}
\end{equation}
Of course, the coordinates we use in the bulk of a dynamical spacetime are immaterial, due to diffeomorphism invariance. The physical meaning of the dilaton-Bondi coordinates $(u,r)$ used above lies in selecting an \emph{asymptotic frame near} $\scri$ \emph{with reference to $\Phi$}, in which the $1/r$ expansion is to be performed. The coefficients of the leading terms in this power series $(\Lambda, \Lambda\phi_h)$, as well as the ``relative Bondi clock'' variable $t_{LR}$ between the two components of $\scri$, e.g. $\scri^-_{L}$ and $\scri^+_R$ or $\scri^+_{L}$ and $\scri^-_R$, and the Wilson line associated to $A$ are then physical degrees of freedom. As we shall see, $\Lambda \phi_h$ measures the Bondi energy. The new feature in our 2D setup, as compared to higher dimensions, is that the asymptotic expansion of the metric is to be performed in the value of the dilaton $r=\frac{\Phi}{\gamma}$. 

Let us also fix the gauge for $A$ and discuss the physical data contained in the function $g(r,u)$. All gauge transformations that die off at infinity $g(r\to \infty,u)\to 0$ are trivial. Moreover, as we will see shortly, the asymptotic value of $A_u$ must be fixed in order for the action variation to vanish on-shell. These two statements imply that $A_u$ is completely fixed up to a gauge choice. In the dilaton-Bondi frame, we choose to express all solutions in the gauge:
\begin{align}
    \partial_r A_u &= -1\,\, , \qquad A_u(r=\gamma^{-1}\phi_h) = 0\ . \label{gaugechoice}
\end{align}
The first condition is chosen so that the equation of motion (\ref{eq:eom}) is satisfied for $\partial_u A_r=0$. The second condition ensures that $A$ will be well-defined at the horizon after Wick rotating $u\to i \tau$, since $r=\gamma^{-1}\phi_h$ is a fixed point of $\frac{\partial}{\partial\tau}$ and, thus, $d\tau$ is singular there. This partially fixes the gauge by restricting $g(u,r)$ in (\ref{LorentzianSol}) to a radial function $g(r)$. The latter, in turn, contains one constant's worth of physical information because its value at the two components $(L,R)$ of the asymptotic boundary controls the spatial Wilson line:
\begin{equation}\label{Wilsonline} 
    W=\int_{\partial {\cal M}_L}^{\partial {\cal M}_R}A_\mu d\ell^\mu = g(r)\Big|^{\partial {\cal M}_R}_{\partial {\cal M}_L} =g_{LR}\ .  
\end{equation}

\subsection*{Asymptotics of off-shell configuration space} 
A necessary requirement of a candidate off-shell configuration domain is that the CJ action be differentiable in it, with vanishing variation $\delta I_{\rm CJ}$ about a solution (\ref{LorentzianSol}). In other words, boundary terms in $\delta I_{\rm CJ}$ must cancel. In our 2D dilaton gravity theory, the variation of the action is computed in Appendix \ref{zapp:1} and gives
\begin{equation}\label{boundaryterms}
\begin{aligned}
\delta I_{\rm CJ}&=({\rm EOM})+
\frac{1}{2}
\int_{\partial \mathcal{M}} dy\sqrt{-h}
\Big[
(2K-n^\alpha \nabla_\alpha)\delta \Phi
-(2 \Psi n_\mu \varepsilon^{\mu \nu}) \delta A_\nu + \\[4pt]
& \hspace{140pt} -
\frac{1}{2}
\Big(
\gamma_{\mu \nu}
(n^\alpha\nabla_\alpha\Phi)+
n_\mu(\delta^\alpha_\nu+\gamma_{\,\,\,\nu}^{\alpha})\nabla_\alpha \Phi 
\Big)\delta g^{\mu \nu}
\Big]\ ,
\end{aligned}
\end{equation}
where the first term is a bulk contribution which vanishes when the equations of motion (\ref{eq:eom}) are satisfied. For the boundary terms to cancel, we need to impose asymptotic boundary conditions on the allowed \emph{off-shell} fluctuations of the fields $(\Phi,g_{\mu \nu},A_\mu)$.

The appropriate boundary conditions can be identified by referring to the on-shell configuration space discussed in the previous section. Suppose we start with a particular solution (\ref{LorentzianSol}) and then consider an arbitrary off-shell fluctuation about it. A natural requirement is that all such fluctuations have the same asymptotic expansion as the classical solutions, in the dilaton-Bondi frame. We may thus constrain the off-shell fields to asymptote to
\begin{align}
    d\bar{s}^2 & = -2(P(u) r +T(u) ) du^2 -2dudr+\mathcal{O}(1/r)\ , \label{metricfalloff} \\[4pt]
    \bar{\Phi}({r},{u}) & = \gamma \left(B(u) r + C(u) \right)+\mathcal{O}(1/r)\ , \label{dilatonfalloff}\\[4pt]
    \bar{A}(r,u) & = (M(u) r+N(u)) du + dg(r,u)+\mathcal{O}(1/r)\ , \label{Afalloff}\\[4pt]
    \bar{\Psi}(r,u) & = L(u) + \mathcal{O}(1/r)\ , 
\end{align}
where $P(u),T(u),B(u),C(u),M(u), N(u)$ and $L(u)$ are arbitrary functions on $\scri$. These expressions should be interpreted as restricting the formal power series of the off-shell fields in $r$ to be of the form (\ref{LorentzianSol}), while allowing the \emph{coefficients} to be arbitrary functions on $\scri$. As we now explain, the asymptotic conditions above are not yet an accurate characterization of the physical configuration space, for two reasons:

\paragraph{(a) Diffeomorphism redundancy:} The first subtlety is that the fluctuations of the dilaton and the metric are not physically distinct. Diffeomorphism invariance implies that only the \emph{relative} frame between the dilaton and the metric in the asymptotic region is physically meaningful. To understand this better, let us, once again, choose an \emph{asymptotic} coordinate system $(\bar{r}(r,u),\bar{u}(r,u))$ that trivializes the general off-shell dilaton configuration (\ref{dilatonfalloff}) via
\begin{align}
    \bar{r}(r,u) &=
    \lim_{r\rightarrow \infty}
    \frac{\bar{\Phi}(r,u)}{\gamma}= B(u) r+C(u)\ , \nonumber\\
    \frac{\partial}{\partial \bar{u}(r,u)} &= 
    \lim_{r\rightarrow \infty}
    \frac{1 }{\gamma}\varepsilon^{\mu\nu} \partial_\nu \bar{\Phi} \partial_\mu = B(u)\partial_u - (B'(u)r +C'(u)) \partial_r\ , \label{generaldilatonframe}
\end{align}
Setting $B(u)=\frac{1}{f'(u)}$ and $C(u) = \frac{h(u)}{f'(u)}$ for convenience and without loss of generality, the solution to this system of equations reads:
\begin{align}
    \bar{u}&= f(u)\, ,
    \qquad  \qquad
    \bar{r}= \frac{r+h(u)}{f'(u)}\ . \label{asymptoticsymmetry0}
\end{align}
Note that in writing this solution, we have implicitly required that the asymptotic $\bar{u}$ coordinate should be ``integrable'' in the sense that
\begin{equation}
    \lim_{r\to\infty} \frac{\partial \bar{u}}{\partial r} = 0\ .
\label{eq:dubar-dr-freedom}
\end{equation}
In other words, we require that the asymptotic $\bar{u}$ coordinate should become arbitrarily close to a pure reparametrization $f(u)$ as we send $r \to \infty$.
It may seem like this is an ad hoc requirement, but in fact it follows directly from the asymptotic boundary conditions, as we now explain.
In order to ensure that there are no $\mathcal{O}(r^2)$ terms generated in the asymptotic metric \eqref{metricfalloff} by writing e.g. \eqref{LorentzianSol} (originally in $(\bar{r},\bar{u})$ coordinates) in the general dilaton frame \eqref{generaldilatonframe}, we find that there are only two possibilities for the function $\frac{\partial \bar{u}}{\partial r}$.
It is either $\mathcal{O}(1/r)$, in which case it is vanishing asymptotically as we have required, or it is a very specific nontrivial function of $u$ that is determined in terms of $f(u)$ and $h(u)$.
The precise form is not important; what is crucial is that, if we choose this nontrivial form, we generate unavoidable $\mathcal{O}(r^2)$ terms in the $du$ component of the asymptotic gauge field \eqref{Afalloff}, violating the asymptotic boundary condition.
So, the metric and gauge field falloff conditions \eqref{metricfalloff} and \eqref{Afalloff} are strong enough to imply \eqref{eq:dubar-dr-freedom} and thus \eqref{asymptoticsymmetry0} uniquely.
In summary, there is a one-to-one correspondence between asymptotic coordinate transformations \eqref{asymptoticsymmetry0} and general asymptotic dilaton frames \eqref{generaldilatonframe} that preserve the boundary conditions, which allows us to always work with a fixed asymptotic dilaton profile even off-shell by merely compensating with an asymptotic symmetry transformation \eqref{asymptoticsymmetry0}.
A more specific version of this argument also applies to our discussion of the classical solutions in the on-shell dilaton frame \eqref{dilframe}, but with the $\mathcal{O}(1/r)$ terms above replaced by exactly zero.

As we have just argued, the coordinate transformation \eqref{generaldilatonframe} is an asymptotic symmetry of the metric, i.e. it preserves the form of its asymptotic expansion in $r$ while changing the coefficients. 
Starting from the on-shell configuration $(P_0,T_0)=\frac{\Lambda}{\gamma}(1,-\frac{\phi_h}{\gamma})$ in the $(\bar{u},\bar{r})$ system, for example, and applying the transformation (\ref{asymptoticsymmetry0}) leads to the class of metrics
\begin{align}
    ds^2 & = -2\left(P(u)r+T(u)\right) du^2 -2dudr\ , \label{offshellmetric}
\end{align}
with the explicit expressions 
\begin{align}
   P(u) &=P_0 f'(u)-\frac{f''(u)}{f'(u)}\ ,\nonumber\\
    T(u)&= T_0 f'(u)^2 +h(u)\left(P_0 f'(u)-\frac{f''(u)}{f'(u)}\right) +h'(u)\ . \label{offshellcharges} 
\end{align}
Asymptotic fluctuations of the dilaton can, thus, be entirely re-absorbed in the metric coefficients $(P(u),T(u))$ of (\ref{metricfalloff}) by the action of a diffeomorphism. We may thus always work in the frame where the dilaton at $r\to \infty$ is trivial, by choosing $B(u) =1$ and $C(u)=0$ in (\ref{dilatonfalloff}) and introducing two functions $(f(u),h(u))$ on each component of $\scri$, controlling the fluctuations of $P(u)$ and $T(u)$ about a classical solution via (\ref{offshellcharges}). It is straightforward to show that with these asymptotic conditions, the boundary terms that depend on $\delta g_{\mu\nu}$ and $\delta \Phi$ in the action variation (\ref{boundaryterms}) vanish.

In summary, both the dilaton value $\Phi$ and its normal derivative $\varepsilon^{\mu\nu}\partial_\nu \Phi$, defining radial and timelike coordinates near $\scri$ respectively, can be \emph{fixed} in the asymptotic region for all allowed configurations, absorbing all physical fluctuations in the asymptotic metric. The latter, in turn, form the orbit of an asymptotic symmetry group of two-dimensional flat space (\ref{asymptoticsymmetry0}).

\paragraph{(b) Gauge field boundary condition:} The second subtlety comes from the requirement that the $\delta A_\mu$ term in (\ref{boundaryterms}) vanishes. For this to be true, we need the component of $A_\mu$ parallel to the boundary to be fixed asymptotically which, as explained above, fixes the entire function $A_u(r,u)$ up to trivial gauge transformations. In our treatment, the asymptotic region is foliated by $\gamma r =\Phi=\frac{1}{\epsilon}$ timelike slices, with the asymptotic boundary at $\epsilon\to 0$, and the time evolution along them is generated by the Killing vector $\partial_u =\zeta^\mu\partial_\mu$ where $\zeta^\mu \to\gamma^{-1}\varepsilon^{\mu\nu}\partial_\nu \Phi$. The pullback of $A_\mu$ on the boundary is then $\zeta^\mu A_\mu|_{r\to \infty}$, and for consistency with the on-shell solution (up to trivial gauge transformations) we fix it to
\begin{equation}
   \varepsilon^{\mu\nu}\partial_\nu \Phi \,A_\mu + \Phi = \phi_h +\mathcal{O}(1/r)\ , \label{asymptoticgauge}
\end{equation}
where $\phi_h$ is the horizon value of the dilaton of the ``seed'' classical solution (\ref{LorentzianSol}). Condition (\ref{asymptoticgauge}) should be interpreted as an asymptotic condition, equating each term in the $r$ expansion of the left hand side  that does not vanish at $r=\infty$, to the corresponding coefficient of the right. This fixes $M(u)=-1$ and $N(u)=\frac{\phi_h}{\gamma}$ in (\ref{Afalloff}).

This condition has an important consequence: The transformations (\ref{asymptoticsymmetry0}) generating the orbit of off-shell fluctuations about the solution (\ref{LorentzianSol}) are further restricted to those that preserve the boundary condition (\ref{asymptoticgauge}). In particular, since $N(u)$ in (\ref{Afalloff}) transforms as $\bar{N}(\bar{u}) \to N(u)$:
\begin{equation}
    N(u)  = f'(u)\bar{N}(f(u))- h(u)\ ,
\end{equation}
in order to preserve the $N=\frac{\phi_h}{\gamma}$ boundary condition we need to accompany the asymptotic symmetry transformation (\ref{asymptoticsymmetry0}) by a gauge transformation $g(u)$ satisfying:
\begin{equation}\label{constraint}
    g'(u) = h(u) +(1-f'(u)) \frac{\phi_h}{\gamma}\ .
\end{equation}
The allowed asymptotic symmetries then take the form:
\begin{align}
    \bar{u}&= f(u)\, ,
    \qquad \quad
    \bar{r}= \frac{r-\frac{\phi_h}{\gamma}(1-f'(u))+g'(u)}{f'(u)}\, ,\qquad \quad \bar{A}=A+dg(u)\ . \label{asymptoticsymmetry}
\end{align}

The physical fluctuating degrees of freedom near $\scri$ are, therefore, a pair of functions $f(u), g(u)$ describing the dynamics of the asymptotic frame and controlling the two leading coefficients of the metric $P(u),T(u)$ via
\begin{align}
   P(u) &=  P_0f'(u)-\frac{f''(u)}{f'(u)}\ ,
   \nonumber\\
   T(u)&=\left(g'(u)-\frac{\phi_h}{\gamma}\right)\left(P_0 f'(u)-\frac{f''(u)}{f'(u)}\right) +g''(u)\ , \label{offshellcharges2}   
\end{align}
where we have used the relation $P_0 \phi_h + T_0 \gamma = 0$ following the ``parent'' solution \eqref{LorentzianSol}. 
Specifically, we have $P_0=\frac{\Lambda}{\gamma}$, $T_0 = -\frac{\Lambda \phi_h}{\gamma^2}$, and $\phi_h$ characterizing the ``parent'' classical solution of reference.

Note that all classical solutions (\ref{LorentzianSol}) of the theory are contained in this functional space: Starting from a ``parent'' solution of the form (\ref{LorentzianSol}), the value of the vacuum energy field can be changed via the asymptotic symmetry $f(u)= \frac{\Lambda'}{\Lambda}u,\, g(u)=0$ while the dilaton location of the horizon $\phi_h$ can be changed by $f(u)=u, \, g(u)= (\phi_h -\phi'_h) u/\gamma$.

\subsection{Boundary action and asymptotic phase space}\label{sec:bdy-action-phase-space}

Summarizing our discussion up to this point, all physical configurations of CJ gravity are restricted asymptotically to behave as
\begin{align}
     ds^2 & = -2(P(u) r +T(u) ) du^2 -2dudr+\mathcal{O}(1/r)\ , \label{metricfalloff2} \\[4pt]
    A(r,u) & = -\Big(r-\frac{\phi_h}{\gamma}\Big) du + d\chi(r)+\mathcal{O}(1/r)\ , \label{Afalloff2}\\[4pt]
     \Phi({r},{u}) & = \gamma \,r
     +\mathcal{O}(1/r)\ ,\label{dilatonfalloff2}
\end{align}
near each component $\scri_{L,R}$ of $\scri$, with $P(u), T(u)$ given by (\ref{offshellcharges2}). The equations of motion in this parametrization of the configuration space become:
\begin{equation}\label{PTeom}
\frac{d}{du}P_{L,R}(u) = 0\ ,
\qquad \qquad
\frac{d}{du}T_{L,R}(u) = 0\ ,
\end{equation}
on each component of $\scri$, whose solutions determine the two pairs of functions $(f_{L,R},g_{L,R})$. Indeed, equations~(\ref{PTeom}) directly follow from the CJ action (\ref{eq:CJaction}) evaluated for the general configuration (\ref{metricfalloff2}-\ref{dilatonfalloff2}) about a given semi-classical background (\ref{LorentzianSol}) with some chosen values of $(\Lambda, \phi_h)$ and their physical role is enforcing conservation of the asymptotic charges of CJ gravity, as we will see below. The action reduces to a sum of two boundary terms
\begin{align}
    I_{\rm CJ}&= \gamma\int_{\scri^-_L}dv\, T_L(v) +\gamma\int_{\scri^+_R}du\, T_R(u) \nonumber\\
    &=\gamma\int_{\scri^-_L}dv\, \left[ 
    \Big(g'_L(v) -\frac{\phi_h}{\gamma}\Big) P_L(v)+ g_L''(v)\right]+\gamma \int_{\scri^+_R}du\, \left[\Big(g'_R(u) -\frac{\phi_h}{\gamma}\Big) P_R(u)+g_R''(u)\right]\ , \label{Lboundaryaction}
\end{align}
with $P_{L,R}(u) = P_0 f'_{L,R} - \frac{f''_{L,R}}{f'_{L,R}}$. Note that this action explicitly depends on the parameters of a given saddle, treating all other configurations as excitations about it. Nevertheless, classically, the choice is immaterial since all other solutions can be accessed by appropriate configurations of $(f(u),g(u))$, as explained below equation~(\ref{offshellcharges2}). Quantum mechanically, on the other hand, the horizon value of the dilaton becomes superselected and describes the ground state entropy of the model, as we will see in Section~\ref{sec:Euclidean}.

It is straightforward to show that each boundary term is invariant under the action of the following infinitesimal symmetry transformations:
\begin{equation}\label{symmetries}
\delta f_{L,R}= \epsilon_{L,R}^0 +\epsilon_{L,R}^1 e^{P_0 f_{L,R}(u_{L,R})}\ ,
\qquad \qquad
\delta g_{L,R} = \sigma_{L,R}^0 +\sigma_{L,R}^{1} e^{-P_0 f_{L,R}(u_{L,R})}\ ,
\end{equation}
with $u_L = v$ and $u_R =u$, which form a Maxwell algebra (\ref{eq:Maxwell}) of symmetries of the classical solutions. The physical phase space of CJ gravity is, then, globally constrained by the requirement of invariance under a ``global'' Maxwell algebra; this is simply the symmetry of global Minkowski space, combined with global gauge transformations. This is enforced by quotienting the saddle points of (\ref{Lboundaryaction}) for the two asymptotic boundaries by the embedded Maxwell subalgebra generated by the coupled infinitesimal variations:
\begin{equation}
\epsilon^0_L = -\epsilon^0_R\ ,
\qquad
\sigma^0_L = -\sigma^0_R\ ,
\qquad
\epsilon^1_L = -\epsilon^1_R\ ,
\qquad
\sigma^1_L = -\sigma^1_R\ .
\label{quotient}
\end{equation}
These particular relations can be deduced by pulling back the global Killing vectors of flat space to the left and right Rindler patches whose dynamics in CJ gravity are described by the combined boundary action \eqref{Lboundaryaction}.

\subsection*{CJ energy}
The energy of CJ gravity can be obtained by performing a Legendre transformation of the action (\ref{Lboundaryaction}). Since the latter involves second derivatives of $f(u)$, however, this transform is non-standard. One needs to first integrate in an extra degree of freedom $\varphi$ via a Lagrange multiplier in order to convert it into a first order action and then the Hamiltonian can be straightforwardly constructed. 

We focus on a single boundary action since the procedure is identical. By defining the field variable $F(u)=P_0^{-1}e^{-P_0 f(u)}$ the action becomes:\footnote{We have dropped the $g''(u)$ term, which is a total derivative that will not affect our discussion of the classical energy or phase space.}
\begin{equation}
    I_{\rm CJ} = \gamma\int du \,\Big(\frac{\phi_h}{\gamma}-g'(u)\Big) \frac{F''(u)}{F'(u)}\ .
\end{equation}
We then introduce in a new additional field $\varphi(u)$ and a Lagrange multiplier $\lambda(u)$ as
\begin{equation}\label{newaction}
    I_{\rm CJ}= \gamma\int du\left[  \Big(g'(u)-\frac{\phi_h}{\gamma}\Big)\varphi'(u) +\lambda(u)\left(  F'(u) +e^{-\varphi(u)}\right)
    \right]\ ,
\end{equation}
yielding a new action whose equations of motion are equivalent to (\ref{PTeom}). Given the canonical momenta
\begin{equation}
    \pi_\varphi = \gamma\Big(g'(u)-\frac{\phi_h}{\gamma}\Big)\ ,
    \qquad \quad
    \pi_F=\gamma\lambda(u) \ ,
    \qquad \quad
    \pi_g =\gamma \varphi'(u) \ ,
    \qquad \quad
    \pi_\lambda =0\ , \label{canonicalmomenta}
\end{equation} 
the Hamiltonian for the new but classically equivalent action (\ref{newaction}) becomes
\begin{align}
    H&= \pi_\varphi \varphi' +\pi_F F' +\pi_g g' -\mathcal{L}=\gamma^{-1}\left(\pi_\varphi +\phi_h\right) \pi_g -\pi_F e^{-\varphi}\ . \label{hamiltonian}
\end{align}
The solutions to the equations of motion are given by
\begin{equation}\label{eq:sol2}
\begin{aligned}
\lambda_*(u)&=p_-\ , 
\hspace{101pt}
\varphi_\ast(u) = \frac{\Lambda}{\gamma}\, u +\varphi_0\ ,\\[4pt]
F_*(u) &= \frac{\gamma}{\Lambda}e^{-\varphi_*(u)} + \frac{p_+}{\gamma}\, \ , 
\hspace{45pt}
g_*(u)= -\frac{\gamma^2 p_-}{\Lambda^2} e^{-\varphi_*(u)} + \frac{\delta\phi_h}{\gamma} \,u +g_0 
\ ,
\end{aligned}
\end{equation}
where $(\delta \phi_h,\Lambda,g_0,\varphi_0,p_\pm)$ are integration constants. Evaluating (\ref{hamiltonian}) on shell, the energy becomes
\begin{equation}\label{onshellenergy}
    E_*= \frac{\Lambda}{\gamma}\, \delta \phi_h \ . 
\end{equation}

The energy depends only on two of the integration constants: $\Lambda$ which measures the vacuum energy of the solution (and temperature of the geometry)\footnote{This can be seen by recalling that $P(u) = -F''(u)/F'(u)$.} and $\delta \phi_h$ which controls the value of the dilaton at the horizon relative to $\phi_h$.\footnote{This follows from the fact that $\delta \phi_h$ is the linear term in the $g(u)$ configuration which by virtue of (\ref{asymptoticsymmetry}) shifts the horizon from $\phi_h$ to $\phi_h-\delta \phi_h$ and, similarly, by virtue of (\ref{offshellcharges2}) it changes the value of $T_0$ appearing in the  metric (\ref{LorentzianSol}).} The remaining four variables that leave the energy invariant parametrize the image of a given $(\Lambda,\delta \phi_h)$ solution under the Maxwell symmetry.
As we will see shortly, $\varphi_0$ is conjugate to a piece of the Bondi-time translation/Minkowski boost charge, $p_-$ and $p_+$ are related to the two null translation charges (which are conjugate to each other as their bracket yields the central element), and $g_0$ is precisely conjugate to the global part of the $\mathbb{R}-$gauge transformation charge, which is the central element of the Maxwell algebra. Notice that by definition $F\geq 0$, implying $p_+\geq 0 $ for consistency.

The total energy of the Lorentzian CJ theory, therefore, is the sum of two copies of (\ref{onshellenergy}) for the left and right systems, respectively. It is important to keep in mind, however, that only configurations invariant under the embedded Maxwell symmetry \eqref{quotient} are physical states of the two-sided system, due to the gauging condition on the global isometries of flat space. The total on-shell energy, therefore, is simply $E_*^{\rm Total} = 2(\Lambda/\gamma) \delta \phi_h$.

\subsection*{Phase space}
The phase space of a single copy of our boundary action is parametrized by the conjugate pairs $(\varphi,\pi_\varphi)$, $(F,\pi_F)$, $(g,\pi_g)$ and $(\lambda, \pi_\lambda)$. This is superficially an eight-dimensional phase space, however, the definition of the canonical momenta imply two constraints that project it down to six dimensions: (a) $\pi_\lambda =0$ and (b) $\pi_F =\gamma\lambda$.
The pullback of the standard symplectic form to the hypersurface defined by these equations yields a non-degenerate reduced symplectic form (actually the standard one for $(\varphi,g,F)$ coordinates) that, combined with the Hamiltonian \eqref{hamiltonian}, defines the classical mechanical theory of one copy of our boundary particle.

We can get a different coordinate system for this reduced one-boundary phase space in terms of the integration constants that appear in \eqref{eq:sol2}: $(\delta \phi_h,\Lambda, g_0,\varphi_0, p_+, p_-)$. By enforcing the standard equal time Poisson bracket relations $\{\varphi,\pi_\varphi  \}=1$, $\{g,\pi_g\}=1$, and $\{F,\pi_F\}=1$ on (\ref{eq:sol2}), we obtain the relations:
\begin{align}
  \{  \varphi_0, \delta \phi_h\} & = 1\, ,   
  \qquad
  \{g_0, \Lambda \} = 1 \ , \qquad \{p_+,p_-\} = 1\ , \label{commutators}
\end{align}
with all other brackets vanishing.
Note that the first and second brackets in (\ref{commutators}) imply that the on-shell Hamiltonian (\ref{onshellenergy}) generates joint Bondi time translations $u\to u+u_0$ and gauge zero mode translations $g \to g+g_0$. The last bracket, in turn, encodes the interaction of in-going and out-going shocks ---the generators of asymptotic null translations. The Maxwell charges written as functions of the phase space variables are\footnote{In the boost charge $K$, the $\phi_h$ quantity is the constant appearing in the action \eqref{newaction}.  We emphasize that it is not $\delta \phi_h$, which is instead a coordinate on the phase space we described in \eqref{eq:sol2}.}
\begin{equation}
    Q=\pi_g/\gamma^2
    \ ,\qquad 
    P_-=\pi_F/\gamma
    \ ,\qquad
    P_+=\pi_g F/\gamma-e^{-\varphi}
    \ ,\qquad
    K=-\pi_\varphi+F\pi_F-\phi_h
    \ ,
\end{equation}
and one finds they satisfy the Maxwell algebra (\ref{eq:Maxwell}) with respect to the standard Poisson bracket described just above \eqref{commutators}:
\begin{equation}
   \{K,P_\pm\}=\mp P_\pm\ ,
   \qquad  \{P_+,P_-\}=Q\ .
\end{equation}
The Hamiltonian is simply the Casimir of the algebra
\begin{equation}
   H=\gamma (P_+ P_- - KQ)\ .
\end{equation}
Evaluated on the solution \eqref{eq:sol2} the charges become
\begin{equation}
     Q=\Lambda/\gamma^2
     \ ,\qquad P_-=p_-
     \ ,\qquad P_+=\Lambda p_+/\gamma^2
     \ ,\qquad 
K= p_+p_- - \delta \phi_h\ .
\end{equation}

Each of the two boundary particle actions comes with its own six-dimensional phase space but the quotient by the embedded Maxwell group (\ref{quotient}) further reduces this 12-dimensional extended space to a four-dimensional global phase space, spanned by the variables that are invariant under the gauged group:\footnote{The gauge reduction is a rather delicate procedure.  The four conditions supplied by the vanishing embedded Maxwell charges \eqref{quotient} define an eight-dimensional hypersurface in the 12-dimensional extended phase space, and the pullback of the symplectic form to this hypersurface has four zero modes.
These zero modes are essentially eigenvector fields of the symplectic form, and we may take any simultaneous level set of these vector fields to be the physical phase space.
Such a level set is described by four conditions on the eight remaining phase space variables.
In our case, a convenient choice is $\varphi_{0,L}+\varphi_{0,R} = g_{0,L}+g_{0,R} = 0$ and $p_{-,L}+p_{-,R}=p_{+,L}+p_{+,R}=0$.
For consistency, we need the zero mode eigenvector fields of the symplectic form to be orthogonal to this hypersurface, and we have checked that this is the case. 
}
\begin{equation}
    \{\varphi_{LR},\delta \phi_h\} = 1 
    \ ,
    \qquad \qquad
    \{g_{LR}, \Lambda\}=1\ ,
\end{equation}
where these coordinates are related to the two-boundary variables by $\varphi_{LR}= \varphi_{0,L}- \varphi_{0,R}$, $g_{LR}= g_{0,L}-g_{0,R}$, $\delta \phi_h = \delta \phi_{h,R} = -\delta \phi_{h,L}$ and $\Lambda = \Lambda_R = -\Lambda_L$. 
These are, of course, the gravitational and gauge Wilson lines discussed earlier in this Section around Figure~\ref{fig:Penrose}, along with their canonical conjugates: the value of the dilaton at the horizon $\Phi_h = \phi_h + \delta\phi_h$ and vacuum energy $\Lambda$, respectively.\footnote{We have chosen to write the phase space in terms of the deviation $\delta \phi_h$ from the background value $\phi_h$ appearing in the boundary action.  We could instead have used the absolute $\Phi_h$ coordinate which would have the same brackets and is the coordinate we earlier called $\phi_h$ in the bulk analysis around \eqref{dilframe}.  The status of $\Lambda$ is similar but the transformation is multiplicative instead of additive (see the discussion under \eqref{offshellcharges}), so by setting the background value to 1 we ensure the $\Lambda$ appearing in these brackets is literally the same quantity as in \eqref{dilframe}. } The latter are the two asymptotic charges of the CJ gravity, appearing as coefficients of the leading terms in the asymptotic expansion of the metric in the dilaton-Bondi coordinates (\ref{LorentzianSol}). The equations of motion (\ref{PTeom}) should, therefore, be understood as enforcing the conservation of those charges in the pure gravitational theory, on-shell.

\subsection{Classical S-matrix}

Pure CJ gravity has a four-dimensional classical phase space of future asymptotic states, denoted here as $\mathcal{P}^+$, equipped with a symplectic form implied by the above Poisson brackets. In local Darboux coordinates the latter formally reads
\begin{equation}
    \omega^+ = dx_1^+ \wedge dp_1^+ + dx_2^+ \wedge dp_2^+ ,
\end{equation}
where $(x^+_1,x^+_2,p^+_1,p^+_2)=(\varphi_{LR}^+, g_{LR}^+,\delta\phi_h^+,\Lambda^+)$. This structure was obtained by expressing the general solution in the outgoing Bondi-dilaton coordinate frame. By similar analysis using an ingoing Bondi-dilaton frame, we can obtain an ingoing phase space $\mathcal{P}^-$ and symplectic form $\omega^-$ with a similar formal expression
\begin{equation}
    \omega^- = dx_1^- \wedge dp_1^- + dx_2^- \wedge dp_2^- .
\end{equation}

The classical S-matrix of CJ gravity is a diffeomorphism $S: \mathcal{P}^- \to \mathcal{P}^+ $ mapping between the two symplectic manifolds
\begin{equation}
    x_i^+(x_1^-,x_2^-,p_1^-,p_2^-) , \quad p_i^+(x_1^-,x_2^-,p_1^-,p_2^-) .
\label{eq:classical-s-matrix}
\end{equation}
By ``classical unitarity'', i.e. conservation of phase space volumes, the function $S$ must preserve the symplectic structure under pullback:
\begin{equation}
    S^*(\omega^+) = \omega^- \ ,
\label{eq:classical-unitarity}
\end{equation}
meaning that the mapping $S$ is a canonical transformation.

Obtaining this map for pure CJ gravity is fairly simple practically because the general solution is known. We start with the general solution (\ref{LorentzianSol}), expressed in the advanced dilaton-Bondi frame $(u,r)$ and parametrized in terms of the future phase space variables $(\Lambda^+,\phi_h^+)$, and re-express it in the retarded coordinate system $(v,r)$, in order to read off the coefficients $(\Lambda^-,\phi_h^-)$ of the past asymptotic expansion of the metric, as functions of the future phase space variables. The map between the two asymptotic Bondi frames is determined up to a Maxwell symmetry transformation (\ref{symmetries}). Nevertheless, this ambiguity, by definition, affects neither the asymptotic form of the metric on $\scri^-$ nor the corresponding coefficients which can be straightforwardly obtained to find:
\begin{equation}
    \Lambda^-=\Lambda^+ \ ,
    \qquad \qquad \quad \phi_h^-= \phi_h^+\ . \label{chargeconservation}
\end{equation}

This has an obvious interpretation: CJ scattering preserves the asymptotic charges. In fact, given that the phase space is 4-dimensional, charge conservation (\ref{chargeconservation}) contains the entire physical content of the S-matrix. There is only one additional subtlety, due to the constraints: The translation frame chosen for $\scri^+$, which is parametrized by $p_\pm$ in the solution (\ref{eq:sol2}) and controls the global translation charges $P_\pm$ of the theory, ought to also be matched with the corresponding frame for $\scri^-$. This matching must imply that a global Minkowski translation affects both $\scri^\pm$ in the expected way. This is achieved by supplementing the prescription for the S-matrix with a matching condition:
\begin{equation}
    P_\pm \big|_{\scri^+} = P_\mp \big|_{\scri^-}.
\end{equation}

This matching of the charges tells us how the future and past momenta are related but it does not tell us how the S-matrix acts on positions. This is because there exists a remaining freedom in how we choose to relate the origins of the past and future Wilson lines that is not a property of the solution but has to be imposed externally. The general form of this matching is constrained by the fact that the full S-matrix needs to be a canonical transformation and we will now prove that such a transformation maps quantum mechanically to a phase that can be reabsorbed by a trivial rotation of the future or past momentum basis.

The most general canonical transformation can be specified by a generating function $G(x_1^-,x_2^-,p_1^+,p_2^+)$ which obeys
 \begin{equation}
     p_i^- = \frac{\partial G}{\partial x_i^-}
     \ , \qquad  \qquad
     x_i^+ = \frac{\partial G}{\partial p_i^+} \ .
     \label{eq:canonical-transformation-G}
 \end{equation}
 Using the constraints $p_i^- = p_i^+$, which are the abstract form of \eqref{chargeconservation}, the first relation above implies $p_i^+ = \frac{\partial G}{\partial x_i^-}$ which after integration gives
 \begin{equation}
     G = p_1^+ x_1^- + p_2^+ x_2^- + k(p_1^+,p_2^+) \ ,
 \end{equation}
 where $k(p_1^+,p_2^+)$ is an arbitrary function of the momenta. From the second formula in \eqref{eq:canonical-transformation-G}, it is subject to the following relation:
 \begin{equation}
     x_i^+ = x_i^- + \frac{\partial k}{\partial p_i^+} \ .
 \end{equation}
 The canonical transformation \eqref{eq:classical-s-matrix} can therefore be written as
 \begin{equation}
     x_i^+ = x_i^- + \frac{\partial k}{\partial p_i^-} 
     \ , \qquad  \qquad
     p_i^+ = p_i^- \ .
     \label{SmatrixClassical}
 \end{equation}
In writing the above, we have used the relation $p_i^- = p_i^+$ to treat the function $k$ as a function of the $p_i^-$ rather than the $p_i^+$. Now let's consider the canonical quantization of $\mathcal{P}^\pm$ so that the S-matrix becomes an operator. The quantum version of \eqref{SmatrixClassical} is
\begin{equation}
     \partial_{p_i^+} = \partial_{p_i^-} +
     i\frac{\partial k}{\partial p_i^-} 
     \ , \qquad \qquad 
     p_i^+ = p_i^- \ ,
     \label{SmatrixQuantum}
 \end{equation}
which, up to global phases, corresponds to a $k$-dependent diagonal phase acting on the momentum basis. We conclude that, whatever the matching condition $k$ is, the S-matrix is the following trivial redefinition of the momenta
\begin{equation}
    \ket{p_1,p_2}^+=\exp[-i k(p_1,p_2)]\ket{p_1,p_2}^-.
\end{equation}

\paragraph{Including backreacting probe matter:} The CJ S-matrix becomes interesting once we include matter and study the gravitational contribution to its scattering. The effect of probe matter can be captured by the inclusion of localized sources for the Maxwell charges of the gravitational theory, along $\scri^+$ and $\scri^-$, respectively. As discussed in the beginning of Section \ref{subsec:abc}, this allows us to explore the much larger asymptotic configuration space (\ref{metricfalloff2}-\ref{dilatonfalloff2}), parametrized by arbitrary variables $(f(u),g(u))$ on each boundary.

To determine the S-matrix, we need a way of relating $\scri^+$ and $\scri^-$ data. The answer is to utilize the smoothness of all bulk configurations near $i^0$: The $\scri^+$ charges as $u\to -\infty$ match the $\scri^-$ charges at $v\to +\infty$\footnote{The left $(L)$ and right $(R)$ versions of this equation are actually not independent. The vanishing of the total Maxwell charges implies that one follows from the other.}
\begin{equation}
    Q_i(u\to -\infty)\big|_{\scri^+_{L,R}}= \bar{Q}_i(v\to +\infty) \big|_{\scri^-_{L,R}} \ ,
    \qquad \forall 
    \,\,\, i=1,2,3,4\ , \label{matching}
\end{equation}
where $Q_i= (P_+,P_-, K, Q)$ and $\bar{Q}_i= (P_-,P_+,K,Q)$. These conditions would entirely determine the gravitational scattering of matter in CJ gravity, if it was not for the presence of horizons in the geometry which requires additional data, related to the horizon states. We postpone the discussion of this issue until the next section, where we explain how a ``horizon'' Hilbert space can be built in the quantum theory using the Euclidean path integral for the boundary action we derived in this section. This Euclidean preparation of horizon states, combined with the Lorentian segments that produce the asymptotic scattering data at $\scri^\pm$ and the matching conditions (\ref{matching}) yield an S-matrix for CJ gravity in presence of probe matter. As a consequence of (\ref{matching}), the S-matrix preserves the total Maxwell charge of the in states since it follows from (\ref{matching}) that
\begin{align}
    \int_{\scri^+_{L,R}} du \frac{d Q_i}{du} + Q_i(u\to +\infty)\big|_{\scri^+_{L,R}} &=  \int_{\scri^-_{L,R}}dv \frac{d Q_i}{dv} + Q_i(v\to -\infty)\big|_{\scri^-_{L,R}}\ ,\nonumber\\[4pt]
    \sum_{i=1}^{n_{out}} q_i\big|_{\scri^+_{L,R}} + Q_i\big|_{\mathcal{H}^+_{L,R}} &=  \sum_{i=1}^{n_{in}} q_i\big|_{\scri^-_{L,R}} + Q_i\big|_{\mathcal{H}^-_{L,R}}\ ,
    \label{SmatrixMatter}
\end{align}
where we used the fact that incoming/outgoing particles at $\scri$ appear as $q_i \delta(u-u_i)$ sources for the Maxwell charges. The resulting dynamics describe the interaction of shockwaves \cite{Dray:1984ha} in the flat two-dimensional black hole background. We leave the detailed investigation of this S-matrix for future work.

\begin{figure}
    \centering
    \includegraphics[scale=0.60]{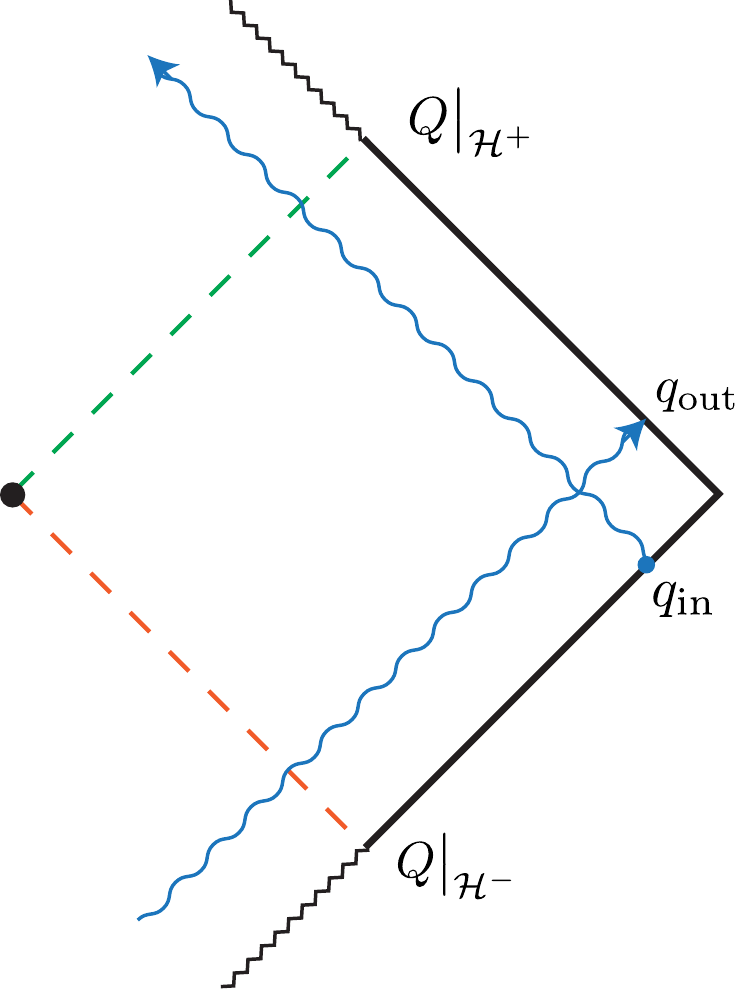}
    \caption{When adding probe matter, the S-matrix is entirely captured by the charge conservation \eqref{SmatrixMatter}. $Q\big|_{\mathcal{H}^-}$ is the past gravitational charge before the incoming shock that carries a charge $q_{\rm in}$, while $Q\big|_{\mathcal{H}^+}$ is the future gravitational charge after the outgoing shock that carries a charge $q_{\rm out}$.}
    \label{fig:Smatrix}
\end{figure}

\section{Euclidean partition function from random matrices}
\label{sec:Euclidean}

\subsection{Gravitational topological expansion}

Until now the analysis was done in Lorentzian signature. We would like now to consider thermal observables and for this purpose we go to Euclidean signature by analytically continuing the Bondi time $u_R$ which parametrizes $\mathscr{I}^+_R$ and whose definition can be found in Section \ref{SectionLorentz}. Before delving into the details of the path integral it is instructive to look at the on-shell solution and in particular its geometry. The analytic continuation of the Rindler patch \eqref{LorentzianSol} is 
\begin{equation}
    ds^2=\frac{4\pi }{\beta}\left( r-\frac{\phi_h}{\gamma}\right)d\tau^2-2id\tau dr \ , 
    \qquad 
    \tau\sim \tau +\beta\ ,
    \label{disk}
\end{equation}
where the temperature is related to the vacuum energy according to $\beta=\frac{2\pi \Lambda}{\gamma}$. The geometry is defined outside the horizon, i.e. for $r>\frac{\phi_h}{\gamma}$ and has the topology of a disk. However the metric is not the usual one, it is complex, a property that follows from the fact that we have analytically continued the retarded time instead of the usual timelike one. The operator that generates retarded time evolution is the Bondi Hamiltonian, therefore, we expect the path integral to compute the thermal trace of this operator. This disk geometry together with the value of the other fields contribute to the on-shell approximation of the path integral. One can also consider analytically continuing the advanced time $v_R$ which will also result in a complex disk where the $\tau,r$-component has opposite sign. These two disks should really be thought of as inequivalent since they come from the analytic continuation of two coordinates that are not related by an isometry.

\begin{figure}
    \centering
    \includegraphics[scale=1]{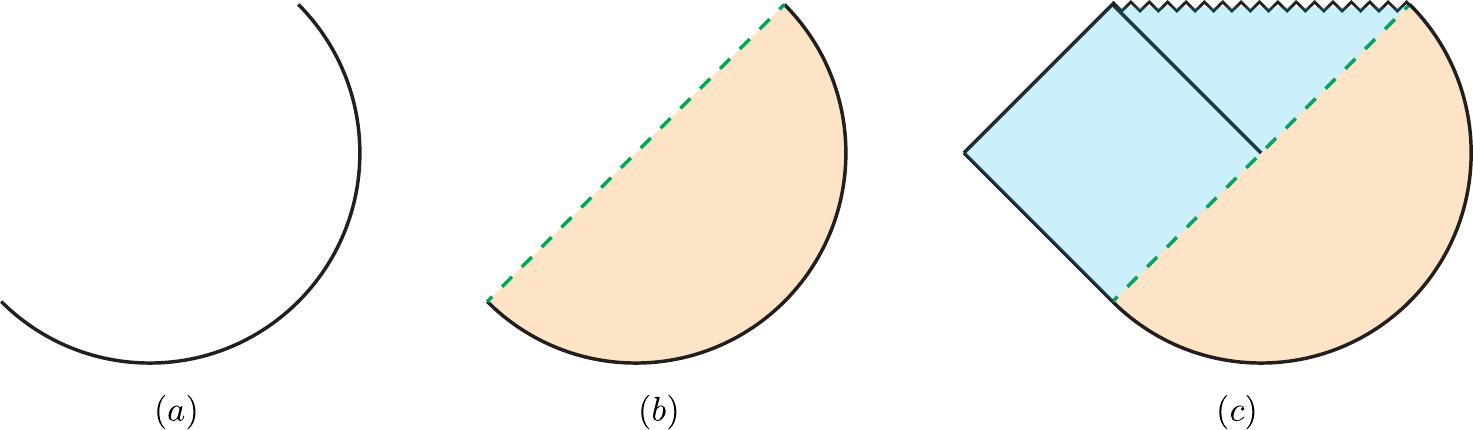}
    \caption{In (a) we represent the boundary path integral that prepares the HH state in the dual theory, it corresponds to a half circle of coordinate length $\beta/2$. This contour is a boundary condition for the bulk geometry that is nothing but a half disk in the on-shell approximation, as represented in (b). Finally we can time-evolve the state to obtain the black hole geometry as shown in (c). The Euclidean and Lorentzian solutions meet at the horizon. This is because we are analytically continuing the retarded time and the horizon is a constant-$u$ line. From the way the half-disk is glued to the Lorentzian geometry we deduce that our HH state belongs to $\mathcal{H}_{\mathscr{I}^-_L\cup\mathscr{I}^+_R }\equiv \mathcal{H}^{-+}$. Equivalently one can construct a HH state that belongs to $\mathcal{H}_{\mathscr{I}^+_L\cup\mathscr{I}^-_R }\equiv \mathcal{H}^{+-}$ by analytically continuing the advanced time and gluing the resulting half-disk to the other horizon.}\label{fig:TFD}
\end{figure}

The full path integral with one asymptotic thermal circle is formally defined as
\begin{equation}\label{PathIntegral}
Z(\beta)=
\int
\mathcal{D}g_{\mu \nu}
\mathcal{D}A_\mu
\mathcal{D}\Phi 
\mathcal{D}\Psi 
e^{-I_E[g_{\mu\nu},A_\mu,\Phi,\Psi]}\ ,
\end{equation}
where $I_E=iI_{\rm CJ}$. Before discussing the precise definition of this expression let us comment on its interpretation. There are two equivalent ways to look at $Z(\beta)$:
\begin{itemize}
    \item As the ``partition function'' of CJ gravity, i.e. the thermal trace of the Bondi Hamiltonian that generates time evolution along $\mathscr{I}^+_R$:
    \begin{equation}
        Z(\beta)=\mathrm{Tr}e^{-\beta H^+_R},
        \label{ThermalTrace}
    \end{equation}
    therefore probing its energy spectrum.
    \bigskip
    \item As the overlap of two Hartle-Hawking states that are prepared by the same half-circle boundary path integrals:
    \begin{align}
        Z(\beta)=^{-+}\langle \mathrm{HH}\vert \mathrm{HH}\rangle^{-+}.
        \label{TFDOverlap}
    \end{align}
    This state belongs to the Hilbert space $\mathcal{H}^{-+}$ and should be thought of as living on the horizon. The half-circle path integral is computed holographically by filling in with the appropriate geometries, see Figure \ref{fig:TFD}.

\end{itemize}

Again, the same construction can be done by analytically continuing the advanced time $v_R$. The interpretation of the resulting path integral is the same only this time the operator whose thermal trace is computed is the Hamiltonian that generates advanced time evolution on $\mathscr{I}^-_R$, i.e. $H^-_R$. This thermal trace is equivalently described as the overlap of HH states that live on $\mathcal{H}^{+-}$. 

For the expression \eqref{PathIntegral} to make sense we need to define the thermal boundary condition, the corresponding field space over which we are summing and which measure it is endowed with. Since the action is linear in $\Phi$, a rotation of its contour of integration along the imaginary axis produces a $\delta(R)$ constraint which forces all metrics to be flat. The sum divides then into a sum over flat connected surfaces and boundary fluctuations. The latter are nothing but the Euclidean version of the off-shell configurations (\ref{metricfalloff2}-\ref{dilatonfalloff2}) at fixed temperature. Fixing the temperature amounts to fixing the span of the asymptotic time defined by the dilaton, i.e. $\partial_\tau=\frac{1}{\gamma}\epsilon^{\mu\nu}\partial_\nu\Phi\partial_\mu$, and imposing the appropriate periodicity on the dynamical variables. The latter restricts the boundary modes $f(\tau)$ and $g(\tau)$ which now depend on the Euclidean time. They need to satisfy the constraints
\begin{equation}
    f(\tau+\beta)=f(\tau)+\beta\ ,
    \qquad \qquad
    g(\tau+\beta)=g(\tau)\ .
\end{equation}
In other words $f(\tau)$ becomes a diffeomorphism and $g(\tau)$ a function of the thermal circle. All off-shell configurations contributing to the Euclidean path integral with fixed topology are obtained by acting on the seed geometry with the large diffeomorphisms
\begin{equation}
\tau \rightarrow f(\tau)\ ,
\qquad \qquad
r \rightarrow  \frac{r+g'(\tau)}{f'(\tau)}\ .
\end{equation}
Interestingly, the periodicity of $g$ in Euclidean signature freezes the classical energy of the seed solution. Indeed as demonstrated in the previous section, the on-shell energy in the one-boundary case is $E_*=\frac{\Lambda}{\gamma}\delta \phi_h$ where $\delta\phi_h$ measures the deviation of the horizon value of the dilaton relative to $\phi_h$. The latter is generated by a linear $g(\tau)$ which is not allowed by the periodicity condition, therefore in Euclidean signature we have $\delta \phi_h=0$.

The $\delta(R)$ constraint reduces drastically the surface topologies that contribute to the path integral. In the one-boundary case there is actually only one surface that is flat, connected and smooth: the disk.\footnote{The classification of geodesically complete Riemann surfaces is very well understood, see Theorem 5.9.1 in \cite{book.surfaces}. While (i) and (v) in that Theorem correspond to Euclidean dS and AdS, cases (ii), (iii) and (iv) are the plane, cylinder and torus, the only three Riemann surfaces. Moreover, since in this case we are interested in surfaces with \textit{asymptotic} boundaries, we are simply left with the disk and cylinder. If we imposed finite cut-off instead of asymptotic boundary conditions, the sphere and cylinder surfaces with an arbitrary number of interior circular boundaries would also contribute.} Therefore all off-shell configurations are obtained by acting with the aforementioned diffeomorphisms on the Euclidean solution \eqref{disk}. The measure on this configuration space is rigorously obtained from the BF formulation of the theory in Appendix \ref{zapp:3} and the result is\footnote{The boundary path integral was computed in \cite{Afshar:2019tvp} while the matching with the gravitational path integral of CJ gravity was done in \cite{Godet:2020xpk}.}
\begin{equation}
Z(\beta)  \simeq 
\frac{2(2\pi)^4}{\pi(\gamma\beta)^2}
e^{S_0},
\label{Disk}
\end{equation}
The approximation sign will become clear in what follows. The power of $\beta$ in the one-loop contribution is consistent with the universal behavior $\beta^{-\frac{\#}{2}}$ where $\#$ is the number of bosonic zero modes in the path integral. In our case they are the four generators of the Maxwell symmetry (\ref{eq:Maxwell}) of the disk. The corresponding density of states is continuous and linear in the energy
\begin{equation}\label{eq:dens}
    \varrho(E)\simeq 
e^{S_0}
\frac{2(2\pi)^4}{\pi\gamma^2}E\ .
\end{equation}

This was for the one-boundary case but one can also consider multi-boundary path integrals. The $\delta(R)$ constraint again plays an important role since there is only one flat connected surface that connects two boundaries which is the cylinder, while there is no flat surface that connects more than two boundaries. The multi-boundary path integral collapses to \cite{Godet:2020xpk} (see Appendix \ref{zapp:3}) 
\begin{equation}\label{Cylinder}
Z(\beta_1,\beta_2) \simeq 
\frac{1}{\gamma(\beta_1+\beta_2)}\ ,
\qquad \qquad
Z(\beta_1,\dots,\beta_n)  \simeq 0\ , \quad n > 2\ .
\end{equation}
One of the advantages of the BF formulation, described in Appendix \ref{zapp:3}, with respect to previous derivations of this result is that it uniquely fixes the relative coefficient between the one and two-boundary contributions, which will
be important when matching with the matrix model. The non-vanishing of the cylinder path integral has important consequences for the interpretation of the gravitational path integral. Indeed it results in a non-factorization of the full two-boundary result since the latter is the sum of the product of two disks and a cylinder. A consistent explanation of this phenomenon is that our gravitational path integral computes the statistical $n$-point function of the operator \eqref{ThermalTrace} (or equivalently \eqref{TFDOverlap}) in a random matrix ensemble that is yet to be defined. The cylinder is then interpreted as a standard connected contribution to the two-point function.

From \eqref{Cylinder} one deduces that only the one and two-point functions of this observable are non-vanishing, meaning that it is a Gaussian variable in the ensemble. Since a Gaussian has support over the entire real axis it also means that the average sums over negative values of $Z(\beta)$ which is unphysical for a thermal trace. In other words, for some members of the ensemble, the density of states is negative, as observed in \cite{Godet:2021cdl}. 

This issue is resolved by the fact that formulae \eqref{Disk} and \eqref{Cylinder} should be understood as approximations of the actual result. We have computed our path integral by writing it as a topological expansion
\begin{equation}\label{Topological}
Z(\beta_1,\dots,\beta_n)\simeq 
\sum_{g=0}^{\infty}
(e^{-S_0})^{2(g-1)+n}
Z_g(\beta_1,\dots,\beta_n)\ ,
\end{equation}
where $Z_g$ is the path integral at fixed genus. The expansion truncates immediately since the only surfaces that are allowed are the disk and the cylinder whose Euler characteristic are respectively minus one and zero. A similar truncation occurs in certain theories of JT supergravity \cite{Stanford:2019vob}, although in that case it is the vanishing of the volumes of some higher genus manifolds that is responsible. Now since this sum is, by definition, perturbative in $e^{-S_0}$ it does not include non-perturbative contributions of order $\mathcal{O}(e^{-e^{S_0}})$ which are doubly non-perturbative in the effective Newton constant. As we will now see, these corrections are captured by a matrix model that we describe in detail. In particular this non-perturbative completion of the gravitational topological expansion restores the positivity of the thermal trace whose correlation functions it computes.

Before moving on let us briefly comment on the fact that the saddles of the path integral required to derive (\ref{Disk}) and (\ref{Cylinder}) involve complex metrics, such as (\ref{disk}). We do not think this is problematic in any way, as these complex metrics necessarily and naturally arise when studying the spectrum of the Bondi Hamiltonian operator through the Euclidean path integral. This is further supported by the observation that the path integral computed in this way is finite and leads to a positive spectral density (\ref{eq:dens}) which can be non-perturbatively completed by a random matrix model. Moreover, we can apply the criteria proposed by Kontsevich and Segal \cite{Kontsevich:2021dmb} (and subsequently extended by Witten \cite{Witten:2021nzp}) whose aim is to determine whether a complex metric is physical and can therefore be used as a saddle when performing a semi-classical expansion. Evaluating this criterion for the analytically continued metric configurations in (\ref{metricfalloff2}), we find that it reduces to the statement that the function $P(\tau)r+T(\tau)$ should be non-negative. 
That condition is indeed satisfied by all the saddles used when computing the CJ gravity partition function (see Appendix \ref{zapp:3}).

\subsection{Matrix model topological expansion}

In this Subsection we construct the appropriate matrix model required to match with the topological expansion~(\ref{Topological}) of the CJ Euclidean partition function to all orders in perturbation theory. Following the presentation in Section 4.1 of \cite{Stanford:2019vob}, we first quickly introduce the ``loop equations" \cite{Migdal:1984gj,Eynard:2004mh,Eynard:2015aea}, a powerful method for computing observables in the matrix model in an asymptotic expansion.

Consider an ensemble of square Hermitian matrices $M$ of dimension $N$, weighted by a probability measure determined by a potential $V(M)$ according to $dM\,e^{-N\,{\rm Tr}\,V(M)}$. The expectation value of any matrix operator $\mathcal{O}=\mathcal{O}(M)$ is defined as
\begin{equation}\label{eq:2}
\langle \mathcal{O} \rangle\equiv 
\frac{1}{\mathcal{Z}}
\int dM\,\mathcal{O}\,e^{-N\,{\rm Tr}\,V(M)}\ ,
\qquad {\rm where} \qquad
\mathcal{Z}=
\int dM\,e^{-N\,{\rm Tr}\,V(M)}\ .
\end{equation}
For observables that only depend on the eigenvalues $\lambda_i \in \mathbb{R}$ of $M$, a standard computation allows us to rewrite this as
\begin{equation}\label{eq:5}
\langle \mathcal{O} \rangle=
\frac{1}{\mathcal{Z}}
\prod_{i=1}^N
\int_{-\infty}^{+\infty}
d\lambda_i\,
\mathcal{O}(\lambda_1,\dots,\lambda_N)\,
\Delta(\lambda_1,\dots,\lambda_N)^2
e^{-NV(\lambda_i)}
\ ,
\end{equation}
where the Vandermonde determinant $\Delta(\lambda_1,\dots,\lambda_N)=\det(\lambda_i^{j-1})$ comes from the Jacobian associated to the change of variables. Two central observables are the eigenvalue spectral density and resolvent, respectively given by
\begin{equation}\label{eq:24}
    \rho(\lambda)=\frac{1}{N}{\rm Tr}\,\delta(\lambda-M)\ ,
    \qquad \qquad
    W(z)={\rm Tr}\frac{1}{z-M}\ .
\end{equation}
The expectation value of the resolvent defines an analytic function in the complex plane $z\in \mathbb{C}$, except at the spectrum of $M$ on the real line. The multi-trace generalization is given by $W(I)=\prod_{i=1}^{n}W(z_i)$ where $I=\lbrace z_1,\,\dots,z_n\rbrace$. The spectral density $\rho(\lambda)$ can be easily obtained from the discontinuity of $W(z)$. 

To derive the loop equations one starts by writing the expectation value of the resolvent in a large $N$ expansion
\begin{equation}\label{eq:35}
\langle W(I) \rangle_c=
\sum_{g=0}^{\infty}
\frac{W_g(I)}{N^{2(g-1)+n}}\ .
\end{equation}
Plugging this in the following simple identity
\begin{equation}
\int_{-\infty}^{+\infty}
d\lambda_1\dots d\lambda_N
\frac{\partial}{\partial \lambda_a}
\left[
\frac{W(I)}{z-\lambda_a}
\Delta(\lambda_1,\dots,\lambda_N)^2
e^{-N\sum_{i=1}^NV(\lambda_i)}
\right]=0\ ,
\end{equation}
and performing some standard manipulations (see Section 4.1 of \cite{Stanford:2019vob} for details) one arrives at a set of closed recursion relations for $W_g(I)$ that we now describe.

\paragraph{Spectral curve:} The leading contribution to the single trace observable $W_0(z)$ is given by
\begin{equation}\label{eq:14}
W_0(z)=\frac{1}{2}\left(
V'(z)-h(z)\sqrt{\sigma(z)}
\right)\ ,
\end{equation}
where $h(z)$ and $\sigma(z)$ are polynomials in $z$, with $\sigma(z)$ only having simple roots. Given the potential $V(z)$, these functions are determined from the knowledge of the analytic structure of $W_0(z)$ that follow from its definition in (\ref{eq:24}). In the large $N$ limit the singularities of the resolvent $W(z)$ in the spectrum of the matrix $M$ condense into a branch-cut square root singularity, going between the branch points at the roots of $\sigma(z)$. When $\sigma(z)=(z-a_-)(z-a_+)$ has only two roots, the matrix model is in a single-cut phase. Computing the eigenvalue spectral density from the discontinuity of $W_0(z)$ in the complex plane gives
\begin{equation}\label{eq:15}
\rho_0(\lambda)=\frac{1}{2\pi}|h(\lambda)|\sqrt{-\sigma(\lambda)}
\times 
\textbf{1}_{\sigma(\lambda)<0}\ ,
\end{equation} 
where $\textbf{1}_{\sigma(\lambda)<0}$ is the indicator function and $\rho_0(\lambda)$ the leading expectation value of the spectral density~(\ref{eq:24}). For any given potential one can easily work out $a_\pm$ and $h(z)$ by requiring $W_0(z)$ in (\ref{eq:14}) has the appropriate large $z$ behavior, i.e. $W_0(z)=1/z+\mathcal{O}(1/z^2)$. When $\sigma(z)$ has more than two roots, the matrix model is in a multicut phase.

A quantity that is closely related to the leading spectral density is the spectral curve $y(z)$
\begin{equation}\label{eq:37}
    y(z)^2=\frac{1}{4}h(z)^2\sigma(z)
    \qquad \Longrightarrow \qquad
    \rho_0(z)=\pm \frac{i}{\pi}y(z\pm i\epsilon)
    \ .
\end{equation}
This defines a two-sheeted Riemann surface, corresponding to the two possible signs of the square root. If we denote $\hat{z}$ as the same point as $z$ but in the second sheet, one has $h(\hat{z})=h(z)$, $\sqrt{\sigma(\hat{z})}=-\sqrt{\sigma(z)}$ and $y(\hat{z})=-y(z)$. It turns out the whole perturbative expansion of the matrix model is entirely fixed by the spectral curve $y(z)$. For this reason, a perturbative definition of the matrix model can be given directly in terms of $y(z)$ instead of the potential $V(z)$. 

\paragraph{A universal observable:} A special role is played by the leading contribution of double trace observables. In the case of $W_0(z_1,z_2)$ in a single-cut matrix model, the following expression can be readily derived from the loop equations
\begin{equation}\label{eq:28}
W_0(z_1,z_2)=
\frac{1}{2(z_1-z_2)^2}
\left[
\frac{a_-a_++z_1z_2-(a_-+a_+)(z_1+z_2)/2}{\sqrt{\sigma(z_1)}\sqrt{\sigma(z_2)}}-1
\right]\ .
\end{equation}
Note this result is independent of the fine grained details of the potential $V(M)$ and spectral curve. It is universal, in the sense that it only depends on the endpoints $a_\pm\in \mathbb{R}$ of the leading spectral density. As $z_1$ goes across the branch-cut, the value of $W_0(z_1,z_2)$ is given by
\begin{equation}\label{eq:26}
    W_0(\hat{z}_1,z_2)+W_0(z_1,z_2)=\frac{-1}{(z_1-z_2)^2}\ .
\end{equation}

\paragraph{General recursion relation:} All other terms in the expansion of the resolvent (\ref{eq:35}) are determined from the following recursion relation
\begin{equation}\label{eq:23}
W_g(z,I)=
\frac{1}{\sqrt{\sigma(z)}}
\sum_{i=\pm }
\,
{\rm Res}\left[
\frac{\sqrt{\sigma(z')}}{2y(z')}
\frac{F_g(z',I)}{(z'-z)}
,z'=a_i
\right] \ .
\end{equation}
The function $F_g(z',I)$ is given by
\begin{equation}\label{eq:21}
F_0(z,z_1,z_2)=
\frac{W_0(z,z_1)}{(z-z_2)^2}+
\frac{W_0(z,z_2)}{(z-z_1)^2}+
2W_0(z,z_1)W_0(z,z_2)
\ ,
\end{equation}
for $g=0$ and $I=\lbrace z_1,z_2 \rbrace$, and
\begin{equation}\label{eq:25}
F_g(z,I)=
W_{g-1}(z,z,I)
+\sum_{k=1}^{|I|}
\left[
2W_0(z,z_k)
+\frac{1}{(z-z_k)^2}
\right]W_g(z,I\setminus z_k)
+\sum_{h,J}'W_{h}(z,J)W_{g-h}(z,I\setminus J)\ ,
\end{equation}
in all other cases. The sum in the third term is over $h=0,\dots,g$ and $J\subseteq I$ that do not contain a factor of $W_0(z)$ or $W_0(z,z_k)$. Since $F_g(z',I)$ only depends on $W_{g'}(I')$ with either $g'<g$ or $I'\subset I$, this gives an expression for $W_g(z,I)$ in terms of the lower order expansion coefficients. 

\subsubsection{Double scaling limit}

Let us now show how a particular choice for the spectral curve implies the matching between the CJ Euclidean partition functions (\ref{Disk}) and (\ref{Cylinder}), and ensemble average of the operator
\begin{equation}\label{eq:Matrixoperator}
\mathbb{O}(\beta_1,\dots,\beta_n)=
\prod_{i=1}^n\mathbb{O}(\beta_i)\ ,
\qquad \qquad
\mathbb{O}(\beta) =
    \int_{-\infty}^{+\infty}
    \frac{dp}{\sqrt{\gamma}}
    {\rm Tr}\,e^{-\beta (\bar{M}^2+p^2)}\ ,
\end{equation}
where we shall shortly explain the meaning of the notation $\bar{M}$. To do so, we need $W_g(z,I)$ in (\ref{eq:23}) to vanish for all values of $g$ and $I$. As noted in \cite{Stanford:2019vob}, this is very easily obtained by considering a single-cut matrix model in the limit in which the branch points $a_\pm$ go to infinity. In that case, the computation of the residue in (\ref{eq:23}) becomes
\begin{equation}\label{eq:13}
\lim_{a_\pm\rightarrow \pm \infty}W_g(z,I)=
{\rm Res}
\left[
\frac{1}{y(1/z')}
\frac{F_g(1/z',I)}{z'(zz'-1)},
z'=0
\right]\ .
\end{equation}
Using $W(z)=1/z+\mathcal{O}(1/z^2)$, (\ref{eq:21}) and (\ref{eq:25}) imply $F_g(z,I)=1/z^2+\mathcal{O}(1/z^3)$, meaning the second factor in (\ref{eq:13}) is regular at $z'=0$. For a spectral curve which diverges for large $z$, the first factor is also regular and the residue in (\ref{eq:13}) vanishes. This shows the full perturbative expansion of the resolvent $W(I)$ vanishes, except for the two special cases in (\ref{eq:14}) and (\ref{eq:28}).

The limit in which a branch point $a_\pm$ goes to infinity is called the ``double scaling limit", a somewhat subtle (but rigorous) procedure that must be treated carefully (the interested reader should consult Appendix \ref{zapp:4} for the proper mathematical treatment). Simply put, for our model it involves taking large $N$ while simultaneously rescaling the eigenvalues of the matrix $\lambda_i$ in the following way
\begin{equation}
\frac{1}{N}=\frac{\hbar}{2t_2}\delta^3\ ,
\qquad \qquad
\lambda_i=\alpha_i\delta\ ,
\end{equation}
with $\delta\rightarrow 0$. Here, $\hbar$ (not Planck's constant) is the scaling parameter which replaces $1/N$, while $\alpha_i$ are the rescaled eigenvalues. The additional parameter $t_2$ is not strictly independent, as it can be set to one without loss of generality, but is useful for bookkeeping reasons. In this limit, the relevant spectral density is no longer normalizable and given by
\begin{equation}\label{eq:82}
    \rho(\alpha)={\rm Tr}\,\delta(\alpha-\bar{M})\ ,
\end{equation}
where $\bar{M}$ is the matrix constructed with the rescaled eigenvalues $\alpha_i$. Roughly speaking, ``zooming in" to the small eigenvalues is equivalent to taking $a_\pm \rightarrow \pm \infty$. This means that, in the double scaling. the expression in (\ref{eq:13}) vanishes and all trace class observables have the following very simple expansion 
\begin{equation}\label{eq:29}
\langle W(z) \rangle \simeq \frac{1}{\hbar}W_0(z)\ ,
\qquad \quad
\langle W(z_1,z_2) \rangle_c \simeq W_0(z_1,z_2)\ ,
\qquad \quad
\langle W(z_1,\dots,z_n) \rangle_c \simeq 0\ ,\quad n > 2\ .
\end{equation}
Appendix \ref{zapp:4} derives the same result, carefully.

We should use this to compute the expectation value of $\mathbb{O}(\beta_1,\dots,\beta_n)$ (\ref{eq:Matrixoperator}) and compare with the Euclidean partition functions in (\ref{Disk}) and (\ref{Cylinder}). The third expression in (\ref{eq:29}) immediately implies ${\langle \mathbb{O}(\beta_1,\dots,\beta_n) \rangle\simeq 0}$ for $n\ge 3$. For the $n=2$ case we can transform the resolvents to $\mathbb{O}(\beta)$ by first using the identity $2x\,{\rm Tr}(x^2-\bar{M}^2)^{-1}=W(x)-W(-x)$ to show
\begin{equation}
     \big\langle
     {\rm Tr}\frac{1}{z_1-\bar{M}^2}
     {\rm Tr}\frac{1}{z_2-\bar{M}^2}
     \big\rangle_c\simeq 
     \frac{1}{2\sqrt{-z_1}\sqrt{-z_2}(\sqrt{-z_1}+\sqrt{-z_2})^2}\ ,
\end{equation}
where we have been careful when dealing with the location of $z_1$ and $z_2$ in each Riemann sheet. After performing an inverse Laplace transform, we can switch the resolvents in $\bar{M}^2$ to exponentials and find
\begin{equation}\label{eq:63}
   \big\langle 
   {\rm Tr}\,e^{-\beta_1 \bar{M}^2}{\rm Tr}\,e^{-\beta_2 \bar{M}^2} \big\rangle_c 
   \simeq 
   \frac{2}{2\pi}\frac{\sqrt{\beta_1\beta_2}}{\beta_1+\beta_2}
   \qquad \Longrightarrow \qquad
   \langle \mathbb{O}(\beta_1,\beta_2) \rangle_c
   \simeq 
\frac{1}{\gamma(\beta_1+\beta_2)}\ ,
\end{equation}
where in the final step we solved the Gaussian $p$ integral in the definition of the operator $\mathbb{O}(\beta)$.

Note that apart from having $a_\pm\rightarrow \pm \infty$, the details of $y(z)$ have not played any role so far. In fact, for this class of double scaled models the specifics of $y(z)$ only determine the behavior of the leading single trace observable (\ref{eq:29}), at least when it comes to their perturbative behavior. Defining a specific matrix model is equivalent to fixing $\rho_0(\alpha)$, which we take as
\begin{equation}\label{eq:38}
    \rho_0(\alpha)=
    \frac{1}{\pi}\left(
    \frac{4\pi}{\sqrt{\gamma}}
    \right)^3
    \alpha^2
    \qquad \Longrightarrow \qquad
    \langle \mathbb{O}(\beta) \rangle \simeq
    \frac{1}{\hbar}
    \frac{2(2\pi)^4}{\pi(\gamma\beta)^2}\ ,
\end{equation}
in order to match (\ref{Disk}) after identifying $\hbar=e^{-S_0}$. All in all, we have shown how a Hermitian matrix model perturbatively defined from $\rho_0(\alpha)$ in (\ref{eq:38}), reproduces the topological expansion of CJ gravity to all orders.

It is important to highlight the matching between the matrix model and CJ gravity is non-trivial and was not guaranteed to work. One of the features that enabled the agreement is an underlying relation between the disk and cylinder partition functions. For instance, if the disk partition function in (\ref{Disk}) had an odd half-integer power of $\beta$ instead of even (while the cylinder partition function is left unchanged), one can easily show there is no consistent matrix model characterized by $\rho_0(\alpha)$ that would reproduce such a result while still ensuring the vanishing of higher order contributions in the perturbative expansion. 

One could also wonder whether there is another random matrix model, built from a different class of matrices that could work. The loop equations of all three Dyson and all seven Altland-Zirnbauer ensembles \cite{1997} are nicely described in \cite{Stanford:2019vob}. One can check that none of these ensembles can be used in order to reproduce the CJ gravity partition function. In particular, while the $(\boldsymbol{\alpha},\boldsymbol{\beta})=(0,2)$ and $(\boldsymbol{\alpha},\boldsymbol{\beta})=(2,2)$ Altland-Zirnbauer ensembles also have the vanishing of the expansion as given in (\ref{eq:29}), they contain an additional contribution (due to the genus ``one-half'' crosscap) that is not present in the CJ gravity topological expansion. These kinds of models are relevant when including non-orientable surfaces in the topological expansion \cite{Harris:1990kc,PhysRevLett.65.2098,Brezin:1990dk,Stanford:2019vob} and are therefore not appropriate for the setup considered in this work.

\section{Non-perturbative flat quantum gravity}
\label{sec:Nonperturbative}

In this Section we go beyond the perturbative analysis in the parameter $\hbar=e^{-S_0}$ and consider non-perturbative corrections. As we shall see, these kinds of contributions play an important role in certain regimes. Implicit in our analysis is the assumption that the matching between observables of CJ gravity and the random matrix model continues to hold beyond perturbation theory. In this way, we are able to compute and analyze the behavior of the fine grained spectrum, spectral form factor and quenched free energy of CJ gravity.

\subsubsection*{Orthogonal polynomials and double scaling}

In Appendix \ref{zapp:4} we carefully describe the method of orthogonal polynomials and the double scaling limit applied to the matrix model required for describing CJ gravity. Here, we summarize its most salient points necessary for understanding the discussion below. The application of this approach to the study JT gravity was first developed in \cite{Johnson:2019eik,Johnson:2020exp,Johnson:2021zuo}.

All observables of the double scaled matrix model can be obtained from the matrix model kernel $K(\alpha,\bar{\alpha})$. For instance, single and double trace observables are computed as
\begin{equation}\label{eq:86}
\begin{aligned}
\langle {\rm Tr}\,F_1(\bar{M})\rangle & =
\int_{-\infty}^{+\infty}d\alpha
K(\alpha,\alpha)F_1(\alpha)\ , \\
\left\langle  {\rm Tr}\,F_1(\bar{M}) {\rm Tr}\,F_2(\bar{M}) \right\rangle_c & = 
\int_{-\infty}^{+\infty}d\alpha d\bar{\alpha}
\left[
\delta(\alpha-\bar{\alpha})
-
K(\alpha,\bar{\alpha})
\right]K(\alpha,\bar{\alpha}) F_1(\alpha )F_2(\bar{\alpha})\ ,
\end{aligned}
\end{equation}
where $F_1(\bar{M})$ and $F_2(\bar{M})$ are arbitrary functions. Note that compared to the loop equations discussed in the previous Section, these expressions do not involve any perturbative expansion of the observables and are therefore able to capture non-perturbative contributions. The kernel $K(\alpha,\bar{\alpha})$ is obtained from a set of polynomials that are orthogonal with respect to the integration measure defined by the potential of the matrix model, which in our case reads
\begin{equation}
V(M)=-M^2+\frac{1}{4}M^4\ .
\end{equation}
In the double scaling limit, the orthogonal polynomials become the functions $\varphi_s(x,\alpha)$, where $x\in \mathbb{R}$ is the scaling part of $n$, the integer order of the polynomial with $s=\pm 1$ indicating whether $n$ is even or odd. The kernel is then obtained by combining and integrating these functions in the following way
\begin{equation}\label{eq:39}
K(\alpha,\bar{\alpha})=
\sum_{s=\pm}\int_{-\infty}^0dx\,\varphi_s(x,\alpha)\varphi_s(x,\bar{\alpha})\ .
\end{equation}
The functions $\varphi_s(x,\alpha)$ can be efficiently computed from the following eigenvalue problem
\begin{equation}\label{eq:31}
\mathcal{H}_s\varphi_{s}(x,\alpha)=\alpha^2\varphi_{s}(x,\alpha)\ ,
\qquad {\rm where} \qquad
\mathcal{H}_s=-\hbar^2\partial_x^2+\left[r(x)^2-s\hbar r'(x) \right]\ ,
\end{equation}
which follows from double scaling a simple recursion relation satisfied by the orthogonal polynomials. The operator $\mathcal{H}_s$ takes the form of a quantum mechanical Hamiltonian, where the potential is determined by a function $r(x)$ which satisfies the following differential equation
\begin{equation}\label{eq:27}
t_2\Big[
r(x)^3-\frac{1}{2}\hbar^2r''(x)
\Big]+r(x)x=0\ ,
\end{equation}
called the ``string equation". Altogether, the double scaled model depends on two parameters: $t_2$ appearing in the string equation (\ref{eq:27}) and $\hbar$, that is the scaling part of $N$.\footnote{As we previously mentioned, the parameter $t_2$ is not strictly necessary, but it is convenient to have for bookkeeping purposes. For definitions, see Appendix~\ref{zapp:4}.} Concretely, the computation of matrix model observables such as (\ref{eq:86}) proceeds as follows. After picking specific values for $(\hbar,t_2)$ one solves the string equation (\ref{eq:27}) and obtains $r(x)$. This allows one to construct the Schrodinger operator $\mathcal{H}_s$ in (\ref{eq:31}), compute its eigenfunctions $\varphi_s(x)$ and obtain the matrix model kernel $K(\alpha,\bar{\alpha})$ by integrating them as indicated in (\ref{eq:39}).

Before showing how this works more explicitly, let us mention we can use the matrix model to also compute fine grained observables. For instance, the probability of having~$k$ eigenvalues in the interval $I=(a,b)$ is given by
\begin{equation}\label{eq:22}
\mathcal{E}_k(a,b)=\left.
\frac{(-1)^k}{k!}\frac{\partial^k}{\partial z^k}
\det({\rm Id}-z\widehat{K}_I)\right|_{z=1}\ ,
\end{equation}
where $\widehat{K}_I(\,\cdot\,)$ is an integral operator acting in the space of functions as
\begin{equation}
\widehat{K}_I(f)=\int_a^b d\bar{\alpha}f(\bar{\alpha})K(\bar{\alpha},\alpha)\ ,
\end{equation}
and ${\rm Id}$ the identity. This means $\mathcal{E}_k(a,b)$ is obtained from the determinant of an operator acting on an infinite dimensional space (\ref{eq:22}), a Fredholm determinant. Although in practice this might seem very difficult to compute, we shall show how it can be obtained explicitly using the approach developed in \cite{Bornemann_2009,Johnson:2021zuo,Johnson:2022wsr}.

\subsubsection*{Perturbative matching with CJ gravity}

For clarity, let us explain how the formalism we just described can also be used to compute observables perturbatively in $\hbar$, matching with CJ gravity observables. In particular, this will allow us to determine the matrix model parameter $t_2$ appearing in the string equation (\ref{eq:27}) in terms of bulk parameters.

The first step is to solve the string equation (\ref{eq:27}) by writing a perturbative expansion $r(x)=r_0(x)+\sum_{n=1}^{\infty}r_n(x)\hbar^n$ and solving it order by order. The leading solution $r_0(x)$ is given by
\begin{equation}\label{eq:81}
r_0(x)=
\begin{cases}
\,\,(-x/t_2)^{1/2}\,\,
\ , \qquad  x\le 0\ , \\
\qquad 0 \qquad \,\,\,\,
\ , \qquad  x\ge 0\ ,
\end{cases}
\end{equation}
which already implies $t_2$ must be positive. This simple expression implicitly determines the leading perturbative behavior of all observables in the matrix model. We can then compute the eigenfunctions $\varphi_s(x,\alpha)$ of the operator $\mathcal{H}_s$ in (\ref{eq:31}) in the leading WKB approximation. As shown in \cite{Johnson:2021owr}, the WKB solution in the classically allowed region is\footnote{As explained in \cite{Johnson:2021owr}, the undetermined constants of the WKB approximation are determined by comparing with the eigenfunctions of a toy model that captures the low energy behavior of the system. Although the eigenfunctions of the toy model are not normalizable, the undetermined constant is fixed here by requiring they form a complete set $\int_{-\infty}^{+\infty}d\alpha\varphi_s(x,\alpha)\varphi_s(x',\alpha)=\delta(x-x')$.}
\begin{equation}\label{eq:57}
\varphi_s^{\rm WKB}(x,\alpha)=
\sqrt{\frac{|\alpha|}{\pi \hbar}}
\frac{\cos\left[\frac{1}{\hbar}\int_{x_{\rm min}}^{x}d\bar{x}\sqrt{\alpha^2-r_0(\bar{x})^2}-\frac{\pi}{4}(s+1)\right]}{(\alpha^2-r_0(x)^2)^{1/4}}\ ,
\qquad \qquad
x\ge x_{\rm min}=-t_2\alpha^2\ .
\end{equation}
From this one finds the following expression for the kernel (\ref{eq:39})
\begin{equation}\label{eq:58}
K_{\rm WKB}(\alpha,\bar{\alpha})=
\frac{1}{\pi}
\frac{\sin\left[2t_2
(|\alpha|^3-|\bar{\alpha}|^3)/3\hbar
\right]}
{|\alpha|-|\bar{\alpha}|}
\ .
\end{equation}
Note that although we have worked to leading order in perturbation theory, the non-diagonal components of the kernel are non-perturbative in $\hbar$. The diagonal components give the leading spectral density $\rho_0(\alpha)$ that can be used to compute the average of $\mathbb{O}(\beta)$ as
\begin{equation}\label{eq:83}
\rho_0(\alpha)
=\frac{2t_2}{\pi}\alpha^2
\qquad \Longrightarrow \qquad
\langle \mathbb{O}(\beta) \rangle \simeq
\frac{1}{\hbar}
\sqrt{\frac{\pi}{\gamma\beta}}
\int_{-\infty}^{+\infty}d\alpha\,\rho_0(\alpha)e^{-\beta \alpha^2}=\frac{ t_2}{\hbar\sqrt{\gamma}\beta^2}\ ,
\end{equation}
where the factor $\sqrt{\pi/\beta}$ comes from solving the Gaussian $p$ integral in the definition of $\mathbb{O}(\beta)$. Enforcing this matches with the single boundary Euclidean partition function $Z(\beta)$ in (\ref{Disk}) fixes the value of $t_2$ to
\begin{equation}\label{eq:85}
t_2=\frac{1}{2}\left(\frac{4\pi}{\sqrt{\gamma}}\right)^3\ .
\end{equation}
Similarly, one can compute the leading behavior of the average of $\mathbb{O}(\beta_1,\dots,\beta_n)$ for $n\ge 2$,
see Appendix A of \cite{Johnson:2021owr} for explicit calculations when $n=2,3$ cases. For arbitrary $n$ it is convenient to use a compact formula derived in \cite{Ambjorn:1990ji,Moore:1991ir}, which adapted to our setup is given by\footnote{To obtain the $x=0^+$ limit, one can take $\kappa$ in (\ref{eq:42}) to scale with $\delta$ as $\kappa=1+\mu \delta^2$, where $\mu$ is an additional parameter of the double scaled model analogous to $t_2$. By taking $\mu\rightarrow 0^+$ on obtains the $x=0^+$ limit in (\ref{eq:53}). See Appendix C of \cite{Rosso:2021orf} for more details.} 
\begin{equation}\label{eq:53}
\langle \mathbb{O}(\beta_1,\dots,\beta_n) \rangle_c=
\frac{1}{\gamma^{n/2}\sum_{i=1}^n\beta_i}
\left[
(\hbar \partial_x)^{n-2}
e^{-r_0(x)^2 \sum_{i=1}^n\beta_i}
\right]_{x=0^+}+\mathcal{O}(\hbar^{n})\ .
\end{equation}
Given that $r_0(x)$ is exactly zero at $x\ge 0$, the connected average of $\mathbb{O}(\beta_1,\dots,\beta_n) $ vanishes to leading order when $n\ge 2$, except for $n=2$ where one matches with the gravitational result $Z(\beta_1,\beta_2)$ in~(\ref{Cylinder}). Higher perturbative $\hbar$ corrections can also be computed in this formalism and explicitly be shown to vanish, e.g. see Appendix~B of \cite{Johnson:2021owr}.

\subsubsection*{Computing non-perturbative contributions}

\begin{figure}
    \centering
    \includegraphics[scale=0.58]{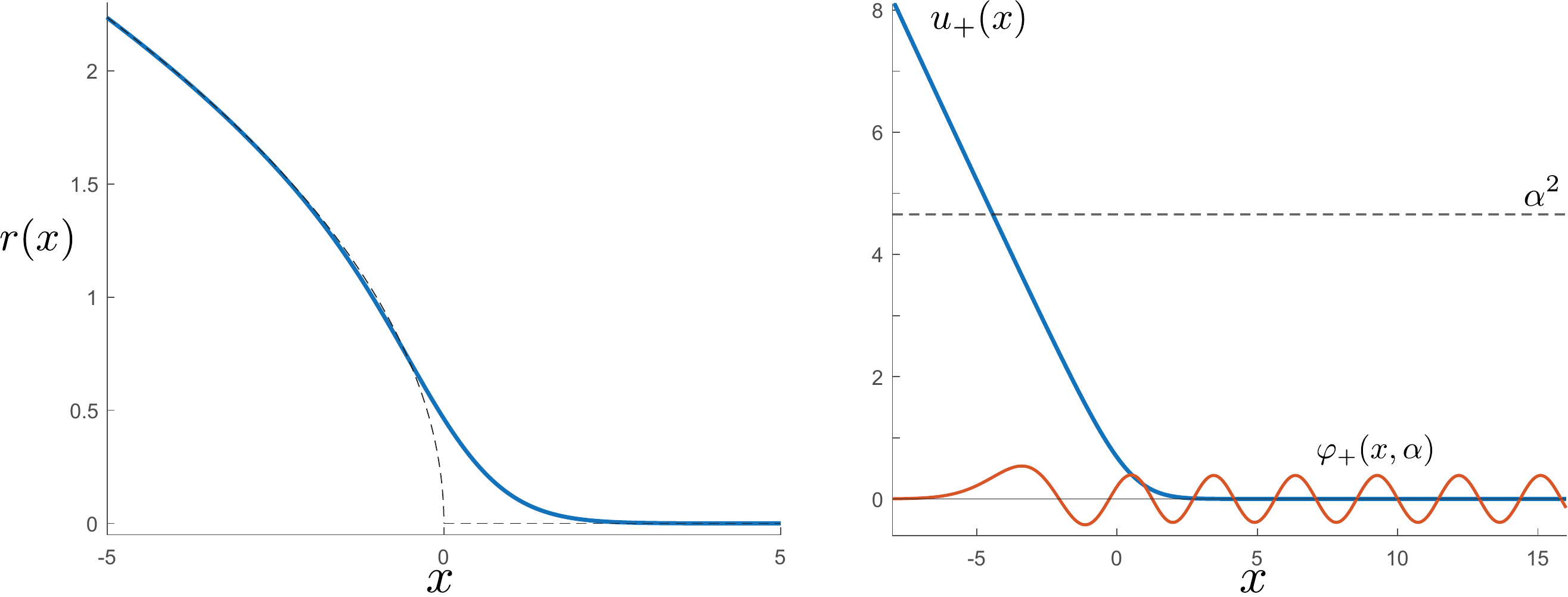}
    \caption{The solid line on the left diagram corresponds to the full numerical solution $r(x)$ to the string equation (\ref{eq:27}) with $\hbar=t_2=1$. The dashed line is the perturbative solution $r_0(x)$ in (\ref{eq:81}). On the right, we plot the potential $u_+(x)=r(x)^2-\hbar r'(x)$ appearing in the operator $\mathcal{H}_+$ in (\ref{eq:31}). The oscillating function corresponds to an eigenstate of $\mathcal{H}_+$.}\label{fig:1}
\end{figure}

We now describe the methodology for computing non-perturbative contributions to the matrix model observables, developed in \cite{Johnson:2019eik,Johnson:2020exp,Johnson:2021zuo} to study JT-like theories. The first step is to solve the string equation (\ref{eq:27}) exactly. Since this is not possible to do analytically, we proceed numerically. In the left diagram of Figure \ref{fig:1} we show the full solution $r(x)$ with $\hbar=t_2=1$. The dashed curve is the solution to all orders in perturbation theory, meaning the difference between them is entirely generated by non-perturbative effects. From this we can construct the potential $u_s(x)=r(x)^2-s\hbar r'(x)$ and numerically solve for the eigenfunctions $\varphi_s(x,\alpha)$ of $\mathcal{H}_s$ (\ref{eq:31}). As an example, in the right diagram of Figure \ref{fig:1} we plot $\varphi_+(x,\alpha)$ with $\alpha^2\sim 4.75$ together with the potential $u_+(x)$. The normalization of the eigenfunctions is determined by comparing with the WKB solution (\ref{eq:57}) in the large $x$ region (see \cite{Johnson:2020exp,Johnson:2021owr} for details). As expected, we observe an oscillatory behavior for $\varphi_+(x,\alpha)$ in the classical region $x\ge x_{\rm min}$ followed by an exponential damping. Although numerical, the kernel $K(\alpha,\bar{\alpha})$ obtained from these eigenfunctions (\ref{eq:31}) includes all perturbative and non-perturbative effects.

All numerical calculations are performed in Matlab. Following \cite{Johnson:2021zuo}, we use the Chebfun package, which enables the computation of the eigenfunctions $\varphi_s(x,\alpha)$ to very high precision. This is particularly important for computing the Fredholm determinant in the expression for $\mathcal{E}_k(a,b)$ in (\ref{eq:22}), that is prone to several numerical instabilities.

\subsection{Fine grained spectrum}
\label{sec:3.1}

Let us start by using this construction to analyze fine grained details of the spectrum. In the non-perturbative completion, one should distinguish between two different spectra: the eigenvalue spectrum of the matrix model, characterized by $\rho(\alpha)$ in (\ref{eq:82}), and the CJ gravity energy spectrum $\varrho(E)$, defined as
\begin{equation}
Z(\beta)=\int_0^{+\infty}dE\,\varrho(E)e^{-\beta E}\ .
\end{equation}
Using the matching between this partition function and $\langle \mathbb{O}(\beta)\rangle$ implies the following relation between their spectral densities
\begin{equation}\label{eq:84}
\varrho(E)=
\int_0^E
\frac{dy}{\sqrt{\gamma}}
\frac{\langle \rho(\sqrt{y}) \rangle}{\sqrt{y(E-y)}}
\ .
\end{equation}
This is a convolution between the free particle and matrix model densities.

\begin{figure}
    \centering
    \includegraphics[scale=0.6]{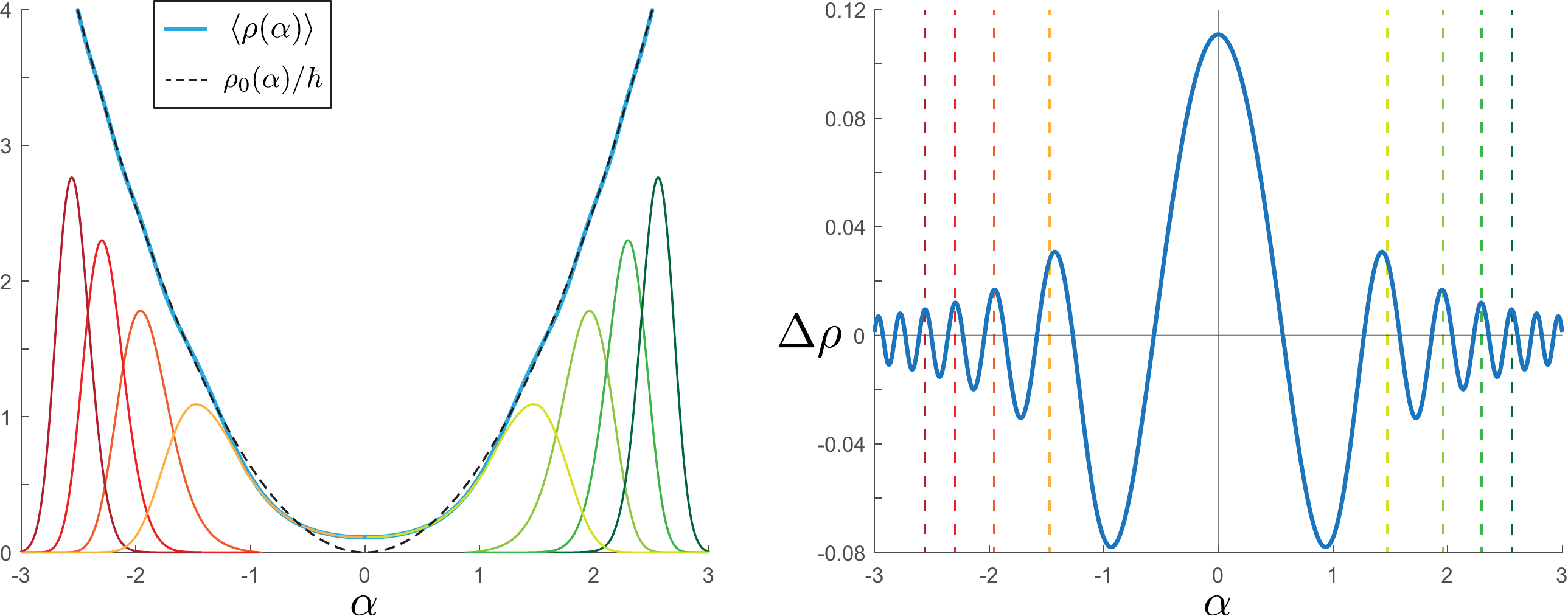}
    \caption{Matrix model spectrum for $\hbar=t_2=1$. On the left diagram, the solid blue line gives $\langle \rho(\alpha) \rangle$, which oscillates around the perturbative result $\rho_0(\alpha)=2\alpha^2/\pi$ indicated by the dashed curve. The remaining green and red curves are the probability density function of the first few individual eigenvalues. The solid curve on the right diagram shows the non-perturbative contributions to the matrix spectrum, obtained from $\Delta \rho=\langle \rho(\alpha) \rangle-\rho_0(\alpha)/\hbar$. The vertical dashed lines indicate the maximums of the green and red curves on the left diagram.}
    \label{fig:4}
\end{figure}

\paragraph{Matrix model:} We now focus on the eigenvalue spectrum $\rho(\alpha)$, using the numerical Kernel obtained from the eigenfunctions, one of which is plotted in Figure~\ref{fig:1}. The blue solid line in the left diagram of Figure~\ref{fig:4} shows the eigenvalue spectral density computed in this way, with the WKB result (\ref{eq:83}) given by the dashed curve. Subleading non-perturbative corrections are non-zero and dominate the behavior for small values of $\alpha$. These are the remnants of the discreteness of the eigenvalues that underliess the random matrix model. To better appreciate the non-perturbative oscillations, we plot $\Delta \rho(\alpha)=\langle \rho(\alpha) \rangle-\rho_0(\alpha)/\hbar$ on the right diagram in Figure \ref{fig:4}.

We would like to go further and characterize more fine grained features of the spectrum. To do so, we follow \cite{Johnson:2021zuo} and use the numerical kernel $K(\alpha,\bar{\alpha})$ to compute the Fredholm determinant in (\ref{eq:22}) and obtain the  probability density function of each individual eigenvalue. The first step in doing so is to discretize the Fredholm determinant as
\begin{equation}\label{eq:68}
\det({\rm Id}-z\widehat{K}_I)
\qquad \longrightarrow \qquad
\det\left(
\delta_{ij}-z\sqrt{w_i}K(e_i,e_j)\sqrt{w_j}
\right)\ ,
\qquad i,j=1,\dots,N_{\rm quad}  \ ,
\end{equation}
where $e_i$ and $w_i$ are the nodes and weights of some quadrature method in the interval $I=(a,b)$ where we are interested in computing $\mathcal{E}_k(a,b)$. The number of quadrature nodes is given by $N_{\rm quad}$. As shown in \cite{Bornemann_2009}, this numerical method efficiently converges to the original Fredholm determinant. Using Clenshaw-Curtis quadrature we can calculate the determinant and obtain $\mathcal{E}_k(a,b)$ for the desired values of $(a,b)$ and $k$. We can then construct the cumulative density function of the $n$-th eigenvalue away from the origin as
\begin{equation}
C_n(b)=\sum_{k=0}^{n-1}\mathcal{E}_k(0,b)\ .
\end{equation}
More explicitly, $C_n(b)$ is the probability of not finding the $n$-th eigenvalue in the interval $(0,b)$. The probability density function is then obtained from its derivative. We show the final result for the first few eigenvalues on the left diagram of Figure \ref{fig:4}, given by the green and red curves.

A smoothing function which filters numerical errors was used in order to obtain these results. To test the filter does not erase actual physics, we have checked the probability density functions after the smoothing are properly normalized with a precision of at least $10^{-3}$. One can also confirm the sum of the individual density functions coincides with $\langle \rho(\alpha) \rangle$. In particular, the maximums of each distribution agree with the non-perturbative oscillations of the spectral density, see the right diagram of Figure \ref{fig:4}.

\begin{figure}
    \centering
    \includegraphics[scale=0.58]{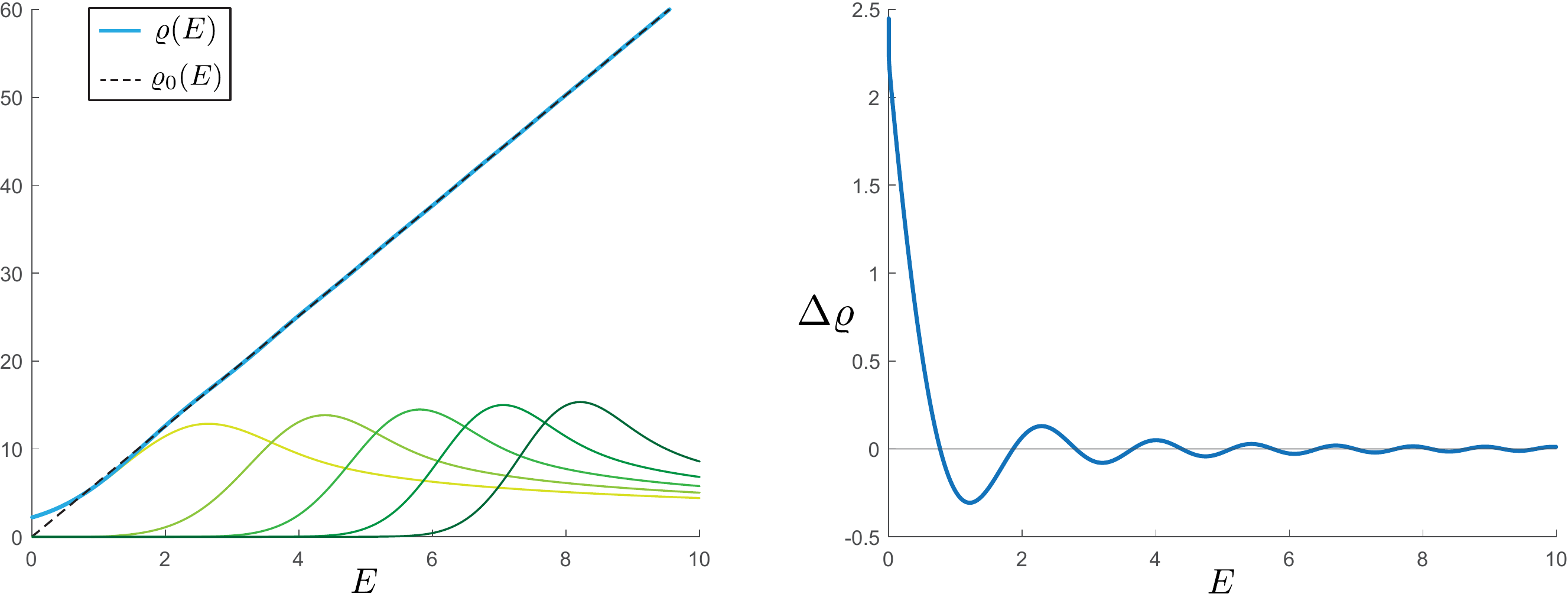}
    \caption{Spectrum of CJ gravity with $\hbar=t_2=1$ (see footnote \ref{foot}). The solid blue line in the left diagram gives the full non-perturbative spectral density $\varrho(E)$ which oscillates around the perturbative linear result. In the right diagram we observe the non-perturbative contribution to the spectral density obtained from $\Delta \varrho(E)=\varrho(E)-\varrho_0(E)/\hbar$.}
    \label{fig:5}
\end{figure}

\paragraph{CJ gravity:} Using the matrix model and the relation in (\ref{eq:84}) we can characterize the full spectrum of CJ gravity $\varrho(E)$, shown in the left diagram of Figure \ref{fig:5}. Non-perturbative effects, which are dominant at low energies, generate oscillations around the leading result $\varrho_0(E)=\frac{t_2}{\sqrt{\gamma}}  E$, given by the dashed line.\footnote{In the plots of Figure \ref{fig:5} the energy has been rescaled $E\rightarrow ({\rm const.})E$ so that when $t_2=1$, the perturbative result in terms of the rescaled energy is $\varrho_0(E)=2\pi E$. This rescaling is conceptually unimportant and was performed for convenience of the numerical calculations.\label{foot}} The oscillations become more visible after we plot $\Delta \varrho=\varrho(E)-\varrho_0(E)/\hbar$ in the right diagram of Figure \ref{fig:5}. Note that unlike $\varrho_0(E)$, the full spectral density does not vanish at the origin.

To better understand the structure of the oscillations we can use the probability density functions of the individual eigenstates of the matrix model. Using the definition of $\rho(\alpha)={\rm Tr}\,\delta(\alpha-M)$ we can rewrite (\ref{eq:84}) as
\begin{equation}\label{eq:87}
    \varrho(E)=\sum_{i=1}^{+\infty}\langle \mu_i(E) \rangle\ ,
    \qquad  {\rm where} \qquad
    \mu_i(E)=
    \frac{2}{\sqrt{\gamma}}\frac{\Theta(E-\alpha_i^2)}{\sqrt{\smash[b]{E-\alpha_i^2}}}\ .
\end{equation}
Each term $\mu_i(E)$ in $\varrho(E)$ is not a probability density function, but characterizes the contribution of each eigenvalue to the CJ gravity spectral density. The discreteness of the matrix model is partially washed away by the continuous contribution to the Hamiltonian $E(p)=p^2$ appearing in the definition of the operator $\mathbb{O}(\beta)$. Using the probability density of the individual eigenvalues, we can compute the ensemble average of $\mu_i(E)$ and obtain the green curves shown in the left diagram of Figure \ref{fig:5}. By summing all these contributions one recovers $\varrho(E)$, as expected.

\subsection{Spectral form factor}
\label{sec:3.2}

We now turn our attention to the spectral form factor, a useful diagnostic of certain universal features of quantum chaos \cite{Cotler:2016fpe,Liu:2018hlr}. In gravity, it is defined in terms of the following analytic continuation of the Euclidean partition function
\begin{equation}\label{eq:71}
S(\beta,t)=
Z(\beta+it,\beta-it)+Z(\beta+it)Z(\beta-it)\ .
\end{equation}
In the non-perturbative completion of CJ gravity, this can be readily written and computed in terms of the operator $\mathbb{O}(\beta_1,\beta_2)$. Just as before, it is convenient to separate and first analyze the contribution coming from the discrete and continuous pieces of $H=\bar{M}^2+ p^2$ appearing in the operator $\mathbb{O}(\beta_1,\beta_2)$. 

\begin{figure}
    \centering
    \includegraphics[scale=0.53]{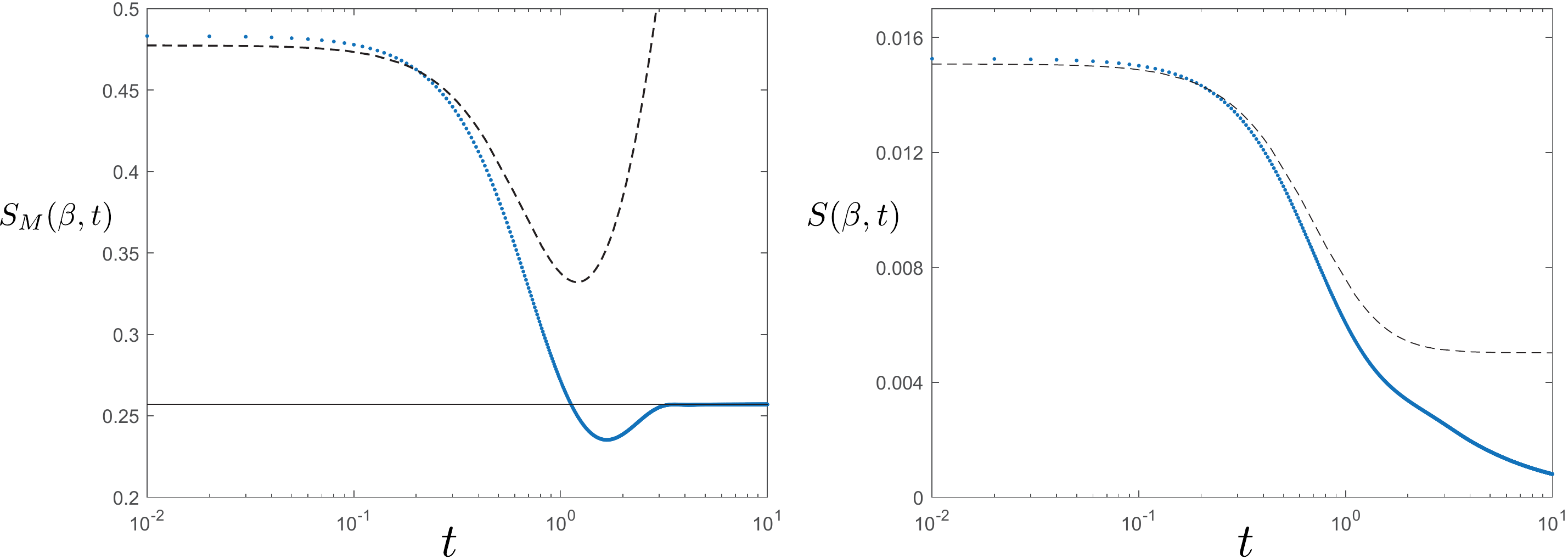}
    \caption{On the left and right diagram we plot the matrix model (\ref{eq:64}) and CJ gravity (\ref{eq:71}) spectral form factors for $\beta=1$ and parameters $\hbar=t_2=\beta=1$. While the dashed lines corresponds to the result to all orders in perturbation theory, the solid curves are the full answer including non-perturbative effects.}
    \label{fig:6}
\end{figure}

\paragraph{Matrix model:} The contribution from the matrix $\bar{M}^2$ to the spectral form factor is given by 
\begin{equation}\label{eq:64}
S_M(\beta,t)=
\langle {\rm Tr}\,e^{-(\beta+it)\bar{M}^2}
{\rm Tr}\,e^{-(\beta-it)\bar{M}^2}\rangle
\simeq 
\frac{t_2^2}{\pi\hbar^2(\beta^2+t^2)^{3/2}}+
\frac{\sqrt{\beta^2+t^2}}{2\pi \beta}
\ ,
\end{equation}
where in the second equality we have evaluated the result to all orders in perturbation theory using (\ref{eq:63}) and (\ref{eq:83}). The first and second terms are the disconnected and connected contributions to the operators appearing in (\ref{eq:64}). To include non-perturbative corrections we use (\ref{eq:86}) together with the numerical kernel $K(\alpha,\bar{\alpha})$. In the left diagram of Figure \ref{fig:6} we plot the final result, the dashed and solid lines corresponding to the perturbative and full result respectively.

The solid line in the left diagram of Figure \ref{fig:6} exhibits three different regimes: after the initial dip at small $t$, there is a ramp that is at later time followed by a plateau. Comparing with the dashed line, the perturbative result roughly captures both the dip and ramp, but fails to reproduce the proper late time behavior. As explained in \cite{Cotler:2016fpe}, both the ramp and plateau are central features that stem from the chaotic spectrum of the underlying microscopic model. While the ramp originates from the universal eigenvalue repulsion of the matrix model, the plateau is a consequence of the discreteness of the $\bar{M}^2$ spectrum. Given that $S_M(\beta,t)$ only includes the random matrix $\bar{M}^2$, it is ultimately no surprise all these features are present in the left diagram of Figure \ref{fig:6}.

\paragraph{CJ gravity:} Computing the spectral form factor (\ref{eq:71}) using the non-perturbative completion provided by the matrix model is very straightforward, since it can be written as
\begin{equation}\label{eq:67}
S(\beta,t)=
\langle \mathbb{O}(\beta+it,\beta-it)\rangle=
\frac{\pi}{\gamma\sqrt{\beta^2+t^2}}
S_M(\beta,t)\ ,
\end{equation}
where the crucial difference with respect to (\ref{eq:64}) is in the overall factor, coming from the $p$ integral in the operator $\mathbb{O}(\beta_1,\beta_2)$. Let us first analyze the behavior of this quantity to all orders in perturbation theory, that is given by
\begin{equation}
S(\beta,t)\simeq 
\frac{t_2^2}{\hbar^2\gamma(\beta^2+t^2)^{2}}+
\frac{1}{2 \gamma\beta}
\,\,
\overset{t\to \infty}{\longrightarrow}
\,\,
\frac{1}{2\gamma\beta}\ .
\end{equation}
From this expression, one might naively conclude its late time behavior is nothing more than the plateau of the spectral form factor, signaling the discreteness of the underlying spectrum \cite{Godet:2021cdl}. However, this is not correct, given that the perturbative expansion breaks down at very late times. This means that in order to properly diagnose the presence of a plateau one needs to compute $S(\beta,t)$ non-perturbatively. In the right diagram of Figure \ref{fig:6} we plot $S(\beta,t)$ for fixed $\beta$, the dashed and solid lines corresponding to the perturbative and full results respectively. We observe no plateau at late times, which is consistent with the fact the spectrum of CJ gravity is not discrete but continuous. Furthermore, the decay to zero at large times is exactly what one should expect for the spectral form factor of a system with a continuous spectrum \cite{Maldacena:2001kr}.

\subsection{Quenched free energy}
\label{sec:3.3}

We now study the free energy of CJ gravity in the canonical ensemble. In terms of the operator $\mathbb{O}(\beta)$ there are in principle two different ways one could define it
\begin{equation}\label{eq:75}
\mathcal{F}_Q(T)=-T\langle \ln \mathbb{O}(1/T) \rangle\ ,
\qquad \qquad
\mathcal{F}_A(T)=-T\ln \langle \mathbb{O}(1/T) \rangle\ ,
\end{equation}
referred as the quenched and annealed free energy respectively. While $\mathcal{F}_Q(T)$ is the actual quantity one is interested in, it involves calculating the ensemble average of the logarithm of $\mathbb{O}(\beta)$, that is in general very challenging to compute. The annealed free energy side steps this technical complication by commuting the logarithm with the ensemble average, so that it can be easily obtained by appropriately integrating the spectral density $\langle \rho(\alpha) \rangle$. The error associated to this modification is expected to be suppressed for high temperatures, meaning $\mathcal{F}_Q(T)\simeq \mathcal{F}_A(T)$ only for large $T$. Using techniques recently developed in \cite{Johnson:2021zuo}, we compute and analyze the low temperature behavior of the quenched free energy for CJ gravity. For an incomplete set of references regarding aspects of the computation of the quenched free energy in similar setups, see \cite{Johnson:2020mwi,Engelhardt:2020qpv,Okuyama:2021pkf,Janssen:2021mek,Godet:2021cdl,Johnson:2021rsh} and references within.

\paragraph{Matrix Model:} We start by analyzing the contribution to the free energy from the matrix $\bar{M}^2$ appearing in the operator $\mathbb{O}(\beta)$. Let us define the matrix model free energies as
\begin{equation}\label{eq:73}
F_Q(T)=-T
\big\langle 
\ln \big({\rm Tr}\,e^{-\bar{M}^2/T} \big)
\big\rangle\ ,
\qquad \qquad
F_A(T)=-T\ln \big\langle {\rm Tr}\,e^{-\bar{M}^2/T} \big\rangle\ ,
\end{equation}
closely related to (\ref{eq:75}). Writing the ensemble average explicitly, it is not hard to show the $T=0$ value of the quenched free energy is $F_Q(0)= \langle {\rm min}(\alpha_i^2) \rangle$, where $\alpha_i$ are the eigenvalues of $\bar{M}$. To compute the $T$ dependence of the quenched free energy we follow the approach of~\cite{Johnson:2021zuo}. Using the probability density functions of each eigenvalue (green and red curves in the left diagram of Figure~\ref{fig:4}) we first generate $b=1,\dots,q$ samples of the smallest magnitude eigenvalues. Using this, we can compute the free energy for a single instance of the ensemble
\begin{equation}\label{eq:74}
\lbrace \alpha_i^b \rbrace^{b=1,\dots,q}_{i=1,\dots,N}
\qquad \Longrightarrow \qquad
F^b(T
)=-T\ln
\bigg[
\sum_{i=1}^{N}e^{-(\alpha_i^b)^2/T}
\bigg]\ .
\end{equation}
Performing the average over the $b=1,\dots,q$ samples either before or after taking the logarithm (\ref{eq:73}), one can compute both the quenched and annealed free energy. 

\begin{figure}
    \centering
    \includegraphics[scale=0.57]{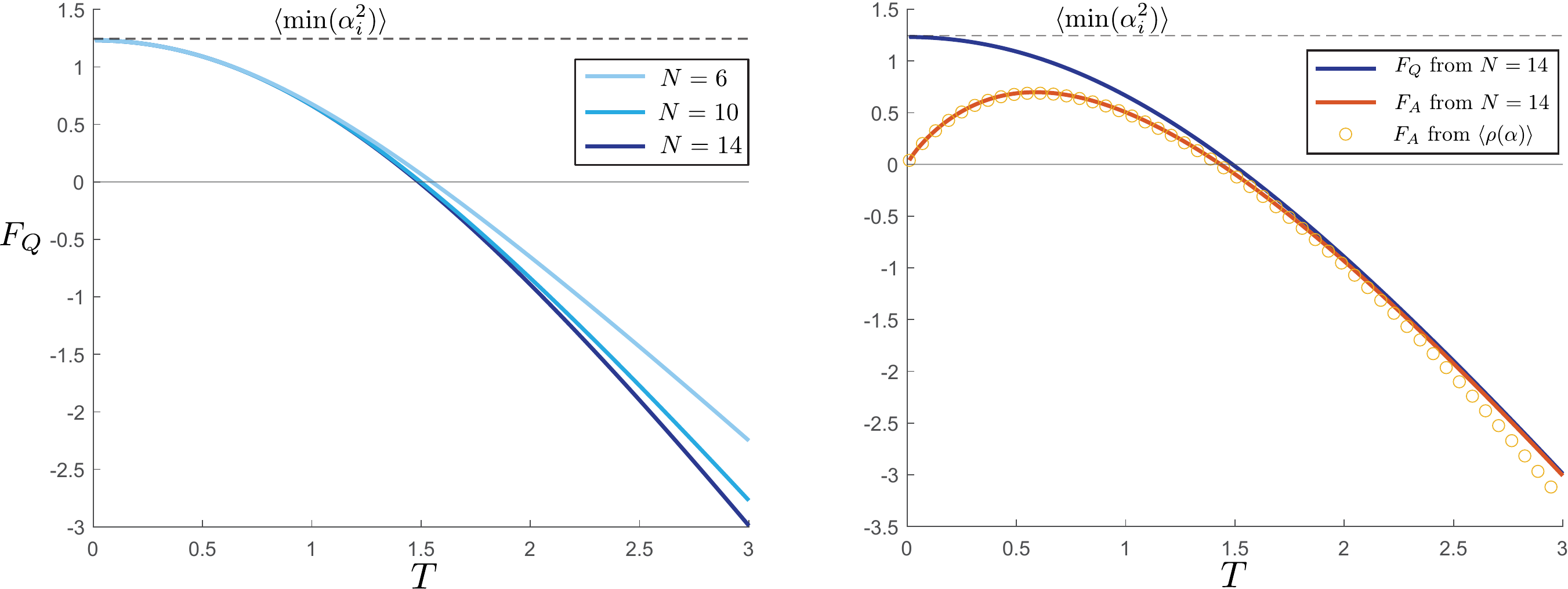}
    \caption{On the left diagram we plot the quenched free energy obtained from creating a sample of $10^6$ of the $N=6,10,14$ smallest magnitude eigenvalues $\alpha_i$ for $\hbar=t_2=1$. On the right, we show the quenched and annealed free energy with $N=14$ together with the annealed free energy directly computed from the spectral density $\langle \rho(\alpha) \rangle$ given by the solid blue line in the left diagram of Figure \ref{fig:5}.}
    \label{fig:8}
\end{figure}

There are two sources of errors in this procedure, given that neither the number of samples $q$ or eigenvalues $N$ can be infinity when doing the computation numerically. For our purposes, we have fixed the number of samples to $q=10^6$, which is enough for the numerical precision required here.\footnote{Using an ordinary laptop, it takes less than an hour to generate $q=10^6$ samples for the smallest $N=14$ eigenvalues, once the probability density function of each eigenvalue is known, given in the left diagram of Figure \ref{fig:4}.} In the left diagram of Figure \ref{fig:8} we plot the quenched free energy obtained from including only the $N=6,10,14$ eigenvalues $\alpha_i$. From that plot we observe the error caused by taking $N$ finite is negligible at low temperatures, i.e. the low energy behavior of the quenched free energy is completely determined by the smallest magnitude eigenvalues. 

The blue and orange curves on the right diagram of Figure \ref{fig:8} show the final results for the quenched and annealed free energies with $N=14$. As expected, both curves agree at large temperatures and have the expected $T=0$ behavior. To check the consistency of our numerical procedure we also plot the annealed free energy (orange circular markers) directly computed from $\langle \rho(\alpha) \rangle$. Perhaps the most important aspect of the right diagram of Figure \ref{fig:8} is that while the quenched free energy is monotonically decreasing with the temperature, the annealed has a local maximum at $T\sim 0.6$. Since the thermodynamic entropy is defined as $S(T)=-F'(T)$, only the quenched free energy gives a positive definite entropy. Any system with a discrete spectrum will have a positive definite entropy as the temperature vanishes, as can be shown from the following simple calculation
\begin{equation}\label{eq:77}
Z(T)=\sum_{n\ge 0}\Omega_ne^{-E_n/T}
\qquad \Longrightarrow \qquad
\lim_{T\rightarrow 0} S(T)=
\lim_{T\rightarrow 0}
\left(1+T \partial_T\right)\ln Z(T)=\ln \Omega_0 \ge 0\ ,
\end{equation}
where $\Omega_n$ is the degeneracy of the states with energy $E_n$. 

\paragraph{CJ gravity:} Using the matrix model free energies (\ref{eq:73}) we can compute the corresponding CJ gravity quantities defined by the non-perturbative completion in (\ref{eq:73}). However, instead of doing that, it is instructive to first analyze what one should expect for the low temperature entropy of a generic system that contains not only a discrete spectrum, but also continuous contribution
\begin{equation}
\rho(E)=\rho_c(E)\Theta(E-E_c)+\sum_{n\ge 0} \Omega_n \delta(E-E_n)\ .
\end{equation}
The continuous spectrum starts at $E_c$ and is characterized by $\rho_c(E)$. Assuming $\rho_c(E)$ is sufficiently smooth and well behaved near $E_c$, one can compute the zero temperature limit of the entropy as in (\ref{eq:77}) and find
\begin{equation}\label{eq:78}
\lim_{T\rightarrow 0}
S(T)=
\begin{cases}
\displaystyle \,\, \ln \Omega_0 \,\,\, \ , \,\, E_0<E_c\ . \\
\displaystyle \,\, -\infty \,\,\,\,\, \ , \,\, E_0>E_c\ .
\end{cases}
\end{equation}
We observe two radically different regimes depending on whether the minimum energy of the spectrum corresponds to the continuous or discrete contributions. This shows the thermodynamic entropy for a system with a continuous spectrum is not necessarily positive definite. 

\begin{figure}
    \centering
    \includegraphics[scale=0.58]{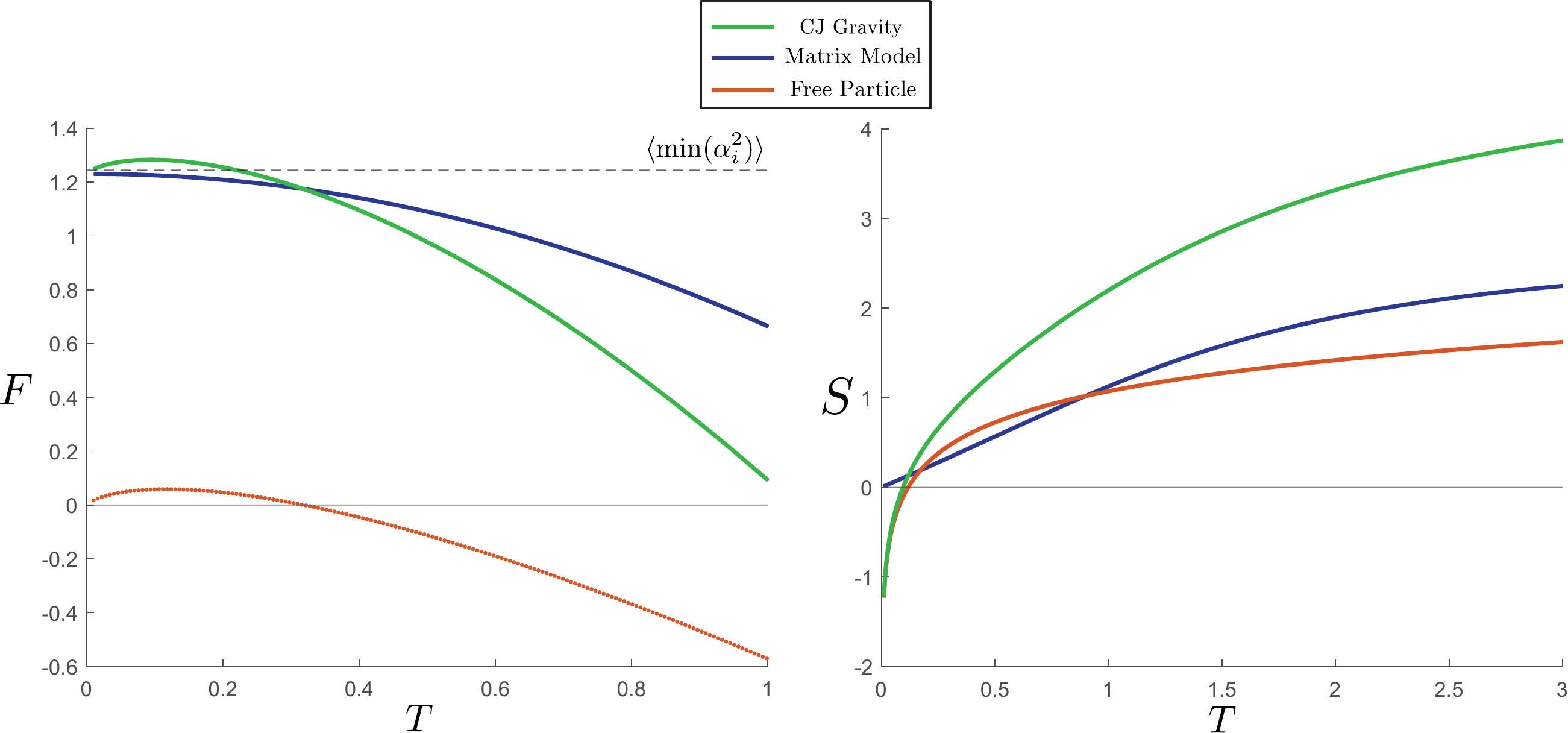}
    \caption{On the left diagram we plot the quenched free energy of the matrix model, free particle and CJ gravity with $\hbar=t_2=1$. On the right, we observe the corresponding thermal entropies, obtained from $S(T)=-F_Q'(T)$, with the CJ gravity result becoming negative for small temperatures due to the free particle contribution to its spectrum.}
    \label{fig:9}
\end{figure}

What are the implications of this general discussion to CJ gravity? The Hamiltonian of the non-perturbative completion provided by the matrix model $H=\bar{M}^2+p^2$ results in the spectrum given in (\ref{eq:87}), which contains continuous contributions. The quenched free energy (\ref{eq:75}) can be written as the sum of two terms
\begin{equation}\label{eq:80}
\mathcal{F}_Q(T)=
F_Q(T)
-\frac{1}{2} T\ln(\pi T/\gamma)\ ,
\end{equation}
arising from the discrete $\bar{M}^2$ and continuous $p^2$ respectively appearing in $H=\bar{M}^2+p^2$. Since $p$ is not affected by the ensemble average, the computation of $\mathcal{F}_Q(T)$ is immediate once $F_Q(T)$ in (\ref{eq:73}) is known. Due to the second term in (\ref{eq:80}), the entropy diverges to minus infinity in the zero temperature limit, in agreement with (\ref{eq:78}). This means that at low enough temperatures the CJ gravity quenched free energy is not monotonic and the thermodynamic entropy diverges to minus infinity. We confirm this by plotting everything explicitly in Figure \ref{fig:9}.

In summary, we have shown the quenched free energy of CJ gravity is not monotonic meaning the entropy goes to arbitrarily negative values for low temperature. This is a consequence of the fact the non-perturbative completion of CJ gravity has a spectrum that contains a continuous component, the ``free'' particle mode. Whether this is a feature of flat space quantum gravity related to the infinite volume of Minkowski space or a bug that needs to be remedied remains to be seen.

\section{Scrambling and short time chaos}
\label{sec:Scrambling}

The ultimate goal of the tools developed in this paper up to this point is the non-perturbative study of gravitational scattering in flat space. For this problem to be non-trivial in two dimensions we need to introduce matter. A careful analysis of how to do this, the construction of the corresponding S-matrix and its matrix model completion is a very interesting problem we wish to return to in a future publication. We offer some preliminary comments on the general strategy for attacking this problem in Section \ref{sec:Discussion}. 

In this Section, we instead switch gears to a slightly simpler problem conceptually: The semiclassical dynamics of CJ gravity coupled to probe matter at a finite cut-off. The finite cut-off allows us to study the dynamics of the system in the usual Hamiltonian evolution picture. The main output of the subsequent discussion is that CJ dynamics exhibit the same maximally chaotic behavior of out-of-time-order correlators (OTOCs) as JT gravity. Translating these findings in the S-matrix language, as in \cite{Polchinski:2015cea} for example, will be left for the future.

\paragraph{Equivalence with charged particle:} In two dimensions the coupling to matter will be facilitated by the fact that the whole gravitational dynamics can be mapped to that of a charged particle. Indeed, using the finite cut-off picture, one can show that the boundary of the spacetime follows the trajectory of a particle under the influence of an constant electric field. 

As explained in Section \ref{SectionLorentz} all solutions are patches of global Minkowski that can be reached through a change of coordinate of the type\footnote{The global Minkowski metric is $ds^2=dx^+dx^-=-dT^2+dR^2$ with $x^\pm=R\pm T$, so that the global retarded time and radial coordinate are $U=-x^-$ and $R=\frac{x^++x^-}{2}$.}
 \begin{equation}
   -x^-= F(u)\ , 
   \qquad \qquad 
   \frac{x^++x^-}{2}= \frac{r+G'(u)}{F'(u)}
   \ ,
\end{equation}
which is nothing but a BMS$_2$ transformation of the global coordinates.\footnote{The relation between $F(u),G(u)$ used here and the boundary modes $f(u),g(u)$ used in Section \ref{SectionLorentz} is $F(u)=\frac{1}{P_0}e^{-P_0f(u)}$ and  $G(u)=g(u) -\frac{1}{4}F^2(u)$.}
Following the logic of Section \ref{SectionLorentz}, the fields $F(u)$ and $G(u)$ satisfy two equations of motion
\begin{equation}\label{eomsfg}
\frac{F''(u)}{F'(u)}+\frac{\Lambda}{\gamma}=0\ ,
\qquad \qquad
\frac{d}{du}\left[F'(u)+\frac{G''(u)}{F'(u)}\right]=0\ .
\end{equation}

Let us consider the location of the physical boundary of the spacetime. In target coordinates it corresponds to a line of fixed radius $r=1/\gamma\varepsilon$, which becomes
\begin{equation}
    (x^+(u),x^-(u))=\left( F(u),  \frac{\frac{1}{\gamma\varepsilon}+G'(u)}{F'(u)}-F(u)\right)\ ,
    \label{Trajectory}
\end{equation}
in global coordinates. Suppose the dynamics for this world line is that of a relativistic particle under the influence of a constant electric field. The equations of motion of such system are
\begin{equation}
    \ddot{x}^\mu+a\,F^\mu_\nu\dot{x}^\nu=0\ ,
    \label{Relativistic}
\end{equation}
with $F_{\mu\nu}= \epsilon_{\mu\nu}$ and $a$ the acceleration of the particle, i.e. its electric charge to mass ratio. One can then check that expanding at small $\varepsilon$ one recovers the two equations in \eqref{eomsfg} with the identification
\begin{equation}
    a=\frac{\Lambda}{\gamma}=\frac{2\pi}{\beta}\ .
\end{equation}
But this is not enough yet to complete the matching of the boundary dynamics with that of a charged particle. We need to take into account the fact that $u$ is the holographic time, according to its definition in Section \ref{SectionLorentz} (the one whose Euclidean continuation is $\beta$-periodic), but it is \emph{not} the proper time of the boundary. Indeed, let us look at the solution space of \eqref{eomsfg}. The most general solution is given by
\begin{equation}
F(u)=e^{-2\pi u/\beta}\alpha_1+\alpha_2
\ ,\qquad 
G(u)=-\frac{1}{4}\,\alpha_1^2\,e^{-4\pi u\beta}-\frac{\alpha_3\,\beta}{2\pi}\,e^{-2\pi u\beta}+\alpha_4+\alpha_5\, u\ .
\label{RindlerDiffeo}
\end{equation}
Note that the mode $\alpha_4$ has no effect on the location of the boundary since only the derivative of $G(u)$ appears in the defining equation \eqref{Trajectory}. For these values of the fields $F(u)$ and $G(u)$, the norm of velocity becomes $\dot{x}^2=-2a/\gamma\varepsilon +\mathcal{O}(\varepsilon^0)$. The holographic time $u$ and proper time $s$ at small $\varepsilon$ are therefore related by a divergent factor
\begin{equation}\label{eq:400}
    s=\sqrt{\frac{2a}{\gamma\varepsilon}}\,u\ .
\end{equation}
Equivalently, this means that the metric induced on the finite cut-off boundary is $ds^2_{\partial}=-\frac{2a}{\gamma\varepsilon}du^2$. Rescaling the time in \eqref{Trajectory} induces the following replacement 
\begin{equation}
    a \quad \longrightarrow \quad a_\partial\equiv a\,\sqrt{\frac{\gamma\varepsilon}{2 a}}=\sqrt{\frac{\varepsilon\Lambda}{2}}=\frac{q}{m}\ .
    \label{NewAcc}
\end{equation}
This new acceleration is the physical acceleration of the boundary particle, i.e. the ratio between its charge and its mass. We observe that when the boundary is sent to infinity, its acceleration vanishes, this is consistent since a trajectory with small acceleration should lie close to null infinity.

\paragraph{Short-term chaos in flat space:}

We would like now to comment on the existence of chaos in the putative dual to CJ gravity. The obvious issue we are facing is that we do not know the dictionary between the bulk and the holographic dual and in particular we do not know how to compute boundary correlation functions using some holographic prescription. Yet the prototypical diagnostic of chaos is the OTOC of local operators. Therefore, to make progress, we will have to make an educated guess for how to compute this OTOC.

In AdS holography this is done using the extrapolate dictionary: the boundary two-point function is matched with the appropriate limit of a bulk two-point function which is itself written as a sum over paths between the two points weighted by the exponential of their length. One can then perform the usual semiclassical WKB approximation of this sum and obtain that the boundary two-point function is given by the geodesic length between the two points:
\begin{equation}
    \langle O(x_1)O(x_2)\rangle \sim e^{-M L_{\mathrm{geodesic}}(x_1,x_2)}.
    \label{Corr}
\end{equation}
This approximation is valid if the CFT operators are dual to heavy bulk fields.
We are going to assume that this formula can be imported to the flat case and that it accurately describes the correlation function of local operators in the holographic dual of CJ gravity. Assuming this dictionary we will be able to conclude on the existence of chaos in the dual theory.

In order to compute boundary correlators we need to couple the pure gravity theory to matter. We have seen that the whole pure gravity theory can be reformulated in terms of the dynamics of a charged particle whose trajectory is nothing but the boundary of the spacetime. But this is only for one boundary. In order to find the location of the second boundary, we need to remind ourselves that the Poincaré symmetry is gauged in the bulk. Indeed acting with an $\mathrm{ISO}(1,1)$ symmetry does not change the resulting cutout geometry, therefore in order to determine the trajectory of the second particle we simply have to ask for the total charge to vanish. The Poincaré charges for a charged relativistic particle are the momentum and boost charges:
\begin{equation}
     P^\mu=m\dot{x}^\mu+q A^\mu\ ,\qquad \quad M^{\mu\nu}=x^\mu\left(P^\nu-\frac{q}{2}A^\nu\right)-x^\nu\left(P^\mu-\frac{q}{2}A^\mu\right)\ .
\end{equation}
The tensor $M^{\mu\nu}$ is anti-symmetric, which in two-dimensions implies there is a single non-trivial component.

\begin{figure}
    \centering
    \includegraphics[scale=0.4]{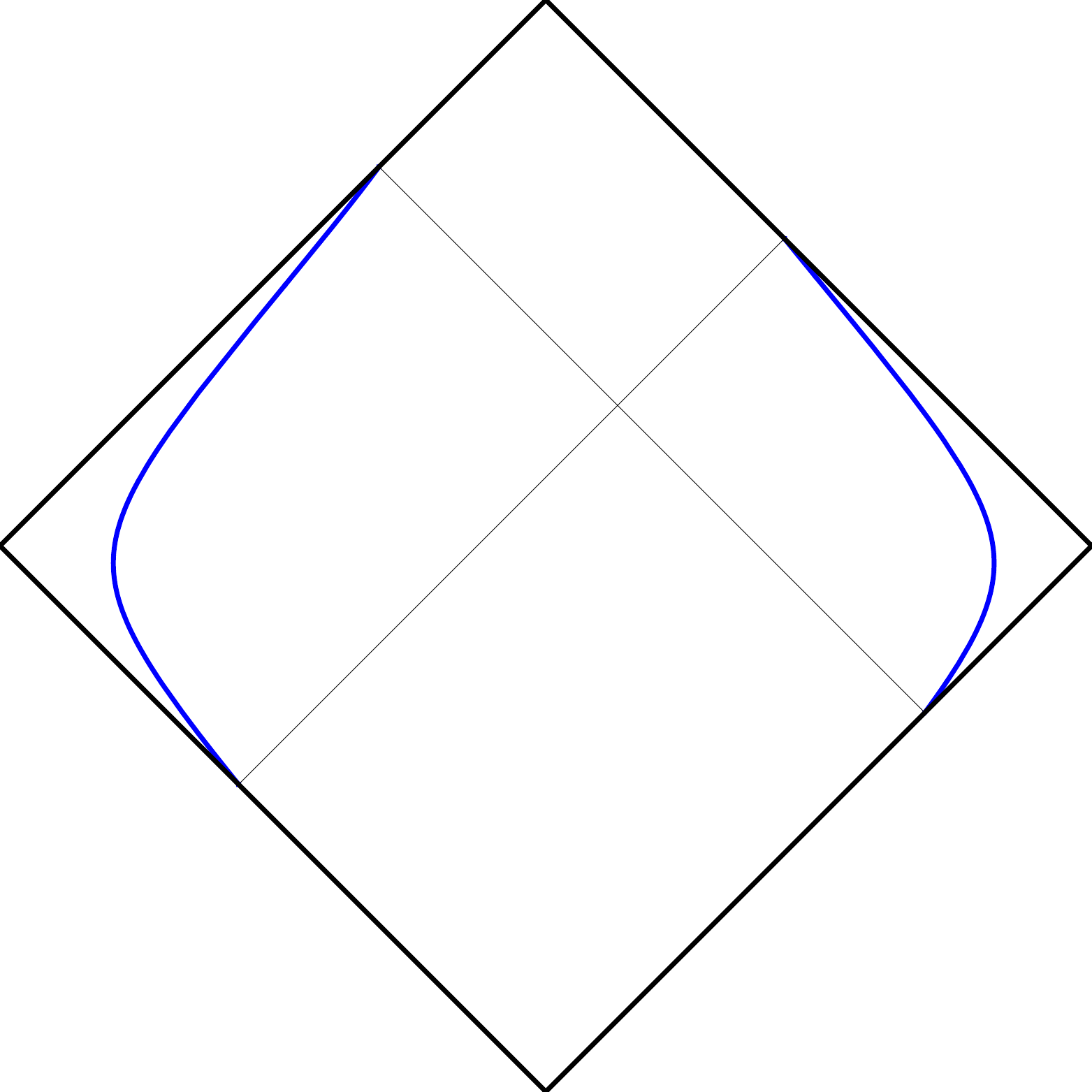}
    \caption{The two boundaries are given by the trajectories of two particles of opposite electric charge and whose total $\mathrm{ISO(1,1)}$ charges vanish.}
    \label{Particle}
\end{figure}

We ask the total charges $P_L+ P_R$ and $M_L+M_R$ to vanish, see Figure~\ref{Particle} for an example of resulting configuration. A consequence of this condition is that the two accelerations are opposites of each other (the total central charge of the Maxwell algebra vanishes) and the bifurcation horizons of the two Rindler spaces match. Also, the distance between both accelerated trajectories and the bifurcate horizon is the inverse of the acceleration so that the distance between the two boundaries is 
\begin{equation}
d=\frac{2 }{\vert a_\partial\vert}=\sqrt{\frac{8}{\Lambda \varepsilon}}\ ,
\label{VacuumDistance}
\end{equation}
where we observe that the distance becomes infinite when $\varepsilon \to 0$.

\begin{figure}
    \centering
    \includegraphics[scale=0.4]{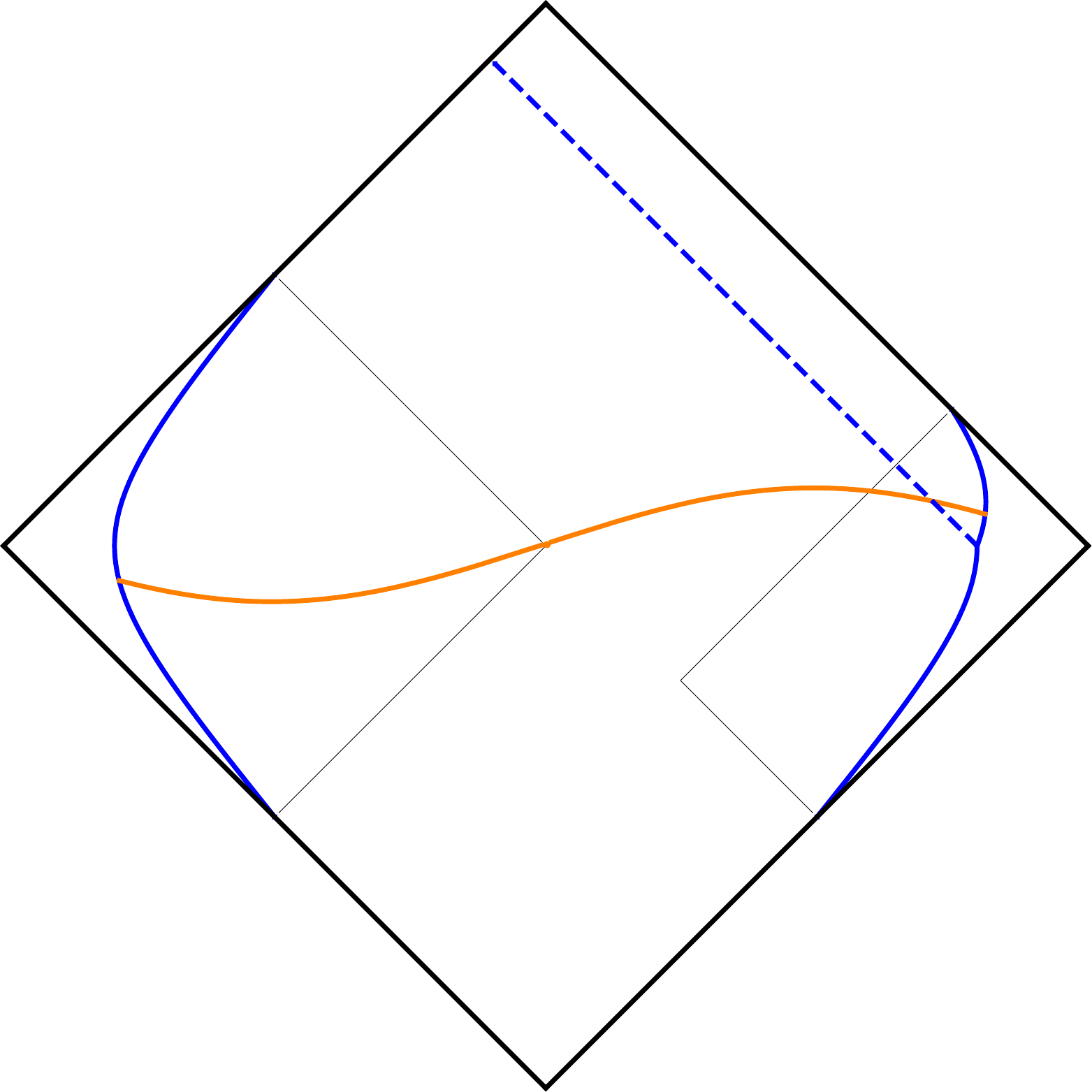}
    \caption{The insertion of an operator on the right boundary corresponds to the emission of a light particle (dashed line). The boundary particle is kicked outward which changes the location of the horizon. The computation of the OTOC involves a geodesic that connects the two boundaries (orange line). This geodesic intersects the left boundary at proper time $-s$ and the right one at $s$ while the insertion of the operator is at $s=0$. }
    \label{ParticlePert}
\end{figure}

With this reformulation of the theory it is now easy to understand how the matter backreacts on the spacetime. The type of physical process we are going to consider are shockwaves which correspond to matter configurations whose stress tensor is localized on a null line. This null trajectory intersects the boundary at a certain time that corresponds to the emission of the shock. The backreaction on the spacetime is encoded in the behaviour of the boundary particle, indeed the equations that govern its dynamics \eqref{eomsfg} are now sourced by a delta-localized stress tensor which translates into a kick of the boundary trajectory. In the particle interpretation it corresponds to a process where a charged particle of mass $m$ emits a light particle. The new charged particle is kicked and its mass has changed. Both the final momentum and mass of the boundary particle can be computed by requiring conservation of local momentum. As a consiequence of the kick, the boundary particle reaches null infinity earlier than its unperturbed counterpart which results in a change of the location of the horizon. See Figure~\ref{ParticlePert} for a representation of the process.

At the location of the emission we have 
\begin{equation}
    m \dot{x}= m_{\mathrm{pert}}\dot{x}_{\mathrm{pert}}+P_{\mathrm{shock}}\ ,
\end{equation}
where $x$ is the trajectory of the unperturbed boundary particle, $\dot{x}_{\mathrm{pert}}$ is the perturbed one and $P_{\mathrm{shock}}=(0,-\delta E)$ is the momentum of the shock (the time derivative is with respect to the proper time) in $x^\pm$ coordinates. Since the electric field is continuous across the emission, we deduce that there is also conservation of the momentum $P_R=P_{R,\mathrm{pert}}+P_{\mathrm{shock}}$. Now we use the fact that before the kick that total charge vanishes: $P_L=-P_R$, to obtain
\begin{equation}
    P_L+P_{R,\mathrm{pert}}+P_{\mathrm{shock}}=0\ .
\end{equation}
From the conservation of the local momentum at the emission we deduce that two of the total $\mathrm{ISO}(1,1)$ charges vanishes. One can also easily derive that the total boost charge vanishes from the fact that the three trajectories intersect at the point of emission. Therefore this process preserves the gauging of the Poincaré symmetry. 

We now have all the ingredients to set up the computation of the OTOC. In the dual theory the introduction of a shock corresponds to the action of an operator on the state. Our analysis in Section \ref{sec:Euclidean} strongly suggests that the right state is the Hartle-Hawking state that is obtained through a Euclidean preparation on the half disk. 

In order to derive the OTOC it is more practical to work with the proper time since it is better adapted to the particle dynamics, eventually we will reintroduce the holographic time to diagnose scrambling in the dual theory. The shock is inserted at $s=0$ on the right boundary and in this new state we want to probe the evolution of the correlations between the two sides. To do so we compute a two-sided two-point function using formula \eqref{Corr}:
\begin{equation}\label{eq:401}
    \frac{\bra{\text{HH}} W^\dagger(0)O_L(-s)O_R(s)W(0)\ket{\text{HH}}}{\bra{\text{HH}} W^\dagger(0)W(0)\ket{\text{HH}}}\sim e^{-M L_{\mathrm{pert}}(s)},
\end{equation}
where $L_{\mathrm{pert}}$ is the length of the geodesic in the shockwave background. This two-point function probes the correlations between the two sides at a proper time $s$ after the shock is inserted. We solve the charge conservation equations and find the new mass of the boundary particle is 
\begin{equation}
    m_{\mathrm{pert}}=m\,\sqrt{1-2\,\frac{\delta E}{m}}\ .
\end{equation}
The full right boundary trajectory in $x^\pm$ coordinates is given by
\begin{equation}
x^\mu(s) =\frac{1}{ 2a_\partial}\times  \left\{
    \begin{array}{ll}
       \big(e^{a_\partial\, s}\, ,\, e^{- a_\partial\,s}\big)\ , & s<0\ , \\[5pt]
        \left(e^{\frac{a_\partial\, s}{y}}\, , \, y^2\,e^{-\frac{a_\partial\, s}{y}}+2\,\frac{\delta E}{m}\right)\ , & s>0\ ,
    \end{array}
    \right.
\end{equation}
where $y=\sqrt{1-2\,\frac{\delta E}{m}}$ is the ratio between the outgoing mass and the ingoing one. The first part corresponds to the unperturbed trajectory, and the second to the kicked boundary whose trajectory depends on the ratio between the energy that is injected and the mass of the ingoing boundary particle. The left boundary mirrors the unperturbed trajectory. Using this result we can calculate our two-point function. Computing the length of the right geodesic (see Figure~\ref{ParticlePert}) and expanding for low-energies approximation $\delta E \ll m$, we obtain
\begin{equation}
\langle O_L(-s)O_R(s)\rangle_W \sim 1-\frac{M\,\delta E}{ a_\partial \,m}\left(e^{ a_\partial\, s}-1\right)\ ,
\end{equation}
where we are working in Planck units. We have normalized by the value of the same two-point function in the unperturbed background, which is given by $e^{-M d}$, where $d$ is the divergent length given in \eqref{VacuumDistance}. 

This is not the end since this formula is not yet written in terms of holographic quantities. Indeed if we introduce the rescaling parameter $\mu(\varepsilon)=\sqrt{\frac{\beta\gamma\varepsilon}{4\pi}}$ we have that the proper time of the boundary and the holographic time are related according to $s=\mu(\varepsilon)^{-1}\,u$ while the acceleration of the boundary is related to the temperature according to $ a_\partial=\mu(\varepsilon)\frac{2\pi}{\beta}$ (see equation \eqref{NewAcc}). The ratio $\delta E/m$ controls the backreaction of the shock on the geometry, it measures the strength of the coupling between gravity and matter. The energy of the shock and the mass of the boundary particle are measured in the same units, therefore if one of the two undergoes a rescaling, the other one will be rescaled accordingly in such a way that the ratio $\delta E/m$ is invariant. We note that $\mu(\varepsilon)\to 0$ when the boundary is sent to infinity. We conclude that in order to obtain a meaningful correlation, i.e. a correlation that does not vanishes in the limit $\varepsilon\to 0$,\footnote{The correlation as it is looks like $1-\#/\sqrt{\varepsilon}\,\sim\, e^{-\#/\sqrt{\varepsilon}}\to 0$. } we need to scale the mass of the probe boundary operator with $\varepsilon$ according to 
\begin{equation}
    M = \mu(\varepsilon)\Delta\ ,
\end{equation}
where $\Delta$ should be understood as the weight of the putative dual operator. We conclude that only operators that are dual to \emph{massless fields} have a non-trivial correlation in the dual theory. This results seems consistent, only massless particles reach $\mathscr{I}^\pm$.

Rewriting the correlation function in terms of holographic quantities we find
\begin{equation}\label{eq:402}
    \langle O_L(-u)O_R(u)\rangle_W \sim 1-\frac{\beta\,\Delta\,\delta E}{ 2\pi \,m}\left(e^{2\pi u /\beta}-1\right)\ .
\end{equation}
The effect of the perturbation becomes of order one after a scrambling time
\begin{equation}
u_\star=\frac{\beta}{2\pi}\log\frac{2\pi\, m}{\beta\,\Delta\,\delta E}\ .
\label{ScramblingTime}
\end{equation}
We conclude that the theory does exhibit chaos for operators that are dual to massless exitations. The argument of the logarithm is large since we are working in a regime of small gravitational coupling (i.e. the backreaction of the shock is small) and should be thought of as the large $N$ parameter of the dual theory. This is also consistent with similar computations of scrambling time in AdS where, in that case, the large parameter is the entropy of the black hole.

Some aspects of this result are puzzling though. We have seen that the gravitational entropy of the solution is given by the value of the dilaton at the horizon, the latter being set by $\phi_h$. Since our formula for the scrambling time does not depend on $\phi_h$ we conclude that it is independent of the gravitational entropy. However one can also look at this result as being the leading order in a $\varepsilon$-expansion and compute ``finite cut-off'' corrections. One can check that the finite cut-off version of \eqref{eomsfg} is exactly the same with the replacement 
\begin{equation}
    \frac{\Lambda}{\gamma}\to  \frac{\Lambda}{\gamma(1-\varepsilon\,\phi_h)}\ .
\end{equation}
This is no surprise, at finite location of the boundary is sensitive to the subleading term in the dilaton expansion. Now all the previous analyses hold with the aforementioned replacement, which amounts to a change of the temperature $\beta\to \beta(1-\varepsilon\,\phi_h)$. Implementing the replacement in \eqref{ScramblingTime} allows us to derive the finite cut-off scrambling time, which now depends on $\phi_h$: The backreaction is now sensitive to horizon physics. The holographic interpretation of this phenomenon is unclear though, it seems that the true UV holographic scrambling time only cares about the asymptotic value of the dilaton which is insensitive to $\phi_h$. This is not the case in AdS, and in particular in JT gravity, where the horizon value of the dilaton appears in the logarithm since it measures the entropy above extremality \cite{Maldacena:2016upp}. In Euclidean AdS, it is the boundary value of the dilaton that sets also its value at the horizon, there is no extra degree of freedom coming from the dilaton. This is one instance of a generic property of thermal solutions in AdS: The value of the vacuum expectation value is determined by the non-normalizable mode. On the other hand in Euclidean flat space, i.e. on the disk, the horizon value remains a free parameter even after fixing the boundary value. It is maybe this decoupling that explains why the scrambling time acquires this degree of universality in CJ gravity.

\section{Discussion}
\label{sec:Discussion}

In this work, we have analyzed CJ gravity, a two-dimensional theory of flat quantum gravity, from both Lorentzian and Euclidean viewpoints.
Here we recall a few of the salient points.

In Lorentzian signature, we motivated boundary conditions using intuition from four-dimensional asymptotically flat quantum gravity.
In a departure from AdS$_2$ quantum gravity, we left the asymptotic form of the metric unfixed in CJ gravity, opting instead to fix the gradient of the scalar dilaton field.
This choice was natural from the four-dimensional perspective, where the asymptotic form of the metric in Bondi gauge is fixed at leading order; as we discussed, the fixing of the scalar dilaton and its gradient enables the construction of an asymptotically fixed coordinate frame in two dimensions.
Under these boundary conditions, we constructed the classical phase space, reviewed the boundary action formulation of the CJ theory \cite{Godet:2020xpk}, computed the energy of on-shell configurations and the classical S-matrix.

In Euclidean signature, we recalled the CJ partition functions with an arbitrary number of boundaries computed in \cite{Godet:2020xpk}.
We then observed that there is a non-perturbatively well-defined system consisting of a Hermitian matrix model and a free particle which precisely reproduces the entire CJ gravity topological expansion. Studying the matrix model through the method of orthogonal polynomials, we constructed several objects of interest in gravity such as the density of states, spectral form factor, and quenched free energy, incorporating fully non-perturbative effects in each calculation.

There are many additional questions concerning flat space quantum gravity which might be addressed using our model or variants thereof.
In what follows we outline several such questions that we consider particularly interesting or within reach.

\subsection*{The free particle sector}
A feature of our proposed non-perturbative completion of CJ gravity that strikes as peculiar at first sight is the existence of a decoupled mode with the dispersion relation of a one-dimensional non-relativistic particle, in addition to the matrix model degrees of freedom. This mode has important physical consequences since it renders the fine grained spectrum continuous, washing off the effects of matrix model eigenvalue repulsion in the late time behavior of the spectral form factor and drastically affecting the behavior of free energy at low temperatures. 

The mathematical origin of this mode is simple to track: The symmetries of the CJ theory are generated by the Maxwell algebra (\ref{eq:Maxwell}) which has four generators and constitutes a central extension of the Poincare algebra. The central charge implies an additional zero mode of the Euclidean partition function which is, in turn, responsible for the temperature dependence that forced us to introduce the free particle sector. Indeed, the disk partition function in a \emph{fixed} central charge sector, i.e. of the CGHS model, has the $\beta$ dependence \cite{Afshar:2021qvi} obtained by just the matrix contribution to the Bondi Hamiltonian in  (\ref{matrixobssummary}). 

Nevertheless, its introduction is necessary for a number of physically significant reasons. Firstly, the central charge controls the temperature of the solution which is otherwise completely fixed in CGHS gravity \cite{Godet:2021cdl}. Perhaps more importantly, the dynamical central charge was important for the existence of a cylinder contribution to the two-boundary partition function which is the cornerstone for our matrix model interpretation ---a property not shared by the CGHS model. And lastly, its presence allowed for the BF formulation of the theory that provided the rigorous tools for constructing the path integration measure. It is, therefore, a feature of our model that requires further understanding, especially given that it does not appear in the same decoupled way in the semi-classical analysis of the theory. The fact that its presence renders the Bondi spectrum continuous suggests a possible connection to the infinite volume of two-dimensional flat space but more research is needed to establish this interpretation.

Further light on the meaning of the free particle sector might be gained by studying deformations and extensions of CJ gravity. A class of deformations of the dilaton potential can be conveniently incorporated by allowing surfaces with defects in the topological expansion \cite{Witten:2020wvy,Maxfield:2020ale} (see also \cite{Rosso:2021orf}). In the case of CJ gravity, the inclusion of defects has a dramatic effect on the topological expansion of its path integral, given that the integral over the dilaton $\Phi$ no longer restricts the possible surfaces to be only the disk and cylinder. The presence of the defects allow for more general surfaces, as there exist (for instance) three-holed sphere geometries with an everywhere-flat metric except at a single conical defect \cite{Flat2} (see also \cite{Flat1}).  

On the other hand, the supersymmetric extension of CJ gravity is analyzed in \cite{Toappear}, where it is shown a very similar procedure can also be applied in that case. The topological expansion of the path integral can be computed exactly and non-perturbatively completed by an appropriately defined random matrix model. Most importantly, the matrix operator required for this matching is exactly the same, $\mathbb{O}(\beta)$ given in (\ref{matrixobssummary}). This suggests the free particle sector is perhaps not an accident of CJ gravity, but a more general feature of a larger class of two-dimensional asymptotically flat theories, whose origin and meaning must be better understood.

\subsection*{The quantum S-Matrix}

The ultimate goal of the quantum theory of gravity in asymptotically flat space times is the non-perturbative definition of the S-matrix. This is a unitary map from an asymptotic past Hilbert space $\mathcal{H}^-$ to an asymptotic future Hilbert space $\mathcal{H}^+$.
These Hilbert spaces are supposed to consist of asymptotic incoming and outgoing states, respectively, and are associated with the null surfaces $\mathscr{I}^-$ and $\mathscr{I}^+$, respectively. As the classical phase space of CJ gravity can be explicitly constructed, we here propose (following \cite{Ashtekar:1981bq}) that the asymptotic Hilbert spaces $\mathcal{H}^-$ and $\mathcal{H}^+$ should be ``identified'' with the Hilbert space resulting from canonical quantization of the respective phase spaces $\mathcal{P}^\pm$. Quantization amounts to constructing the Hilbert spaces and commutation relations\footnote{In order to ensure the quantization procedure has an essentially unique outcome we assign affine linear structures to $\mathcal{P}^\pm$ by considering $(x_i^\pm, p_j^\pm)$ to be preferred coordinate systems on $\mathcal{P}^\pm$ up to affine linear transformations.}
\begin{equation}
    \mathcal{H}^\pm = L^2(x_1^\pm, x_2^\pm)\ ,
    \qquad \qquad
    [x_j^\pm, p_{j'}^\pm] = i \delta_{jj'}\ ,
\end{equation}
with all other commutators between the canonical coordinates vanishing.\footnote{It may seem at this point that there is a natural isomorphism between $\mathcal{H}^+$ and $\mathcal{H}^-$, especially if $S$ is a canonical transformation. However, this is not necessarily so.  In general, there is no natural equivalence between the Hilbert spaces obtained by quantizing $\mathbb{R}^4$ with different affine linear structures, even if the symplectic structures are the same.  So, in CJ gravity, there is only a natural equivalence between $\mathcal{H}^+$ and $\mathcal{H}^-$ if the change of coordinates mapping $(x_i^-,p_j^-) \to (x_i^+, p_j^+)$ is actually an affine linear transformation.}

The classical S-matrix is a canonical transformation discussed in Section \ref{sec:bdy-action-phase-space} and we may quantize it to obtain a canonical quantum S-matrix as follows.
Because the CJ phase space is topologically trivial, any canonical transformation is actually an exact canonical transformation, i.e.  generated by a Hamiltonian vector field.
Hence there exists a generating function $G : \mathcal{P}^- \to \mathbb{R}$ such that the vector field $V_G$ defined by the one-form equation
\begin{equation}
    dG = \omega^-(V_G,\cdot ) \ ,
\end{equation}
can be exponentiated to yield a one-parameter family of canonical transformations $S_t : \mathcal{P}^- \to \mathcal{P}^+$ with
\begin{equation}
    \frac{dS_t}{dt} = V_G(S_t) \ , \qquad t \in [0,1]\ ,
\end{equation}
where $S_1=S$. The real function $G$, upon quantization, is supposed to become a Hermitian operator $\mathcal{G}$.
Then, the canonical quantum S-matrix, which we denote $\mathcal{S}$, is the exponential
\begin{equation}
    \mathcal{S}_t = \mathcal{T} \exp \left( -i \int_0^t dt\; \mathcal{G} \right) \ ,
\end{equation}
where $\mathcal{S}_1 = \mathcal{S}$ and $\mathcal{T}$ is the time-ordering operator.

With the canonical quantum S-matrix $\mathcal{S}$ and the asymptotic Hilbert spaces $\mathcal{H}^\pm$, we can finally discuss our proposal for a non-perturbative completion for certain matrix elements of $\mathcal{S}$.
In order to create nontrivial incoming and outgoing states in $\mathcal{H}^\pm$ that are primed for non-perturbative completion, we will act on Euclidean path integral states with probe matter operators that preserve the corresponding Hilbert spaces.
Namely, we act with operators $\mathcal{O}_\text{in}$ and $\mathcal{O}_\text{out}$ which preserve
\begin{equation}
    \mathcal{O}_\text{in} : \mathcal{H}^- \to \mathcal{H}^- \ , \qquad \quad \mathcal{O}_\text{out} : \mathcal{H}^+ \to \mathcal{H}^+ \ .
\end{equation}
At this point, we must address a subtlety concerning the superselection effect for the variable $\phi_h$ we observed in Euclidean signature.
Our prescription for a non-perturbative completion of CJ gravity involves the Euclidean path integral, where the variable $\phi_h$ is fixed.
However, we saw that the Lorentzian phase space contains a family of solutions with varying $\phi_h$.
As such, quantum mechanically, the Hilbert spaces $\mathcal{H}^+$ and $\mathcal{H}^-$ will admit operators which modify the value of $\phi_h$.
Our non-perturbative completion via the Euclidean path integral is not enough to describe the S-matrix elements of such operators due to the superselection effect in Euclidean signature.
We will comment further on possible extensions of our analysis to incorporate these operators, but for now we focus on operators which preserve the value of $\phi_h$ and act only on the gauge field sector of the Hilbert spaces.

With this subtlety in mind, to obtain the non-perturbative matrix elements of $\mathcal{S}$, we write Euclidean path integral preparations for the desired incoming and outgoing states generated with operators like $\mathcal{O}_\text{in}$ or $\mathcal{O}_\text{out}$.
We join the Euclidean half-disk generating the incoming state to the Euclidean half-disk generating the outgoing state by using the canonical quantum operator $\mathcal{S}$, which acts on the Hilbert space of CJ gravity on an interval Cauchy slice.
All together, this is a Euclidean path integral with some pairwise boundary operator insertions, and can be expressed purely as a boundary condition for the Euclidean path integral together with some canonical matrix elements of the operators $\mathcal{S}$ and $\mathcal{O}$.
Subsequently, instead of using the disk approximation to evaluate the Euclidean overlap, we use the matrix model density of states.
The matrix elements of $\mathcal{S}$ and any creation/annihilation operators $\mathcal{O}$ are kept fixed in this process: only the density of states is corrected by the matrix model.\footnote{Note that this procedure is directly inspired by the prescription for computing non-perturbative correlation functions in JT gravity \cite{Saad:2019pqd,Iliesiu:2021ari}.}

We leave a detailed analysis of the non-perturbative CJ gravity S-matrix for the future.

\subsection*{Black holes and celestial holography}

In asymptotically AdS spacetimes, the AdS/CFT correspondence allows for an explicit quantum mechanical description of black hole microstates.
The microstates, which are non-perturbatively defined as the high-energy eigenstates of the CFT Hamiltonian, have a coarse-grained density which matches the expectation for that quantity implied by the Bekenstein-Hawking entropy formula.
The fact that black hole microstates can be thought of as energy eigenstates is due partially to their stability in AdS spacetimes.
A black hole which is large enough in AdS is thermodynamically stable and can equilibrate with its own Hawking radiation.

In flat space, black hole thermodynamics is quite different.
Flat space black holes cannot equilibrate with their own radiation, no matter how large or small they are.
As such, a putative holographic dual to asymptotically flat quantum gravity must supply a description of the microstates of flat black holes which does not make reference to energy eigenstates or other quantities with trivial time evolution.
Finding a precise description of flat space black hole microstates may be quite difficult without a general theory of how to non-perturbatively complete flat space string theory, but a more modest question is the following: given a non-perturbative completion, how can we construct a black hole?

In AdS/CFT, a very simple black hole with a known non-perturbative description is the eternal two-sided AdS-Schwarzschild black hole.
This geometry is dual to the thermofield double entangled state between two holographic CFTs.
In flat space, because we have only asymptotic Hilbert spaces and a preferred map between them in the S-matrix, we can try to use these ingredients to construct a flat space black hole in a similar manner.

The two-sided flat space Schwarzschild black hole is unstable, but understanding its construction may lead to more insights about flat space black holes.
The only gauge-invariant data we can specify is a state in the past Hilbert space $\mathcal{H}^-$, as this will be mapped by the S-matrix to a state in $\mathcal{H}^+$.
Our goal is to choose a state in $\mathcal{H}^-$ which will pass through the S-matrix in such a way that the geometric semiclassical description of this scattering process includes a bulk Cauchy slice that matches e.g. the time-reflection symmetric slice in the two-sided Schwarzschild black hole geometry.
Importantly, we do not assume anything about the geometry at very early or very late times in the scattering, as we expect any semiclassical description of a flat space black hole to break down in these regimes.\footnote{Even in AdS spacetimes where black holes are stable, the semiclassical description can break down at late times due to non-perturbative effects \cite{Iliesiu:2021ari}.}

In CJ gravity, we discussed the fact that the half-circle boundary condition for the Euclidean path integral creates the Hartle-Hawking state in e.g. $\mathcal{H}^-$. The norm of this state gives the CJ partition function, and therefore it is essentially analogous to the thermofield double state but created with the Bondi Hamiltonian.
We also noted that the half-disk path integral is joined, in the saddle-point approximation, to the horizon in the two-sided CJ black hole spacetime.
The disk computed using the past Bondi-dilaton frame was joined to one Lorentzian horizon, and the disk computed using the future Bondi-dilaton frame was joined to the other horizon.
From this, we concluded that the Hilbert space $\mathcal{H}^-$ ought to be identified with the union of null surfaces $\mathscr{I}^+_R \cup \mathscr{I}^-_L$, and similarly for $\mathcal{H}^+$.
This mixed past-future null surface suggests that, if we identify the Bondi Hamiltonian thermofield double as the relevant state to create a two-sided flat space black hole, we must act with a ``one-sided'' S-matrix to recover a complete past formulation of the black hole.

In CJ gravity, because of the low-dimensional nature of the theory, the Hilbert space $\mathcal{H}^-$ does not factorize between the two null surfaces whose asymptotic states it describes.
So, it is not actually possible in that theory to formulate a ``purely incoming'' description which refers only to $\mathscr{I}^-_L \cup \mathscr{I}^-_R$.
In higher dimensions and in more rich theories, we expect that the principles of celestial holography imply the existence of two separate celestial theories, in which we must construct an entangled initial state to create the two-sided Schwarzschild black hole.
Let these theories be $L$ and $R$, with past and future Hilbert spaces $\mathcal{H}_L^\pm$ and $\mathcal{H}_R^\pm$ and S-matrices $S_L$ and $S_R$ respectively.
The facts we described in the previous paragraph point to a natural proposal: a two-sided Schwarzschild black hole is created with a thermofield double entangled state between $\mathcal{H}_R^+$ and $\mathcal{H}_L^-$:
\begin{equation}
    | \text{TFD} \rangle = \frac{1}{Z(\beta)} \sum_{n=0}^\infty e^{-\beta E_n/2} \ket{n}_L^- \otimes \ket{n}_R^+ \ .
\end{equation}
The state $\ket{n}$ is an eigenstate of the Bondi Hamiltonian.
To obtain a purely incoming description of the black hole, we act with the matrix $S_R^\dagger$:
\begin{equation}
    | \text{Schwarzschild} \rangle = \frac{1}{Z(\beta)} \sum_{n=0}^\infty e^{-\beta E_n/2} \ket{n}_L^- \otimes S_R^\dagger \ket{n}_R^+ \ .
\end{equation}
In this way, our analysis of CJ gravity suggests that flat space black holes may be non-perturbatively described using a state with a rather complicated entanglement structure.
As in AdS, the entanglement is manifested in the geometric connection between the two sides of the black hole, but the purely incoming description of the state requires acting with a one-sided S-matrix, and this complicates the structure of the state while preserving the entanglement entropy.

\noindent \paragraph{Acknowledgements:} We thank Panos Betzios, Aidan Chatwin-Davies, Laura Donnay, Victor Godet, Clifford Johnson, Dominik Neuenfeld, Ana-Maria Raclariu, Romain Ruzziconi, and Clara Weill for discussions.  AK and LL are supported by the Simons Foundation via the It from Qubit Collaboration. CM and FR acknowledge support from NSERC. FR is also supported in part by the Simons Foundation.

\appendix
\addtocontents{toc}{\protect\setcounter{tocdepth}{1}}

\section{Variational problem of CJ gravity}
\label{zapp:1}

The aim of this Appendix is to analyze the variational problem of the CJ gravity action (\ref{eq:CJaction}) when subject to the asymptotic boundary conditions in (\ref{metricfalloff2}-\ref{dilatonfalloff2}). To do this, let us first compute the variation of the action in full generality. While the variation of the topological term (\ref{eq:CJterms}) vanishes, for the bulk contribution one finds
\begin{equation}
\begin{aligned}
\delta I_{\rm bulk} & =\frac{1}{2}
\int d^2x\sqrt{-g}
\Big[ 
R\delta \Phi+2(1-\varepsilon^{\mu \nu}\partial_\mu A_\nu)\delta \Psi
+2\varepsilon^{\mu \nu}(\partial_\mu \Psi)\delta A_\nu+\\
& \hspace{145 pt}
-\Psi g_{\mu \nu}\delta g^{\mu \nu}
+\Phi g^{\mu \nu}\delta R_{\mu \nu}-\nabla_\mu \left(2 \Psi\varepsilon^{\mu \nu} \delta A_\nu\right)
\Big]
\ .
\end{aligned}
\end{equation}
where we have used $\delta \sqrt{-g}=- \sqrt{-g}g_{\mu \nu}\delta g^{\mu \nu}/2$ and that in two dimensions the Ricci tensor satisfies ${R_{\mu \nu}=Rg_{\mu \nu}/2}$. The contribution $\delta R_{\mu \nu}$ can be worked out using Palatini's identity together with the explicit expression of the connection $\Gamma^{\mu}_{\nu \rho}$ in terms of the metric
\begin{equation}\label{eq:140}
\begin{aligned}
\delta \Gamma^{\mu}_{\alpha \beta}& =
-\frac{1}{2}\left[ 
g_{\rho \alpha}\nabla_\beta
+g_{\rho \beta}\nabla_\alpha
-g_{\alpha \beta}\nabla_\rho
\right] \delta g^{\mu \rho}\ ,\\[3pt]
\delta R_{\mu \nu}& =
\nabla_\rho \delta \Gamma^{\rho}_{\mu \nu}-\nabla_\mu \delta 
\Gamma^\rho_{\nu \rho }\ , \\[3pt]
g^{\mu \nu}\delta R_{\mu \nu}& =
\nabla_\rho \left[\left(g_{\mu \nu}g^{\alpha \rho }-\delta^\alpha_\mu \delta^\rho_\nu\right)
\nabla_\alpha \delta g^{\mu \nu}\right]
\equiv \nabla_\rho B^\rho\ ,
\end{aligned}
\end{equation}
where we have conveniently defined the vector $B^\mu$. Using these identities we can write the variation of the action explicitly and rearrange it as
\begin{equation}
\begin{aligned}
\delta I_{\rm bulk} & =\frac{1}{2}
\int d^2x\sqrt{-g}
\Big[ 
R\delta \Phi+2(1-\varepsilon^{\mu \nu}\partial_\mu A_\nu)\delta \Psi
+2\varepsilon^{\mu \nu}(\partial_\mu \Psi)\delta A_\nu+\\
& \hspace{112 pt}
-\delta g^{\mu \nu}\left(
\nabla_\mu \nabla_\nu \Phi
-g_{\mu \nu} \nabla^2 \Phi
+\Psi g_{\mu \nu}
\right)
+\nabla_\mu \Theta^\mu 
\Big]
\ .
\end{aligned} \label{Svariation}
\end{equation}
where we have define $\Theta^\mu$ as
\begin{equation}
    \Theta^\mu=
\Phi B^\mu
-(g_{\alpha \beta }g^{ \mu \rho}-\delta^\mu_\alpha \delta^\rho_\beta ) (\nabla_\rho\Phi)  
\delta g^{\alpha \beta}
-2 \Psi\varepsilon^{\mu \nu} \delta A_\nu\ , \label{boundaryterm}
\end{equation}
which can be integrated into a boundary term. Enforcing the vanishing of the bulk terms in (\ref{Svariation}) for arbitrary variations, results in the equations of motion (\ref{eq:eom}). However, the on-shell variation of the bulk action does not vanish due to the boundary term controlled by $\Theta^\mu$. We want to rewrite the boundary contribution (\ref{boundaryterm}) in a nicer way that allows one to better understand each of its contributions. 

To do this, let us first recall some basic notions about non-null hypersurfaces, assuming the boundary is specified by a constraint $f(x^\mu)=0$, implicitly meaning $f(x^\mu)$ is non-zero away from the boundary. Since $\partial_\mu f$ is non-zero at any point on the boundary we can define a normal vector $n^\mu$ as
\begin{equation}\label{eq:89}
n_\mu =\frac{(\partial_\mu f)}{[(\partial_\alpha f)(\partial^\alpha f)]^{1/2}}\ ,
\end{equation}
with $n^2=g^{\mu \nu}n_\mu n_\nu=1$ corresponding to a timelike boundary. While a priori the normal vector is only defined at the boundary, since $n^2=1$ is the correct normalization for the tangent vector to an affinely parametrized timelike or spacelike geodesic, we can extend $n^\mu$ by considering the geodesics which emanate from the boundary. This means the normal vector satisfies the following two identities
\begin{equation}\label{eq:120}
n^\beta \nabla_\alpha n_\beta=0\ ,
\qquad \qquad
n^\alpha \nabla_\alpha n_\beta=0\ .
\end{equation}
While the first one follows from $n^2=1$, the second is nothing more than the geodesic equation. Let us now define a projector $\gamma_{\mu \nu}$ to the boundary manifold
\begin{equation}\label{eq:88}
    \gamma_{\mu \nu}=g_{\mu \nu}-n_\mu n_\nu
    \qquad \Longrightarrow \qquad
    n^\mu \gamma_{\mu \nu}=0\ ,
    \qquad 
    t^\mu \gamma_{\mu \nu}=t_\nu \ ,
\end{equation}
where $t^\mu$ is a vector tangent to the boundary. This projector is very much related to the induced metric $h_{ab}$
\begin{equation}
h_{ab}=\frac{\partial x^\mu }{\partial y^a}
\frac{\partial x^\nu }{\partial y^b}g_{\mu \nu}=
\frac{\partial x^\mu }{\partial y^a}
\frac{\partial x^\nu }{\partial y^b}\gamma_{\mu \nu}\ ,
\end{equation}
where $y^a$ are coordinates on the boundary manifold. The extrinsic curvature of the boundary is defined as $K^{\mu \nu}=\gamma^{\mu \alpha}\gamma^{\nu\beta}\nabla_\alpha n_\beta$, whose trace can be written as $K=\nabla_\alpha n^\alpha$ (we have used ${n^\beta\nabla_\alpha n_\beta=0}$). Extending the normal vector in terms of geodesics, we can further write $K_{\mu \nu}=\nabla_\mu n_\nu$.

Using all this, we can rewrite the boundary term that originates from (\ref{boundaryterm}) in terms of the variation of the extrinsic curvature. To compute $\delta K$ we first derive the following useful relations
\begin{equation}\label{eq:141}
\delta n_\mu=-\frac{1}{2}
n_\mu n_\alpha n_\beta \delta g^{\alpha \beta}\ ,
\qquad \quad
\delta n^\mu =
\frac{1}{2}(\delta^\mu_\alpha+\gamma^{\mu}_{\,\,\,\alpha})n_\beta \delta g^{\alpha \beta}\ .
\end{equation}
Using this, one can compute the variation of $K$ and obtain
\begin{equation}
\delta K
=-
\frac{1}{2}n_\mu B^\mu
+\frac{1}{2}
K_{\mu \nu}
\delta g^{\mu \nu}+
\frac{1}{2}\nabla_\mu (\gamma^{\mu}_{\,\,\,\alpha} n_\beta \delta g^{\alpha \beta})\ ,
\end{equation}
where note the appearance of $B^\mu$ previously defined in (\ref{eq:140}). This expression is quite useful for expressing the boundary term $n_\mu \Theta^\mu$ that arises from integrating (\ref{boundaryterm}) as
\begin{equation}
n_\mu\Theta^\mu=
-2\Phi \delta K
+
\left[ \Phi K_{\mu \nu}
-(n^\alpha\nabla_\alpha\Phi)
\gamma_{\mu \nu}
\right]\delta g^{\mu \nu}
-2 \Psi n_\mu \varepsilon^{\mu \nu} \delta A_\nu
+D_\mu c^\mu\ ,
\end{equation}
where we have defined $c^\mu=\Phi \gamma^{\mu}_{\,\,\, \alpha} n_\beta \delta g^{\alpha \beta}$, which is a vector tangent to the boundary, i.e. $n_\mu c^\mu=0$. The covariant derivative on the boundary can be written as $D_\mu c^\nu=\gamma_{\,\,\,\mu}^\alpha \gamma_{\,\,\,\beta}^\nu \nabla_\alpha c^\beta$, so that using (\ref{eq:120}) we get $\nabla_\mu c^\mu=D_\mu c^\mu$. Thus, we arrive at the final expression for the variation of the bulk action
\begin{equation}
\begin{aligned}
\delta I_{\rm bulk}=
({\rm EOM})+\frac{1}{2}
\int_{\partial \mathcal{M}} dy\sqrt{-h}
\Big[
\big( \Phi K_{\mu \nu} 
-(n^\alpha\nabla_\alpha\Phi)
\gamma_{\mu \nu}
\big)  \delta g^{\mu \nu}
-2\Phi \delta K
-2 \Psi (n_\mu \varepsilon^{\mu \nu} \delta A_\nu)
+D_\mu c^\mu
\Big]
\end{aligned}
\end{equation}
where $({\rm EOM})$ correspond to the bulk terms in (\ref{Svariation}) that vanish when the equations of motion (\ref{eq:eom}) are satisfied. Since $D_\mu$ is the covariant derivative along the boundary, the term $D_\mu c^\mu$ can be integrated by parts and dropped, assuming there is no variation of the fields at the boundary of $\partial \mathcal{M}$, which in two dimensions corresponds to a collection of points.

Apart from the bulk term, the total CJ gravity action (\ref{eq:CJaction}) includes a boundary term given in (\ref{eq:CJterms}). When computing its variation, it is useful to remind ourselves the following identities which relate the variations of $g_{\mu \nu}$, $\gamma_{\mu \nu}$ and $h_{ab}$ in the following way
\begin{equation}\label{eq:92}
\gamma_{\mu \nu}\delta g^{\mu \nu}=\gamma_{\mu \nu}\delta \gamma^{\mu \nu}=h_{ab}\delta h^{ab}\ .
\end{equation}
The first equality can be shown using (\ref{eq:88}), while the second is obtained by choosing coordinates $x^\mu=(r,y^a)$ such that the boundary is at fixed $r$. Using this, the variation of the boundary term (\ref{eq:CJterms}) becomes
\begin{equation}
\begin{aligned}
\delta I_{\partial} & =
\frac{1}{2}
\int_{\partial \mathcal{M}}
dy\sqrt{-h}\Big[ 
2\Phi \delta K
+(2K-n^\alpha \nabla_\alpha)\delta \Phi
+\\[4pt]
& \hspace{90pt} -
\frac{1}{2}
\Big(
2\Phi K_{\mu \nu}
-\gamma_{\mu \nu}n^\alpha \nabla_\alpha \Phi
+n_\mu(\delta^\alpha_\nu+\gamma_{\,\,\,\nu}^{\alpha})\nabla_\alpha \Phi 
\Big)\delta g^{\mu \nu}
\Big]\ ,
\end{aligned}
\end{equation}
where we have used that in two dimensions $K_{\mu \nu}=\gamma_{\mu \nu}K$ as well as (\ref{eq:141}). Putting everything together, we arrive at the following expression for the variation of the CJ gravity action (\ref{eq:CJaction})
\begin{equation}\label{eq:CJ-variation}
\begin{aligned}
\delta I_{\rm CJ}&=({\rm EOM})+
\frac{1}{2}
\int_{\partial \mathcal{M}} dy\sqrt{-h}
\Big[
(2K-n^\alpha \nabla_\alpha)\delta \Phi
-(2 \Psi n_\mu \varepsilon^{\mu \nu}) \delta A_\nu + \\[4pt]
& \hspace{140pt} -
\frac{1}{2}
\Big(
\gamma_{\mu \nu}
(n^\alpha\nabla_\alpha\Phi)+
n_\mu(\delta^\alpha_\nu+\gamma_{\,\,\,\nu}^{\alpha})\nabla_\alpha \Phi 
\Big)\delta g^{\mu \nu}
\Big]\ .
\end{aligned}
\end{equation}

To have a well-defined variational principle, this expression must vanish when evaluated on solutions (\ref{LorentzianSol}) to the equations of motion. While the bulk terms go to zero by construction, the boundary contributions do not, given that variations of the fields, such as $\delta g_{\mu \nu}$, have no reason to vanish when evaluated at the boundary. To fix this, we constraint ourselves to variations which satisfy the asymptotic behavior specified in (\ref{metricfalloff2}-\ref{dilatonfalloff2}). Fixing the radial coordinate to some constant $r=r_0$, evaluating the coefficients of (\ref{eq:CJ-variation}) on the solution (\ref{LorentzianSol}) and evaluating the variations of the fields using (\ref{metricfalloff2}-\ref{dilatonfalloff2}), one finds the boundary terms vanish as $r=r_0\rightarrow  \infty$.\footnote{The subleading variations of the metric must satisfy $\delta g_{rr}=\mathcal{O}(1/r^3)$ and $\delta g_{ur}=\mathcal{O}(1/r^2)$, which is not obviously implied by (\ref{metricfalloff2}), unless one assumes the subleading corrections in (\ref{metricfalloff2}) must also be in the Bondi gauge, which means $\delta g_{rr}=\delta g_{ur}=0$.} This ensures the variational problem for CJ gravity is well defined.

\section{Path integral measure from BF formulation}
\label{zapp:3}

The central aim of this Appendix is to present a bulk computation of the measure used in computing the bulk Euclidean path integral of CJ gravity. We follow the same approach as in Section 3.3 of \cite{Saad:2019lba}, where the corresponding measure for JT gravity was derived using its formulation as a BF gauge theory.

We start by rewriting the bulk action of CJ gravity in first order formulation, where instead of working with the metric $g_{\mu \nu}$ one consider the one-form frame $e^a=e^a_\mu dx^\mu$ so that $g_{\mu \nu}=e^a_\mu e^b_\nu \delta_{ab}$. The Latin indices $a,b=0,1$ are raised and lowered with $\delta^a_b$. Apart from the zweibeins $e^a$ one has the spin connection $w_{ab}$, which in two dimensions is entirely determined by a single one-form component $w$ according to $w^{a}_{\,\,\,b}=\epsilon^a_{\,\,\,b} w$ with $\epsilon_{01}=1$. In terms of these quantities, the torsion $T^a$ and curvature tensor $R_{ab}$ are written as
\begin{equation}
T^a=de^a+w^{a}_{\,\,\,b}\wedge e^b\ ,
\qquad \qquad
R^a_{\,\,\,b}=\frac{1}{2}R^a_{\,\,\,bcd}e^c\wedge e^d=
dw^a_{\,\,\,b}+w^a_{\,\,\,c}\wedge w^c_{\,\,\,b}\stackrel{(2d)}{=}dw^a_{\,\,\,b}\ .
\end{equation}
Since we are ultimately interested in gravitational theories with vanishing torsion, the $T^a=0$ condition provides two equations that determine the spin connection $w_{ab}$ in terms of $e^a$. Putting everything together, we can write the bulk action (\ref{eq:CJaction}) of Euclidean CJ gravity in the first order formalism as
\begin{equation}\label{eq:117}
\begin{aligned}
I_{\rm bulk} & =-\frac{1}{2}\int_{\mathcal{M}}
d^2x\sqrt{g}(\Phi R+2\Psi-2\Psi \varepsilon^{\mu \nu}\partial_\mu A_\nu)
\\
& = 
-\int_{\mathcal{M}}
\left[ 
\eta_a(de^a+\epsilon^{a}_{\,\,\,b}w\wedge e^b)
-\Psi(dA-e^0\wedge e^1)
+\Phi dw 
\right]\ ,
\end{aligned}
\end{equation}
where we have used $e^0\wedge e^1=\sqrt{g}dx^0\wedge dx^1$ and $dw=\frac{R}{2} \sqrt{g} dx^0\wedge dx^1$.
The last two terms give the rewriting of the bulk action, while for the first one we have introduced the additional fields $\eta_a=(\eta_0,\eta_1)$ which act as Lagrange multipliers enforcing the vanishing of the torsion. All in all, the field content in the first order formalism is given by four scalars $(\eta_a,\Phi,\Psi)$ and four one-forms $(e^a,w,A)$.

One can show the action (\ref{eq:117}) is equivalent to a BF gauge theory \cite{Cangemi:1992bj}. To define a BF theory, we consider a Lie group $G$ with an associated Lie algebra $\mathfrak{g}$, so that an arbitrary group element can be obtained through the exponential map $\mathfrak{g}\rightarrow \exp(\mathfrak{g})$. Given the generators $T_A$ with $A=1,\dots,\dim(\mathfrak{g})$, we define the dynamical fields of the BF theory as $\boldsymbol{B}$ and $\boldsymbol{A}$, respectively given by a space-time scalar and a one-form, both valued in the algebra. Under the group action, these fields transform as
\begin{equation}
\boldsymbol{B}  \longrightarrow G^{-1}\boldsymbol{B}G\ ,
\qquad \qquad
\boldsymbol{A}  \longrightarrow  G^{-1}(d+\boldsymbol{A})G\ ,
\end{equation}
which means $\boldsymbol{B}$ transforms in the adjoint representation. From the gauge connection $\boldsymbol{A}$ one obtains the two-form field strength defined in the usual way $\boldsymbol{F}=d\boldsymbol{A}+\boldsymbol{A}\wedge \boldsymbol{A}$, which transforms in the adjoint representation $\boldsymbol{F}\rightarrow G^{-1}\boldsymbol{F}G$. The bulk term appearing in the action of a BF theory is most easily written by introducing the bilinear form ${\rm Tr}(T_AT_B)\equiv h_{AB}$, where $h_{AB}$ is obtained from the quadratic Casimir $C_2=h^{AB}T_AT_B$ satisfying $[C_2,T_A]=0$. The BF action is then simply defined by integrating the gauge invariant two-form ${\rm Tr}(\boldsymbol{B}\boldsymbol{F})$ over a manifold $\mathcal{M}$. 

To recover (\ref{eq:117}) we consider the central extension of the two-dimensional Euclidean Poincare group ${\rm ISO}(2)$, usually called the Maxwell group. Its algebra contains four bosonic generators $(P_a,J,Q)$ with the following non-vanishing commutators and quadratic Casimir
\begin{equation}\label{eq:142}
[P_a,J]=\epsilon_{a}^{\,\,\,b}P_b\ ,
\qquad [P_a,P_b]=-\epsilon_{ab}Q
\qquad \Longrightarrow \qquad
C_2=P_aP^a+2QJ\ ,
\end{equation}
where $Q$ is the central charge $[Q,\mathfrak{g}]=0$. Expanding the scalar and gauge connection in terms of the algebra generators 
\begin{equation}
i\boldsymbol{B}=\eta^a P_a-\Psi J+\Phi Q\ ,
\qquad \qquad
\boldsymbol{A}=e^aP_a+wJ+AQ\  ,
\end{equation}
one can easily show the matching of the bulk actions
\begin{equation}
-i\int_{\mathcal{M}}{\rm Tr}(\boldsymbol{B}\boldsymbol{F})=
-\frac{1}{2}\int_{\mathcal{M}}
d^2x\sqrt{g}(\Phi R+2\Psi-2\Psi \varepsilon^{\mu \nu}\partial_\mu A_\nu)=I_{\rm bulk}\ .
\end{equation}
To appropriately capture the gravitational dynamics at the boundary of the manifold we add an additional boundary term \cite{Saad:2019lba}, so that the full BF action is given by
\begin{equation}
I_{\rm BF}[\boldsymbol{B},\boldsymbol{A}]=
-i\int_{\mathcal{M}}{\rm Tr}(\boldsymbol{B}\boldsymbol{F})+\frac{i}{2}\int_{\partial \mathcal{M}}{\rm Tr}(\boldsymbol{B}\boldsymbol{A})\ .
\end{equation}
Assuming the boundary is parameterized by a $\beta$-periodic coordinate $\tau$, a well defined variational problem can be formulated with the following boundary condition $(\boldsymbol{B}+i\gamma \boldsymbol{A}_\tau)\big|_{\partial \mathcal{M}}=0$, where $\gamma$ is the same dimensionfull constant used in the main text. The equations of motion derived in this way are given by
\begin{equation}\label{eq:121}
\delta I_{\rm BF}[\boldsymbol{B},\boldsymbol{A}]=0
\qquad \Longrightarrow \qquad
\boldsymbol{F}=0\ ,
\qquad
d\boldsymbol{B}+[\boldsymbol{A},\boldsymbol{B}]=0\ ,
\end{equation}
which can be shown to be equivalent to the original CJ gravity equations of motion. Further details on the behavior of the fields at the asymptotic boundary that one must impose to match with gravity are discussed below.

\subsection{Symplectic form}

Consider the partition function of the BF gauge theory, which can be written as
\begin{equation}\label{eq:123}
Z=\int \mathcal{D}\boldsymbol{A}\mathcal{D}\boldsymbol{B}\,
e^{-I_{\rm BF}[\boldsymbol{B},\boldsymbol{A}]}=
\int \mathcal{D}\boldsymbol{A}
\delta(\boldsymbol{F})\,
e^{-I_\partial[\boldsymbol{A}]}\ .
\end{equation}
The integral over $\boldsymbol{B}$ was trivially solved, giving rise to the Dirac delta which constraints the remaining path integral over flat connections $\boldsymbol{A}$. This path integral was studied in \cite{Witten:1991we}, where it was shown that by appropriately gauge fixing using the Fadeed-Popov method the resulting measure is the one obtained from the symplectic form in the space of gauge connections
\begin{equation}\label{eq:124}
\Omega(\delta_1\boldsymbol{A},\delta_2\boldsymbol{A})=c_0\int_{\mathcal{M}} {\rm Tr}(\delta_1\boldsymbol{A} \wedge\delta_2 \boldsymbol{A})\ ,
\end{equation}
with $c_0$ an arbitrary proportionality constant. Here, $\delta_i\boldsymbol{A}$ are one-forms in the space of flat connections, i.e. variations of the gauge connection which preserve $\boldsymbol{F}=0$ to first order. The symplectic form (\ref{eq:124}) gives an explicit measure for the path integral (\ref{eq:123}), indirectly providing a measure for CJ gravity.

One needs to compute the symplectic form for two different cases: the asymptotic boundary and the gluing of two half-cylinders. Below we consider each of these cases separately.

\subsubsection*{Cylinder partition function from gluing half-cylinders}

The Euclidean equations of motion for the metric $R=0$ admit a solution with cylindrical topology, i.e. the flat cylinder obtained by an identification of the plane
\begin{equation}\label{eq:131}
    ds^2=d\rho^2+b^2d\varphi ^2 \ , 
    \qquad \qquad \varphi \sim \varphi +1\ ,
\end{equation}
where $b$ is the proper circumference. The general solution for the dilaton (\ref{vacsolution}) is consistent with such a quotient only when $\Lambda=0$ and two of the integration constants are adjusted to remove the time dependence of the solution yielding
\begin{equation}
    \Phi= \gamma \, c \,\rho+ \phi_0 \ .
\end{equation}
where $c, \phi_0$ integration constants.

The Euclidean version of the asymptotic boundary conditions discussed in Section \ref{SectionLorentz} are $\Phi|_{\partial {\cal M}}=\frac{1}{\epsilon}$ and $\frac{1}{\sqrt{g}}\partial_\rho \Phi|_{\partial {\cal M}} = \frac{\gamma}{\beta}$. The second condition concisely summarizes the thermal boundary conditions, if we recall that the Bondi boundary time flow is generated by  $\partial_\tau \overset{\rho\to \infty}{\longrightarrow} \gamma^{-1}\varepsilon^{\mu\nu} \partial_\mu \Phi \partial_\nu = \frac{c}{b}\,\partial_{\varphi}$ and, hence, fixing the Bondi temperature to $\beta$ via $\tau \sim \tau + \beta$ which implies $c=\frac{b}{\beta}$. Since $c$ is a dynamical variable on the cylinder, the proper circumference $b$ of a cylinder at fixed Bondi temperature $\beta$ is a fluctuating variable that is also the circumference of the asymptotic boundary ---in contrast to the more familiar JT gravity story in AdS. 

It is obvious from the dilaton solution that we cannot satisfy these conditions on both asymptotic boundaries of an infinite length cylinder. There is, therefore, no classical saddle of CJ theory with cylindrical topology. This situation is familiar from the JT gravity case and the treatment is known. We use the constrained instanton method, whereby we construct a ``half-cylinder'', i.e. a cylinder with only one asymptotic boundary satisfying our boundary conditions and another one at $\rho_c=0$ with $\Phi=\phi_0$. Then we take a pair of those ``half-cylinders'' ---possibly with different Bondi temperatures--- and glue them together at a fixed value of $\rho=0$, matching their dilaton values and their proper lengths $b$ and integrating over it with the appropriate volume measure. Note that the resulting dilaton configuration solves the equations of motion locally everywhere, except at the interface where the discontinuity in $\partial_\rho \Phi$ implies a source term. 

Fluctuations of Euclidean CJ gravity about such a constrained instanton are described again by a pair of boundary modes $f(\tau)$ and $g(\tau)$ with an action that can be derived in the same way as (\ref{Lboundaryaction}) in the main text and reads:
\begin{equation}
    I_{\partial}^{\rm cylinder}= \int_0^\beta d\tau \left( T_0 f'(\tau)^2 -g'(\tau)\frac{f''(\tau)}{f'(\tau)}\right)\ ,
\end{equation}
where $T_0= \frac{b^2}{2\beta^2}$ corresponds to the constant mode appearing in the Euclidean metric (\ref{eq:136}). Note that the parameter $T_0$ appearing in the action depends on the proper circumference of the cylinder which can have arbitrary values for the same Bondi temperature $\beta$.

When gluing the two half-cylinders there are actually two moduli one has to integrate over: the length of the gluing geodesic $b$ and the relative twist $\vartheta\in[0,b]$ associated to rotations along the symmetry axis of one of the half-cylinders. Let us now compute the symplectic form (\ref{eq:124}) which determines the integration measure along $(b,\vartheta)$. Following \cite{Saad:2019lba}, the gluing of the half-cylinders with a twist is described by the following metric
\begin{equation}\label{eq:125}
ds^2=d\rho^2+(bd\varphi+\vartheta \delta(\rho)d\rho)^2\ ,
\end{equation}
which for $\vartheta=0$ becomes (\ref{eq:131}). To see $\vartheta\neq 0$ corresponds to a twist one notes the coordinate $\tilde{\varphi}=b\varphi+\vartheta \Theta(\rho)$ gives the usual cylinder metric again, the difference being that $\tilde{\varphi}$ is not a continuous coordinate in the manifold but jumps at $\rho=0$. 

The gauge connection $\boldsymbol{A}$ associated to (\ref{eq:125}) can be easily computed and written as
\begin{equation}
\boldsymbol{A}=d\rho P_0+(bd\varphi+\vartheta \delta(\rho)d\rho)P_1+AQ\ ,
\end{equation}
where we have not written the component $A$ explicitly, given that we shall find it does not contribute to the symplectic form. The variation of the connection with respect to the moduli is given by
\begin{equation}
\delta \boldsymbol{A}=
(\delta b d\varphi+\delta \vartheta \delta(\rho)d\rho)P_1+\delta AQ\ .
\end{equation}
Using this we can easily write the symplectic form (\ref{eq:124}), noting the contribution from $\delta A$ drops out
\begin{equation}\label{eq:130}
\Omega(\delta_1\mathcal{A},\delta_2\mathcal{A}) =c_0
\int_\mathcal{M}
\left[ 
\delta_1 b
\delta_2 \vartheta
-\delta_1 \vartheta
\delta_2 b
\right]
\delta(\rho)
d\varphi\wedge d\rho
\qquad \Longrightarrow \qquad
\Omega=c_0 \delta b \wedge \delta \vartheta\ , 
\end{equation}
where the Dirac delta localizes the integral over $\mathcal{M}$ on the gluing interface at $\rho=0$. This simple result for the symplectic form implies the measure in the moduli $(b,\vartheta)$ is flat, so that the cylinder partition function is obtained by gluing two half-cylinders in the following way 
\begin{equation}\label{eq:129}
Z_{\rm cylinder}(\beta_1,\beta_2)=
c_0
\int_0^{\infty}db\int_0^b d\vartheta 
Z_{\rm half \textendash cyl}(\beta_1,b,\vartheta)
Z_{\rm half \textendash cyl}(\beta_1,b,\vartheta)\ .
\end{equation}
This is the measure assumed in \cite{Godet:2020xpk} when computing the cylinder partition function of CJ gravity. Note the result is completely analogous to the one obtained for JT gravity \cite{Saad:2019lba}.

\subsubsection*{Asymptotic boundary}

To compute the symplectic form associated to the degrees of freedom at the asymptotic boundary we must further constraint the asymptotic value of the gauge connection $\boldsymbol{A}$. Our guide for doing so is the solution to the metric and gauge field equations of motions of CJ gravity 
\begin{equation}\label{eq:136}
ds^2=2(P(\tau)r+T(\tau))d\tau^2+2id\tau dr\ ,
\qquad \qquad
A=rd\tau\ .
\end{equation}
In the BF formulation, we shall denote the gauge connection associated to a particular solution of this form as $\boldsymbol{a}[P(\tau),T(\tau)]$, given by
\begin{equation}\label{eq:134}
\boldsymbol{a}=
\frac{dr}{\sqrt{g_{\tau \tau}}}P_0
+\left[\frac{idr}{\sqrt{g_{\tau \tau}}}+\sqrt{g_{\tau \tau}}d\tau\right]P_1
-\frac{1}{2}
\big[d\tau
\partial_r g_{\tau \tau} 
+i(d\tau \partial_\tau+dr \partial_r)\ln(g_{\tau \tau})
\big]J
+rd\tau Q\ ,
\end{equation}
where $g_{\tau\tau}=2(P(\tau)r+T(\tau))$. We constraint ourselves to gauge connections which have this form at the asymptotic boundary located at large $r$. An arbitrary $\boldsymbol{A}$ is obtained by acting on $\boldsymbol{a}$ with a gauge transformation $G$
\begin{equation}\label{eq:135}
\boldsymbol{A}=G^{-1}(d+\boldsymbol{a})G\ .
\end{equation}
Of course, an arbitrary group element $G$ will not give a gauge connection that has the same form as (\ref{eq:134}). Writing $G=e^{\Theta}$ with $\Theta$ in the algebra, the allowed gauge transformations $\Theta$ are obtained by solving the constraint that results from expanding both sides of (\ref{eq:135}) to linear order
\begin{equation}
\delta \boldsymbol{A}=
d\Theta+[\boldsymbol{a},\Theta]=
\left[
\frac{\delta \boldsymbol{A}}{\delta \bar{P}}
\delta \bar{P}
+\frac{\delta \boldsymbol{A}}{\delta \bar{T}}
\delta \bar{T}
\right]_{(\bar{P},\bar{T})=(P,T)}\ ,
\end{equation}
where $\bar{P}(\tau)$ and $\bar{T}(\tau)$ are the functions appearing in $\boldsymbol{A}$ when written as (\ref{eq:134}). From this condition one can explicitly solve for $\Theta$, as well as $\delta \bar{P}$ and $\delta \bar{T}$. One finds the gauge transformation is parametrized by two arbitrary periodic functions $\varepsilon(\tau)$ and $\sigma(\tau)$ in the following way
\begin{equation}\label{eq:126}
\begin{aligned}
\Theta[\varepsilon(\tau),\sigma(\tau)&]=-
\left[ \frac{\varepsilon'(\tau)r-\sigma'(\tau)}{\sqrt{g_{\tau \tau}}}\right]P_0+
\left[\varepsilon(\tau)
\sqrt{g_{\tau \tau}}
-i\frac{\varepsilon'(\tau)r-\sigma'(\tau)}{\sqrt{g_{\tau \tau}}}\right]P_1
+\big[
\varepsilon(\tau)r-\sigma(\tau)
\big] Q \\[5pt]
& \hspace{15pt}
-\left[ 
\frac{1}{2}\varepsilon(\tau) \partial_r g_{\tau \tau}
+\frac{\varepsilon''(\tau)r-\sigma''(\tau)}{g_{\tau \tau}}+
\frac{i}{2}
\frac{\varepsilon(\tau)\partial_\tau g_{\tau \tau}
-
(\varepsilon'(\tau)r-\sigma'(\tau))\partial_r g_{\tau \tau}
}{g_{\tau \tau}}
\right]J    \ ,
\end{aligned}
\end{equation}
while the corresponding infinitesimal variations of $\bar{P}(\tau)$ and $\bar{T}(\tau)$ are
\begin{equation}\label{eq:128}
\begin{aligned}
\delta \bar{P} & = \varepsilon(\tau)P'(\tau)+\varepsilon'(\tau)P(\tau)-i \varepsilon''(\tau)\ , \\[4pt]
\delta \bar{T} & = \varepsilon(\tau)T'(\tau)+2\varepsilon'(\tau)T(\tau)+\sigma'(\tau)P(\tau)+i\sigma''(\tau)\ .
\end{aligned}
\end{equation}

One can use these expressions to compute the symplectic form. Since the variations $\delta_i\boldsymbol{A}$ correspond to infinitesimal gauge transformations, they can be written as
$\delta_i\boldsymbol{A}=d\Theta_i+[\boldsymbol{A},\Theta_i]$, which allows us to write the symplectic form as a boundary integral
\begin{equation}\label{eq:127}
\Omega(\delta_1\boldsymbol{A},\delta_2\boldsymbol{A})=c_0\int_{\partial \mathcal{M}} {\rm Tr}\big(
\Theta_1\wedge(d\Theta_2+[\boldsymbol{A},\Theta_2])
\big)\ .
\end{equation}
where we used that $(d\Theta_1+[\Theta_1,\boldsymbol{A}])\wedge \delta_2\boldsymbol{A}=d(\Theta_1\wedge\delta_2\boldsymbol{A})$ for gauge transformations which preserve ${\boldsymbol{F}=0}$. Using (\ref{eq:126}) and (\ref{eq:128}) this becomes
\begin{equation}
\begin{aligned}
\Omega(\delta_1\boldsymbol{A} ,\delta_2\boldsymbol{A}) & =c_0
\int_0^\beta d\tau \bigg[ 
T(\tau)
\big(
\varepsilon_1(\tau)\varepsilon_2'(\tau)-\varepsilon_1'(\tau)\varepsilon_2(\tau)
\big)
+P(\tau)
\big(
\varepsilon_1(\tau)\sigma_2'(\tau)-\sigma_1'(\tau)\varepsilon_2(\tau)
\big) \\[4pt]
& \hspace{220pt}
-i\big(
\varepsilon_1'(\tau)\sigma_2'(\tau)-\sigma_1'(\tau)\varepsilon_2'(\tau)
\big)
\bigg]\ .
\end{aligned}
\end{equation}
This expression becomes more transparent when written in terms of the one-forms $\delta(\,\cdot\,)$ acting on the space parametrized by the periodic functions $(\varepsilon(\tau),\sigma(\tau))$
\begin{equation}\label{eq:133}
\Omega=
c_0\int_0^\beta d\tau \bigg[ 
T(\tau)
\delta \varepsilon\wedge \delta\varepsilon'
+P(\tau)
\delta\varepsilon \wedge \delta\sigma'
-i\delta\varepsilon' \wedge \delta\sigma' 
\bigg]\ .
\end{equation}
This agrees with the symplectic form of the warped Virasoro computed from a group theoretic description in \cite{Afshar:2019tvp}. The partition function of CJ gravity computed in \cite{Godet:2020xpk} used the measure implied by this symplectic form.

\subsection{One loop computation}

We finish this Appendix by using the path integral measures obtain from the BF formulation to compute the disk and cylinder partition functions in the one loop approximation, which in this case turns out being exact due to the Duistermaat-Heckman theorem \cite{Duistermaat:1982vw,Stanford:2017thb,Afshar:2019tvp,Godet:2020xpk}. It is important that we carry out this computation in order to keep track of the relative proportionality constant between the disk and cylinder partition function. While this constant was arbitrarily fixed in \cite{Godet:2020xpk}, here it can be computed explicitly using the relation between the measures implied by (\ref{eq:130}) and (\ref{eq:133}).

As a first step, let us explicitly write the boundary action appearing in (\ref{eq:123}) using the boundary condition given above (\ref{eq:121})
\begin{equation}
I_\partial =\frac{\gamma}{2}\int_0^\beta {\rm Tr}\left[(d\Theta_\tau+e^{-\Theta}\boldsymbol{a}_\tau e^{\Theta})^2 \right]\ , 
\end{equation}
where $a_\tau$ and $\Theta$ respectively given in (\ref{eq:134}) and (\ref{eq:126}). To identify the saddles of this action, we first expand to linear order in $(\varepsilon,\sigma)$ and find
\begin{equation}
I_\partial =\gamma\int_0^\beta d\tau\, T(\tau)
-\gamma \int_0^\beta d\tau(\varepsilon(\tau)T'(\tau)+\sigma(\tau)P'(\tau))+\dots 
\end{equation}
This shows the saddles correspond to having constant values of the functions $(P(\tau),T(\tau))=(P_0,T_0)$,\footnote{Here and below, the notation $P_0$ refers to a fixed contstant, not to be confused with $P_0$ the element of the Maxwell algebra (\ref{eq:142}), which does not appear explicitly in the remainder of this Appendix.} so that the linear term in the expansion vanishes. From the metric in (\ref{eq:136}) one can identify particular values for these constants that correspond to the disk and half-cylinder topologies \cite{Godet:2020xpk}
\begin{equation}\label{eq:137}
{\rm Disk}: \quad (P_0,T_0)=\frac{2\pi}{\beta}(1,0)\ , 
\qquad \qquad
{\rm Half \textendash cylinder}: \quad (P_0,T_0)=\frac{b^2}{2\beta^2}(0,1)\ .  
\end{equation}
Keeping for the moment $(P_0,T_0)$ arbitrary, one can compute the contribution to the boundary action to the quadratic order and obtain
\begin{equation}
I_{\partial}^{(2)}[\varepsilon,\sigma]=\gamma\int_0^\beta d\tau \left[ 
T_0\varepsilon'(\tau)^2
+P_0\varepsilon'(\tau)\sigma'(\tau)
+i\varepsilon'(\tau)\sigma''(\tau)
\right]\ ,
\end{equation}
so that the one loop partition function is given by
\begin{equation}\label{eq:139}
Z_{P_0,T_0}=e^{-\gamma \beta T_0}\int \frac{d\varepsilon d\sigma}{{\rm Vol}(\mathcal{G}_0)} 
{\rm Pf}(\Omega)e^{-I_{\partial}^{(2)}[\varepsilon,\sigma]}\ ,
\end{equation}
where ${\rm Pf}(\Omega)$ the Pfaffian of the symplectic form (\ref{eq:133}). In the measure we have divided by the volume of $\mathcal{G}_0$ which corresponds to the degenerate directions of $\Omega$ that one should not integrate over. To compute the relevant integrals, it is convenient to expand the function $\varepsilon(\tau)$ and $\sigma(\tau)$ in a Fourier series as
\begin{equation}
\varepsilon(\tau)=\sum_{n\in \mathbb{Z}}\varepsilon_n e^{in(\frac{2\pi}{\beta})\tau}\ ,
\qquad \qquad
\sigma(\tau)=\sum_{n\in \mathbb{Z}}\sigma_n e^{in(\frac{2\pi}{\beta})\tau}\ .
\end{equation}
Enforcing these functions are real fixes half of the coefficients to $\varepsilon_n^\ast=\varepsilon_{-n}$ and $\sigma_n^\ast=\sigma_{-n}$, so that the independent complex directions are $(\varepsilon_n,\sigma_n)$ with $n\ge 0$. The symplectic form and quadratic boundary action become
\begin{equation}\label{eq:138}
\begin{aligned}
\Omega & = -2\pi i c_0 \sum_{n\ge 0}n\left[ 
2T_0\delta \varepsilon_n\wedge \delta \varepsilon_{n}^\ast
+\Big(P_0+\frac{2\pi}{\beta}n\Big)\delta\varepsilon_n\wedge \delta \sigma_{n}^\ast
+\Big(P_0-\frac{2\pi}{\beta}n\Big)\delta\sigma_n\wedge \delta \varepsilon_{n}^\ast
\right]\ , \\[4pt]
I_{\partial}^{(2)} &=
\gamma
\frac{(2\pi)^2}{\beta}
\sum_{n\ge 0}
n^2
\left[
2T_0
|\varepsilon_n|^2
+\Big(P_0+\frac{2\pi}{\beta}n\Big)\varepsilon_n\sigma_{n}^\ast
+\Big(P_0-\frac{2\pi}{\beta}n\Big)\sigma_{n}\varepsilon_{n}^\ast
\right]\ .
\end{aligned}
\end{equation}

\paragraph{Half-cylinder:} For the values of $(P_0,T_0)$ corresponding to the half-cylinder (\ref{eq:137}) we identify two degenerate directions of the symplectic form $\mathcal{G}_0=(\varepsilon_0,\sigma_0)$. Taking this into account one can compute the Pfaffian and find the measure of the integral is given by
\begin{equation}
\frac{d\varepsilon d\sigma}{{\rm Vol}(\mathcal{G}_0)}{\rm Pf}(\Omega)=
\prod_{n\ge 1}
d^2\varepsilon_n
d^2\sigma_n
\left[
\frac{c_0(2\pi  n)^2}{\beta}
\right]^2\ ,
\end{equation}
where $d^2\varepsilon_n=d\varepsilon_nd\varepsilon_n^\ast$. From this, it is straightforward to compute the Gaussian path integral that determines the one loop contribution and find
\begin{equation}
Z_{\rm half \textendash cylinder}(\beta,b)=
e^{-\frac{ \gamma b^2}{2\beta}}
\prod_{n\ge 1}\left(\frac{c_0\beta}{\gamma n}\right)^2=
\frac{\gamma}{2\pi c_0 \beta}
e^{-\frac{ \gamma b^2}{2\beta}}\ ,
\end{equation}
where the infinite product is regularized using the Riemann zeta function. Finally, we can glue the two half-cylinders (\ref{eq:129}) to obtain the cylinder partition function
\begin{equation}
Z_{\rm cylinder}(\beta_1,\beta_2)=
\frac{1}{\gamma(\beta_1+\beta_2)}\ ,
\end{equation}
where for convenience we have redefined the proportionality constant $c_0$ in terms of the arbitrary inverse length scale $\gamma$ as $c_0=\gamma^2/(2\pi)^2$. Note that in this way the partition function is dimensionless.

\paragraph{Disk:} The situation is slightly different for the disk, given that for the values of $(P_0,T_0)$ required in (\ref{eq:137}) there are two additional degenerate directions (\ref{eq:138}) in the symplectic form ${\mathcal{G}_0=(\varepsilon_0,\sigma_0,\varepsilon_{1}^\ast,\sigma_1)}$. Taking this into account, we can compute the Pfaffian and write the integration measure as
\begin{equation}
\frac{d\varepsilon d\sigma}{{\rm Vol}(\mathcal{G}_0)} 
{\rm Pf}(\Omega)=
d\varepsilon_1d\sigma^\ast_1
\left[
2c_0\frac{(2\pi)^2}{\beta}
\right]
\prod_{n\ge 2}d^2\varepsilon_n d^2\sigma_n
\left[c_0n\frac{(2\pi)^2}{\beta}\right]^2
(n^2-1)\ ,
\end{equation}
where the first term between square brackets is what remains from the quotient over $\mathcal{G}_0$. Solving the integrals, being particularly careful with the $\varepsilon_1$ and $\sigma_1^\ast$ contributions, one finds the following expression for the disk partition function
\begin{equation}
Z_{\rm disk}(\beta)=
\frac{4c_0\beta}{\gamma}
\prod_{n\ge 2}
\left(\frac{c_0\beta}{\gamma n}\right)^2
=
\frac{2}{\pi} 
\left(\frac{\gamma}{c_0\beta}\right)^2
=\frac{2(2\pi)^4}{\pi(\gamma\beta)^2}
\ ,
\end{equation}
where we have used Zeta regularization and $c_0=\gamma^2/(2\pi)^2$.

\section{Orthogonal polynomials and double scaling}
\label{zapp:4}

In this Appendix we describe the method of orthogonal polynomials. Its main advantage with respect to the loop equations is that it allows for explicit computation of observables beyond the $1/N$ perturbative expansion, see \cite{Bessis:1979is,Itzykson:1979fi,Bessis:1980ss} for early references and \cite{Eynard:2015aea} for a review. Using this formalism, the double scaling limit of the model that allows us to describe CJ gravity becomes much clearer and enables the computation of non-perturbative effects in $\hbar$.

\subsection{Finite $N$ analysis}

Let us start by considering the method of orthogonal polynomials for a finite $N$ Hermitian random matrix model, defined by a potential $V(M)$. One can define a set of polynomials $P_n(\lambda)$ labeled by $n\in \mathbb{N}_0$
\begin{equation}\label{eq:56}
P_n(\lambda)\equiv 
\frac{1}{\mathcal{Z}_n}
\prod_{j=1}^n
\int_{-\infty}^{+\infty}
d\lambda_j e^{-NV(\lambda_j)}
\Delta(\lambda_1,\dots,\lambda_n)^2
\prod_{i=1}^n(\lambda-\lambda_i)\ ,
\end{equation}
where $P_0(\lambda)=1$ and $\mathcal{Z}_n$ is given by the numerator in $P_n(\lambda)$ but without the $\prod_{i=1}^n(\lambda-\lambda_i)$ insertion in the integral. The normalization ensures $P_n(\lambda)=\lambda^n+\mathcal{O}(\lambda^{n-1})$ is a monic polynomial. From its definition one can show the polinomials satisfy the following orthogonality relation \cite{Itzykson:1979fi}
\begin{equation}\label{eq:32}
\int_{-\infty}^{+\infty}
d\lambda\,e^{-NV(\lambda)}
P_n(\lambda)
P_{m}(\lambda)=
h_n\delta_{n,m}\ ,
\qquad \qquad
h_n=\frac{1}{n+1}\frac{\mathcal{Z}_{n+1}}{\mathcal{Z}_n}\ ,
\end{equation}
where $h_n$ is the norm of $P_n(\lambda)$. To obtain a set of orthonormal functions on the real line, it is useful to define the functions $\psi_n(\lambda)$ as
\begin{equation}\label{eq:43}
\psi_n(\lambda)=
\frac{P_n(\lambda)}{\sqrt{h_n}}
e^{-\frac{N}{2} V(\lambda)}
\ ,
\qquad \qquad
\braket{n|m}\equiv
\int_{-\infty}^{+\infty}d\lambda\,
\psi_n(\lambda)
\psi_m(\lambda)
=\delta_{n,m}\ ,
\end{equation}
where the bra-ket notation is sometimes convenient to indicate the inner product between the vectors $\psi_n(\lambda)$ in the Hilbert space $\mathcal{H}=L^2(\mathbb{R})$. The multiplication by $\lambda$ is effectively an operator in this Hilbert space, whose action can be worked out as
\begin{equation}\label{eq:36}
\lambda \psi_n(\lambda)=
\sqrt{R_{n+1}}\psi_{n+1}(\lambda)
+\sqrt{R_n}\psi_{n-1}(\lambda)\ ,
\end{equation}
where $R_{n+1}=h_{n+1}/h_n> 0$ and $R_0=0$. To derive this, we have assumed the potential $V(\lambda)$ is even, which from (\ref{eq:56}) implies $P_n(-\lambda)=(-1)^nP_n(\lambda)$. 

\begin{figure}
\centering
\includegraphics[scale=0.65]{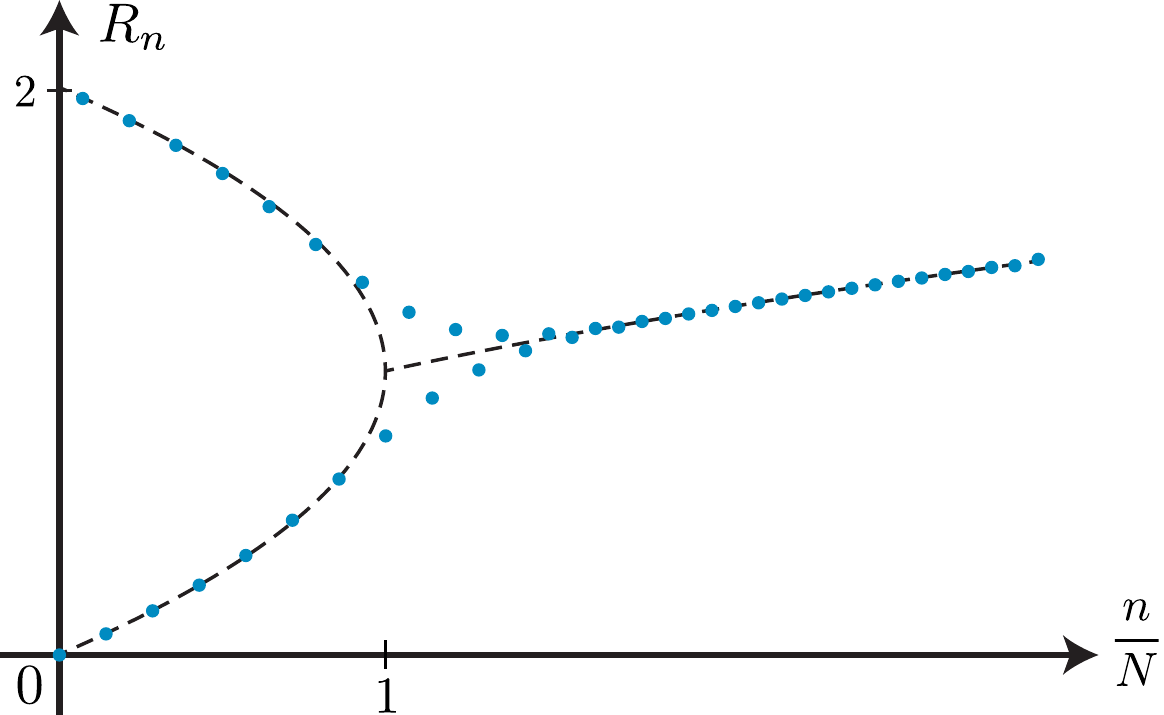}
\caption{Solution $R_n$ to the discrete string equation $\mathcal{S}$ in (\ref{eq:49}) for $N=14$. The dashed line corresponds to the leading solution given in (\ref{eq:51}).}\label{fig:3}
\end{figure}

As we shall see, the expectation value of all observables can be written in terms of the coefficients $R_n$. For a particular model these coefficients are determined from a recursion relation called the ``string equation"
\begin{equation}
    \int_{-\infty}^{+\infty}d\lambda\,
    \partial_\lambda\big(
    \psi_n(\lambda)
    \psi_{n-1}(\lambda)
    \big)=0
    \qquad \Longrightarrow \qquad
    \mathcal{S}\equiv \sqrt{R_n}\braket{n|V'(\lambda)|n-1}-\frac{n}{N}=0\ .
\end{equation}
Given any potential $V(\lambda)$ one can use (\ref{eq:36}) to write $\mathcal{S}$ explicitly. For instance, for the potential (\ref{eq:pot}) relevant for CJ gravity, one finds the following string equation
\begin{equation}\label{eq:49}
    \mathcal{S}=-2R_n+R_n\left(
    R_{n-1}+R_n+R_{n+1}\right)-\frac{n}{N}=0\ ,
\end{equation}
with initial conditions $R_0=0$ and $R_1=\int d\lambda\,\lambda^2e^{-NV(\lambda)}/\int d\lambda\,e^{-NV(\lambda)}$. In Figure \ref{fig:3} we solve the string equation for a fixed value of $N$. For small $n/N$, $R_n$ jumps between two different branches, which end up merging around $n/N\sim 1$. The leading behavior in the large $N$ limit was computed in~\cite{10.2307/121101}
\begin{equation}\label{eq:51}
R_n \simeq
\begin{cases}
\displaystyle
\,\,1-(-1)^n\sqrt{1-\frac{n}{N}}\ , \qquad \frac{ n}{N}\le 1\ , \\[8pt]
\displaystyle
\,\, 
\,\,\,\,\frac{1}{3}+
\frac{1}{3}\sqrt{1+\frac{3n}{N}}
\,\,\,\,\, \ , \qquad
\frac{n}{N}\ge 1\ ,
\end{cases}
\end{equation}
see dashed line in Figure \ref{fig:3}.

\paragraph{Computing observables:} We now show how the formalism can be used to compute the ensemble average of observables in the matrix model. The simplest case is the partition function $\mathcal{Z}$ of the matrix model (\ref{eq:2}), which can be easily obtained from (\ref{eq:32})
\begin{equation}\label{eq:33}
\mathcal{Z}=N!\prod_{n=0}^{N-1}h_n=
N!\,h_0^N\prod_{n=1}^{N-1}
R_n^{N-n}\ ,
\end{equation}
where $\mathcal{Z}_N=\mathcal{Z}$. Since $h_0$ can be fixed to one by a constant shift in the potential $V(\lambda)$, the partition function is entirely determined by $R_n$.

The main complication in computing more involved observables comes from the fact that the $N$ integrals over $\lambda_i$ are coupled due to the Vandermonde determinant in (\ref{eq:5}). This is where the orthogonal polynomials become useful, as the determinant can be rewritten as
\begin{equation}\label{eq:34}
    \Delta(\lambda_1,\dots,\lambda_N)^2=
    \det(\lambda_i^{j-1})^2=
    \det(P_{j-1}(\lambda_i))^2=
    \frac{\mathcal{Z}}{N!}
    \prod_{k=1}^{N}e^{NV(\lambda_k)}
    \det(\psi_{j-1}(\lambda_i))^2\ .
\end{equation}
In the second equality we have used the determinant is invariant under linear combinations of its columns to rewrite it directly in terms of the polynomials, while in the last one we expressed the determinant in terms of the orthonormal functions $\psi_n(\lambda)$, using (\ref{eq:33}). Using this, the expectation value (\ref{eq:5}) of an arbitrary observable becomes
\begin{equation}\label{eq:59}
\langle \mathcal{O} \rangle =
    \sum_{\sigma\in S_N}
    (-1)^{\pi_\sigma}
    \prod_{k=1}^N
    \int_{-\infty}^{+\infty}d\lambda_k
    \psi_{\sigma(k-1)}(\lambda_k)
    \mathcal{O}(\lambda_1,\dots,\lambda_N)
    \psi_{k-1}(\lambda_k)\ ,
\end{equation}
with $\sigma$ an element of the permutation group $S_N$ with parity $\pi_\sigma$. We have expanded both determinants in (\ref{eq:34}) and then relabeled the indices to get rid of one summation, which results in an additional factor of $N!$ (the order of $S_N$) cancelling the factor in (\ref{eq:34}). For the special case of single and double trace observables one obtains the following expressions
\begin{equation}\label{eq:41}
\begin{aligned}
& \langle {\rm Tr}\,F_1(M) \rangle    =
    \sum_{\sigma\in S_N}
    (-1)^{\pi_\sigma}
    \sum_{n=1}^N
    \braket{\sigma(n-1)|F_1(\lambda)|n-1}
    \prod_{k\neq n}
    \braket{\sigma(k-1)|k-1}
    =
    \sum_{n=0}^{N-1}\braket{n|F_1(\lambda)|n}\ , \\
 & \left\langle 
{\rm Tr}\,F_1(M)
{\rm Tr}\,F_2(M)
\right\rangle_c  =
\sum_{n=0}^{N-1}
\braket{n|F_1(\lambda)F_2(\lambda)|n}
-
\sum_{n,m=0}^{N-1}
\braket{n|F_1(\lambda)|m}
\braket{m|F_2(\lambda)|n}\ ,
\end{aligned}
\end{equation}
where $F_i(M)$ are arbitrary functions. Both quantites are determined by the matrix $\braket{n|F_i(\lambda)|m}$, which can be explicitly computed using (\ref{eq:36}) and written in terms of $R_n$. Both of these observables are determined by a single object, called the matrix model kernel
\begin{equation}\label{eq:44}
K(\lambda,\bar{\lambda})=\sum_{n=0}^{N-1}
\psi_n(\lambda)\psi_n(\bar{\lambda})
=
\sqrt{R_N}
\frac{\psi_{N}(\lambda)\psi_{N-1}(\bar{\lambda})-\psi_{N-1}(\lambda)\psi_N(\bar{\lambda})}{\lambda-\bar{\lambda}}\ .
\end{equation}
The second equality, sometimes called the Christoffel-Darboux formula, shows the kernel is integrable and can be derived using (\ref{eq:36}). In terms of the kernel, the expressions in (\ref{eq:41}) can be written as
\begin{equation}\label{eq:55}
\begin{aligned}
 \langle {\rm Tr}\,F(M)\rangle &=
\int_{-\infty}^{+\infty}d\lambda\,
K(\lambda,\lambda)F(\lambda)\ , \\
\left\langle  {\rm Tr}\,F(M) {\rm Tr}\,G(M) \right\rangle_c & = 
\int_{-\infty}^{+\infty}d\lambda d\bar{\lambda}
\left[
\delta(\lambda-\bar{\lambda})
-
K(\lambda,\bar{\lambda})
\right]K(\lambda,\bar{\lambda}) F(\lambda)G(\bar{\lambda})\ .
\end{aligned}
\end{equation}

Using $K(\lambda,\bar{\lambda})$ one can compute any observable, including the statistics of individual eigenvalues. Given an interval $I=(a,b)$ consider
\begin{equation}
\mathcal{E}_k(a,b)=
\binom{N}{k}
\frac{1}{\mathcal{Z}}
\prod_{n=1}^k \int_I d\lambda_n
\prod_{m=k+1}^N \int_{\mathbb{R}\setminus I} d\lambda_m
\Delta(\lambda_1,\dots,\lambda_N)^2
e^{-N\,{\rm Tr}\,V(M)}\  ,
\end{equation}
which gives the probability of finding exactly $k$ eigenvalues in $I$. The binomial prefactor is required to account for the relabeling of $k$ integration indices. By defining the following generating function
\begin{equation}
G(a,b;z)=
\big\langle \prod_{\ell=1}^N(1-z \chi_I^{\ell}) \big\rangle\ ,
\qquad {\rm where} \qquad
\chi_I^\ell=
\begin{cases}
\,\,\, 1 \,\,\,  \ , \,\, \lambda_\ell\in I=(a,b)\ , \\
\,\,\, 0 \,\,\,  \ , \,\, \lambda_\ell\notin I=(a,b)\ , \\
\end{cases}
\end{equation} 
one can easily derive
\begin{equation}
\mathcal{E}_k(a,b)=\left.
\frac{(-1)^k}{k!}\frac{\partial^k}{\partial z^k}
G(a,b;z)\right|_{z=1}\ .
\end{equation}
Using (\ref{eq:59}) the generating function can be written as the determinant of an $N$ dimensional matrix
\begin{equation}
G(a,b;z)=
\underset{0\le n,m \le N-1}{\det}
\left(
\delta_{n,m}
- z \braket{n|\chi_I|m}
\right)\ .
\end{equation}
To get a handle of this determinant in the large $N$ limit, it is convenient to define the integral operator $\widehat{K}_I(\,\cdot\,)$ acting on the space $\lbrace \psi_n(\lambda) \rbrace_{n=0}^{N-1}$ according to
\begin{equation}
    \widehat{K}_I(f)=\int_a^b d\bar{\lambda}f(\bar{\lambda})K(\bar{\lambda},\lambda)\ .
\end{equation}
The generating function is then given by $G(a,b;z)=\det({\rm Id}-z\widehat{K}_I)$, where ${\rm Id}$ is the identity operator in $\lbrace \psi_n(\lambda) \rbrace_{n=0}^{N-1}$. Putting everything together we arrive at the final expression for $\mathcal{E}_k(a,b)$
\begin{equation}\label{eq:61}
\mathcal{E}_k(a,b)=\left.
\frac{(-1)^k}{k!}\frac{\partial^k}{\partial z^k}
\det({\rm Id}-z\widehat{K}_I)\right|_{z=1}\ ,
\end{equation}
that will be very useful to study the non-perturbative completion of CJ gravity provided in this paper.

\subsection{Critical behavior and double scaling}

\begin{figure}
    \centering
    \includegraphics[scale=0.35]{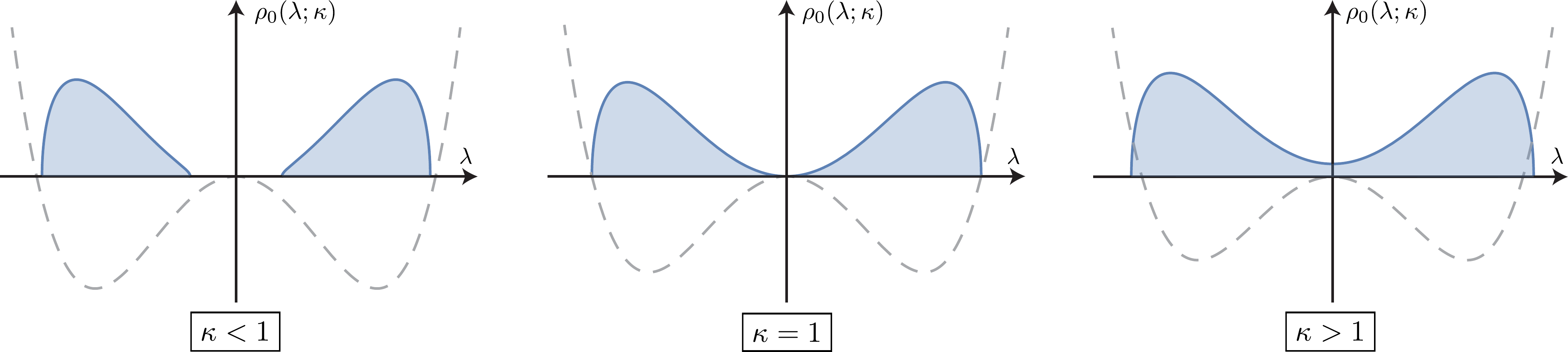}
    \caption{Leading spectral density associated to the potential (\ref{eq:42}), with $\kappa=1$ corresponding to the potential (\ref{eq:pot}) needed to describe CJ gravity. Depending on whether $\kappa$ is larger or small than one, the matrix model is in a single or double-cut phase.}
    \label{fig:2}
\end{figure}

We now go beyond the finite $N$ case and consider the double scaling limit. To explain what this is we must first introduce a notion of critical behavior. It is instructive to consider the following slight variation of the potential in (\ref{eq:pot})
\begin{equation}\label{eq:42}
V(\lambda;\kappa)=
\frac{1}{\kappa}
\left(-\lambda^2+\lambda^4/4\right) \ ,
\end{equation}
so that $\kappa=1$ corresponds to the case we are ultimately interested. The large $N$ spectral density (\ref{eq:15}) can be easily computed by requiring the resolvent (\ref{eq:14}) to have the appropriate large $z$ behavior $W_0(z)\simeq 1/z$
\begin{equation}
\rho_0(\lambda;\kappa)=
\frac{1}{2\pi \kappa}
\times
\begin{cases}
\displaystyle
\,\,
\left(\lambda^2+(m_0^2-4)/2\right)
\sqrt{m_0^2-\lambda^2}\ , \qquad \kappa \ge 1\ ,\\[4pt]
\displaystyle
\quad \,\,
|\lambda|
\sqrt{(m_+^2-\lambda^2)(\lambda^2-m_-^2)}
\,\,\,\,\, ,
\qquad \kappa\le 1\ ,
\end{cases}
\end{equation}
where $m_0^2=4\left(1+\sqrt{1+3\kappa} \right)/3$ and $m_\pm^2=2(1\pm \sqrt{\kappa})$. Depending on the value of $\kappa$ the matrix model is in a single or double-cut phase, see Figure \ref{fig:2}. Precisely at $\kappa=1$ there is a phase transition which signals the criticality of the model, that is entirely characterized by the behavior of the leading spectral density $\rho_0(\lambda)\sim \lambda^2$. The double scaling limit involves zooming in to the eigenvalues with $\lambda\sim 0$, meaning the details away from this region are non-universal and therefore irrelevant for the double scaled model. It is important to note there is an infinite class of critical potentials one could write down which produce the desired $\rho_0(\lambda)\sim \lambda^2$ behavior. For this reason, we should think of the potential in (\ref{eq:42}) as a representative of this class.\footnote{For this double scaled model, the independence on the details of the potential was rigorously shown in \cite{Claeys}.}

Fixing $\kappa=1$ we can take the double scaling limit of this model, first worked out in \cite{Douglas:1990xv,Crnkovic:1990mr} and then rigorously studied in \cite{10.2307/121101,Bleher:2002ys}. To do so, consider the following ansatz for the matrix model parameters
\begin{equation}\label{eq:52}
\frac{1}{N}=\frac{\hbar}{2t_2}\delta^{2+1}\ ,
\qquad \qquad
\frac{n}{N}=1+\frac{x}{t_2}\delta^2\ ,
\qquad \qquad
\lambda= \alpha \delta\ .
\end{equation}
where $\delta\rightarrow 0$ and $(\hbar,x,\alpha)$ are the scaling parameters associated to each of these quantities. The parameter $t_2$ is an additional constant parameter that is not strictly necessary but is convenient when doing the matching to gravity. Figuring the right power of $\delta$ in each case is a matter of trying different values until one gets a useful ansatz. When $\delta$ goes to zero, $N$ becomes large, $n/N$ approaches one (where the two branches in $R_n$ meet, see Figure \ref{fig:3}) and $\lambda$ approaches zero, where $\rho_0(\lambda)\sim \lambda^2$.

To take the double scaling limit of the string equation $\mathcal{S}$ in (\ref{eq:49}) we also need an ansatz for the coefficients $R_n$. Building on the numerical solution in Figure \ref{fig:3} and the leading behavior (\ref{eq:51}), let us consider
\begin{equation}\label{eq:54}
R_n=1-(-1)^nr(x)\delta+
\sum_{i=2}^{3}\left[
f_i(x)+(-1)^ng_i(x)
\right]\delta^i\ .
\end{equation}
The factors of $(-1)^n$ reproduce the small $n$ behavior shown in Figure \ref{fig:3}. While $r(x)$ controls leading scaling behavior, the functions $\big(f_i(x),g_i(x)\big)$ determine the subleading contributions with and without the $(-1)^n$ insertion. Using this together with (\ref{eq:52}) it is straightforward to expand the string equation (\ref{eq:49}) in a power series in $\delta$
\begin{equation}
\begin{aligned}
\mathcal{S}=
\left[
4f_2(x)-(x/t_2+r(x)^2)
\right]\delta^2
+&\bigg\lbrace
\left[
4f_3(x)+2g_2(x)r(x)
\right]
+\\[4pt]
&\left.+
\frac{(-1)^n}{2}
\left[
-4f_2(x)r(x)+\frac{1}{2} \hbar^2r''(x)
\right]
\right\rbrace
\delta^3
+\mathcal{O}(\delta^4)\ .
\end{aligned}
\end{equation}
Requiring $\mathcal{S}=0$ order by order, one can easily solve the algebraic equations for the functions $f_2(x)$ and $g_2(x)$. This is not the case for $r(x)$, which instead satisfies a differential equation
\begin{equation}\label{eq:72}
\lim_{\delta \rightarrow 0}
\frac{1}{\delta^3}
\mathcal{S}=0
\qquad \Longleftrightarrow \qquad
t_2\Big[
r(x)^3-\frac{1}{2}\hbar^2r''(x)
\Big]
+r(x)x
=0\ .
\end{equation}
The recursion relation for the coefficients $R_n$ becomes an ordinary differential equation for $r(x)$. The boundary conditions are obtained by matching with the leading solution (\ref{eq:51})
\begin{equation}\label{eq:65}
{\rm Boundary\,\,conditions:}
\qquad
\lim_{x\rightarrow -\infty}r(x)=\sqrt{-x/t_2}
\qquad \qquad
\lim_{x\rightarrow +\infty}r(x)=0\ .
\end{equation}

\paragraph{Computing observables:} To compute the expectation value of observables one needs to study the behavior of the $L^2(\mathbb{R})$ functions $\psi_n(\lambda)$ in the double scaling limit. Same as for the recursion coefficients $R_n$, we need an ansatz that distinguishes between even and odd $n$ 
\begin{equation}
\psi_{2n}(\lambda)=(-1)^n
\sqrt{\frac{\hbar}{2}}\varphi_+(x,\alpha)\ ,
\qquad \qquad
\psi_{2n+1}(\lambda)=
(-1)^{n}\sqrt{\frac{\hbar}{2}}\varphi_-(x+\hbar \delta/2,\alpha)\ .
\end{equation}
The countable vectors $\psi_n(\lambda)$ in the Hilbert space $\mathcal{H}$ are replaced by the uncountable $\varphi_s(x,\alpha)$ with $x\in \mathbb{R}$ and $s=\pm 1$. The orthonormality condition (\ref{eq:43}) becomes
\begin{equation}
\braket{s,x|x',s'}=
\int_{-\infty}^{+\infty}
d\alpha\,\varphi_s(x,\alpha)\varphi_{s'}(x',\alpha)=
\delta(x-x')\delta_{s,s'}\ ,
\end{equation}
while the multiplication by $\lambda$ (\ref{eq:36}) is now given by
\begin{equation}
\alpha \varphi_s(x,\alpha)=\left(-s\hbar\partial_x+r(x)\right)\varphi_{-s}(x,\alpha)\ .
\end{equation}
If we apply $\alpha^2$ instead, we obtain a very useful eigenvalue problem
\begin{equation}\label{eq:45}
\mathcal{H}_s\varphi_{s}(x,\alpha)=\alpha^2\varphi_{s}(x,\alpha)\ ,
\qquad {\rm where} \qquad
\mathcal{H}_s=-\hbar^2\partial_x^2+\left[r(x)^2-s\hbar r'(x) \right]
\end{equation}
that can be used to solve for $\varphi_s(x,\alpha)$. The matrix model kernel as given in (\ref{eq:44}) becomes
\begin{equation}\label{eq:47}
K(\alpha,\bar{\alpha})=
\int_{-\infty}^{0}dx\,
\sum_{s=\pm}
\varphi_s(x,\alpha)\varphi_s(x,\bar{\alpha})=
\hbar^2\sum_{s=\pm}
\frac{\varphi_s(x,\alpha)\overset{\leftrightarrow}{\partial_x}\varphi_s(x,\bar{\alpha})}{\alpha^2-\bar{\alpha}^2}\bigg|_{x=0}
\ ,
\end{equation}
where $\overset{\leftrightarrow}{\partial_x}=\overset{\rightarrow}{\partial_x}-\overset{\leftarrow}{\partial_x}$. The simple formulas in (\ref{eq:55}) are given by
\begin{equation}\label{eq:46}
\begin{aligned}
\langle {\rm Tr}\,F_1(M)\rangle & =
\int_{-\infty}^{+\infty}d\alpha
K(\alpha,\alpha)F_1(\alpha \delta)\ , \\
\left\langle  {\rm Tr}\,F_1(M) {\rm Tr}\,F_2(M) \right\rangle_c & = 
\int_{-\infty}^{+\infty}d\alpha d\bar{\alpha}
\left[
\delta(\alpha-\bar{\alpha})
-
K(\alpha,\bar{\alpha})
\right]K(\alpha,\bar{\alpha}) F_1(\alpha \delta)F_2(\bar{\alpha}\delta)\ .
\end{aligned}
\end{equation}
Note the factor of $\alpha \delta$ that remains on the right hand side after the change of integration variable $\lambda\rightarrow \alpha \delta$. This means that in the double scaled model one should work with observables that are functions of the rescaled matrix $\bar{M}=M/\delta$ instead of the bare matrix $M$.

Let us comment on $\mathcal{E}_k(a,b)$ in (\ref{eq:61}). For the finite $N$ matrix model, the operator $\widehat{K}_I(\,\cdot\,)$ acts on the finite dimensional space $\lbrace \psi_n(\lambda) \rbrace_{n=0}^{N-1}$. After double scaling limit the space becomes infinite dimensional $\lbrace \varphi(x,\alpha) \rbrace_{x \in \mathbb{R}}$, meaning the determinant in (\ref{eq:61}) is now a Fredholm determinant. Although in practice this might seem a very difficult quantity to compute, it has been recently shown to be tractable \cite{Johnson:2021zuo,Johnson:2022wsr}.

\addcontentsline{toc}{section}{References}
\bibliography{References}
\bibliographystyle{JHEP}

\end{document}